\documentclass[a4paper,11pt]{article}
\pdfoutput=1
\usepackage{jheppub}
\usepackage{amsmath}
\usepackage{amssymb}
\usepackage{color}
\usepackage{graphicx}
\usepackage{hyperref}
\usepackage{xspace,slashed}
\usepackage[utf8]{inputenc}
\usepackage{subcaption}
\usepackage{enumitem}
\usepackage{stackengine}
\usepackage{algorithm}
\usepackage{algorithmic}
\usepackage{soul}
\usepackage{booktabs}

\DeclareMathOperator{\sech}{sech}
\DeclareMathOperator{\csch}{csch}

\makeatletter
\g@addto@macro\bfseries{\boldmath}
\makeatother

\newcommand{\order}[1]{\mathcal{O}\left(#1\right)}
\newcommand{\ord}[1]{\mathcal{O}_{#1}}

\newcommand{\MeV}{\;\mathrm{MeV}}
\newcommand{\GeV}{\;\mathrm{GeV}}
\newcommand{\TeV}{\;\mathrm{TeV}}
\newcommand{\as}{\alpha_s}

\newcommand{\arccosh}{\operatorname{arccosh}}

\newcommand{\elll}{{\ell\ell}}

\newcommand{\DL}{\text{\sc dl}}
\newcommand{\LL}{\text{\sc ll}}
\newcommand{\CS}{\text{\sc cs}}

\newcommand{\Higgs}{\text{\sc h}}
\newcommand{\ptH}{p_{t,\Higgs}}
\newcommand{\yH}{y_{\Higgs}}
\newcommand{\mH}{m_{\Higgs}}
\newcommand{\cut}{\text{cut}}

\newcommand{\CBIH}{CBI$_\text{H}$\xspace}
\newcommand{\CBIHDY}{CBI$_\text{H,DY}$\xspace}
\newcommand{\CBIHR}{CBI$_\text{HR}$\xspace}
 
\newcommand{\ptll}{p_{t,\elll}}

\usepackage{xcolor}
\definecolor{light-gray}{gray}{0.8}

\definecolor{semiblue}{rgb}{0.3,0.3,0.8}
\newcommand{\logbook}[2]{}

\preprint{OUTP-21-16P}

\newcommand{\OXaff}{Rudolf Peierls Centre for Theoretical Physics,
  Clarendon Laboratory, Parks Road, Oxford OX1 3PU, UK}
\newcommand{\ASCaff}{All Souls College, Oxford OX1 4AL, UK}

\title{Cuts for two-body decays at colliders}
\author[a,b]{Gavin P.\ Salam}
\affiliation[a]{\OXaff}
\affiliation[b]{\ASCaff}
\author[a]{and Emma Slade}

\abstract{
  Fixed-order perturbative calculations of fiducial cross sections for
  two-body decay processes at colliders show disturbing sensitivity to
  unphysically low momentum scales and, in the case of
  $H\to \gamma \gamma$ in gluon fusion, poor convergence.
  Such problems have their origins in an interplay between the
  behaviour of standard experimental cuts at small transverse momenta
  ($p_t$) and logarithmic perturbative contributions.
  We illustrate how this interplay leads to a factorially divergent
  structure in the perturbative series that sets in already from the
  first orders.
  We propose simple modifications of fiducial cuts to eliminate their
  key incriminating characteristic, a linear dependence of the
  acceptance on the Higgs or $Z$-boson $p_t$, replacing it with
  quadratic dependence.
  This brings major improvements in the behaviour of the perturbative
  expansion.
  More elaborate cuts can achieve an acceptance that is independent of
  the Higgs $p_t$ at low $p_t$, with a variety of consequent
  advantages.
}

\begin{document}

\maketitle

\section{Introduction}

The starting point for almost any analysis at high-energy colliders is
a set of requirements, or ``cuts'', on the transverse momenta and
pseudorapidities of the objects that enter the analysis.
Long ago, it was pointed
out~\cite{Klasen:1995xe,Harris:1997hz,Frixione:1997ks} that the choice
of these cuts is delicate when studying final states that involve two
back-to-back objects.
Many collider analyses fall into this category: for example a dijet
system, a $t \bar t$ system, or the two-body decay of a resonance such
as a $Z$ or Higgs boson.
Refs.~\cite{Klasen:1995xe,Harris:1997hz,Frixione:1997ks} noted that
the common practice at the time, of applying identical minimum
thresholds on the transverse momenta of the two objects (``symmetric
cuts''), led to sensitivity to configurations with a small transverse
momentum imbalance between the two objects, where perturbative
calculations could be affected by enhanced (though integrable)
logarithms of the imbalance.
Ultimately, the discussions in those papers resulted in the widespread
adoption of so-called ``asymmetric'' cuts whereby one chooses
different transverse-momentum thresholds for the harder and softer of
the two jets.

In recent years, QCD calculations have made amazing strides in
accuracy (for a review, see Ref.~\cite{Heinrich:2020ybq}), reaching
N3LO precision for key $2\to1$ processes, both
inclusively~\cite{Anastasiou:2015vya,Anastasiou:2016cez,Duhr:2020seh,Duhr:2020sdp}
and differential in the rapidity~\cite{Dulat:2018bfe,Cieri:2018oms}
and in the full decay
kinematics~\cite{Chen:2021isd,Billis:2021ecs,Camarda:2021ict}.
As the calculations have moved forwards, an intriguing situation has
arisen in the context of gluon-fusion Higgs production
studies, where the calculations are arguably the most advanced.
For this process, inclusive cross sections and cross sections
differential in the Higgs boson rapidity show a perturbative series
that converges well at N3LO.
However, calculations for fiducial cross sections, which include
asymmetric experimental cuts on the photons from $H \to \gamma \gamma$
decays, show poorer convergence and significantly larger scale
uncertainties~\cite{Chen:2021isd,Billis:2021ecs}.
Furthermore, it turns out that to obtain the correct N3LO prediction,
it is necessary to integrate over Higgs boson transverse momenta that
are well below a GeV, which is physically unsettling (albeit
reminiscent of the early observations in
Ref.~\cite{Klasen:1995xe,Harris:1997hz,Frixione:1997ks}). 

Refs.~\cite{Billis:2021ecs,Alekhin:2021xcu} have noted
that such problems (which appear to be present to a lesser extent
also in the context of Drell-Yan studies) are connected with the fact
that both asymmetric and symmetric cuts yield an acceptance for
$H \to \gamma \gamma$ decays, $f(\ptH)$, that has a linear
dependence on the Higgs boson transverse momentum
$\ptH$~\cite{Ebert:2019zkb,Ebert:2020dfc}:
%
\begin{equation}
  \label{eq:f-f0-f1}
  f(\ptH) = f_0 + f_1 \cdot \frac{\ptH}{\mH} +
  \order{ \frac{\ptH^2}{\mH^2}}\,.
\end{equation}
In section~\ref{sec:existing-cuts}, concentrating on the
$H\to \gamma\gamma$ case, we will review how this linear dependence
arises and we will also examine its impact on the perturbative series
with a simple resummation-inspired toy model for its all-order
structure.
That model implies that any power-law dependence of the acceptance for
$\ptH \to 0$ results in a perturbative series for the fiducial cross
section that diverges $(-1)^n \as^n n!$, i.e.\ an alternating-sign
factorial divergence, coming predominantly from very low $\ptH$
values.

Factorial growth implies that, however small the value of $\as$, the
perturbative series will never converge.
Non-convergence of the series is a well known feature of QCD, notably
because of the same-sign factorial growth induced by infrared QCD
renormalons~\cite{Beneke:1998ui}.
In that context, the smallest term in the series is often taken as a
fundamental non-perturbative ambiguity.
The alternating-sign factorial growth that we see is different, in
that the sum of all terms can be made meaningful, with the help of
resummation.
However, fixed-order perturbative calculations still cannot reproduce that
sum to better than the smallest term in the series.
As is commonly done with infrared renormalon calculations, one can
express the size of the smallest term in the series as a power of
$(\Lambda / \mH)$, where $\Lambda\equiv \Lambda_\text{\sc qcd} \sim 0.2\GeV$ is the fundamental infrared
scale of QCD.
The power that emerges with standard $H\to \gamma\gamma$ fiducial cuts
is $\sim (\Lambda / \mH)^{0.2}$, i.e.\ no fixed-order calculation can
attain an accuracy better than that.%
\footnote{In contrast,
  the infrared renormalon-induced fundamental non-perturbative ambiguity for
  inclusive~\cite{Beneke:1995pq} and
  rapidity-differential~\cite{Dasgupta:1999zm} cross sections for
  colour-singlet objects at hadron colliders is widely believed to be
  $\Lambda^2 / m^2$ (though open questions still remain on this point~\cite{Beneke:1998ui}), and there are indications that the same might
  hold true for cross sections differential in colour-singlet
  kinematics~\cite{FerrarioRavasio:2020guj,Caola:2021kzt}.}
This provides a simple explanation for the significant uncertainties
seen recently in N3LO fiducial calculations for $H \to \gamma \gamma$
decays~\cite{Chen:2021isd,Billis:2021ecs}.
It is also related to the observation that perturbative calculations
for Drell-Yan fiducial cross sections appear to be particularly
sensitive to the details of the treatment of low $p_t$ values in the
fixed-order calculations~\cite{Alekhin:2021xcu}.
One approach to resolving this issue is to give up on the use of
fixed-order perturbation theory for Higgs boson fiducial cross
sections and other two-body processes, and instead calculate fiducial
cross sections using suitably matched resummed plus fixed order
calculations (an approach that was explored long ago for dijet
calculations~\cite{Banfi:2003jj} and advocated recently for the Higgs
case~\cite{Billis:2021ecs}).
We believe this to be a valid approach, and it is probably the only
robust option available for interpreting fiducial results measured
with today's widespread cut choices.
However, in general, it seems less than ideal to give up on the use of
pure fixed order calculations for predicting hard cross sections,
especially as fixed order calculations are conceptually simpler than
resummation, and in many respects more flexible (though if a
sufficiently accurate resummation can be achieved through a parton
shower~\cite{Dasgupta:2020fwr,Hamilton:2020rcu,Karlberg:2021kwr,Forshaw:2020wrq,Holguin:2020joq,Nagy:2020dvz,Nagy:2020rmk}
together with high-order matching, cf.\ the approaches of
Refs.~\cite{Monni:2019whf,Alioli:2021qbf,Prestel:2021vww}, this might
alleviate the flexibility issue).%
\footnote{ Another approach~\cite{Glazov:2020gza}, is essentially to
  give up on fiducial cross sections themselves and instead to
  ``defiducialise'' the cross sections, i.e.\ to divide out fiducial
  acceptances calculated perturbatively as a function of $p_{t}$ and
  then compare the results to more inclusive perturbative cross
  sections.
  The computational approaches that we develop here also lend
  themselves to use in the context of defiducialisation.
  This is further discussed in Appendix~\ref{sec:defid-remark}.
}

Here, instead, we take the approach of re-examining the cuts used to
select two-body final states.
We will argue (section~\ref{sec:good-simple-cuts}) that the
experiments should choose cuts designed to provide an acceptance that
depends at most quadratically on the net transverse momentum of the decaying
heavy boson (e.g.\ Higgs) and its two-body decay system.
The advantages of quadratic dependence have been noted recently also
in Ref.~\cite{Alekhin:2021xcu}.
One simple approach to achieving quadratic dependence is to replace a
cut on the transverse momentum of the harder photon (in the
$H\to \gamma\gamma$ case) with a cut on the scalar sum of the photon
transverse momenta.
Using the sum for cuts and/or binning in the context of perturbative
calculations was examined at least as early as
Refs.~\cite{Adloff:1998st,Carli:1998zr} in a dijet context, and has
seen sporadic use since, with evidence of improved perturbative
stability also in Refs.~\cite{Adloff:2000tq,Rubin:2010xp}.
An equally simple and arguably even better approach is to cut on the
product of the two photon transverse momenta.
Both sum and product cuts alleviate the most damaging part of the
factorial divergence in the perturbative series.
The underlying quadratic dependence remains robust, with some minor,
avoidable caveats, also in combination with rapidity cuts
(section~\ref{sec:rapidity-cuts}).

In section~\ref{sec:cbi-cuts}, we shall see that it is even possible
to design cuts such that the acceptance has no dependence at all on
$\ptH$ for small $\ptH$ (at least within an approximation where we can
neglect aspects related to photon isolation).
Choosing cuts where the acceptance depends little on $\ptH$ is
advantageous not only for fiducial measurements and calculations, but,
potentially, also for measurements where experiments quote results
in selected regions of Higgs phase space (independently of the
decays), for example STXS~\cite{Berger:2019wnu} cross sections, or for determining total
cross sections.
In particular, the less the acceptance depends on the kinematics of
the Higgs boson, the less reliant the experiments are on a knowledge
of the kinematic distributions (and their potential modifications by
BSM effects~\cite{Bishara:2016jga,Soreq:2016rae}) in order to quote a
final result.

In section~\ref{sec:Drell-Yan} we will give a brief discussion of the
Drell-Yan process, commenting notably on the additional
characteristics that arise associated with the spin degrees of freedom
of the decaying boson.
In general, acceptance effects have a reduced impact on the
perturbative series for resonant Drell-Yan production, owing to the
smaller $C_F$ rather than $C_A$ colour factor appearing in the $p_t$
resummation.
However, there is still a benefit to be had from judicious choices of
cuts, especially considering the high experimental accuracies of
Drell-Yan studies.

Finally, we will close with an overview of our findings and comments
on possible future work (section~\ref{sec:conclusions}).

\paragraph{Reading guide}
This manuscript contains material of varying degrees of technicality.
For readers who wish to familiarise themselves with the problems with
existing cuts, we hope that section~\ref{sec:existing-cuts} should be
relatively accessible, with the essence of our arguments to be found
in section~\ref{sec:symmetric-cuts}.
For readers interested in how we can assemble simple solutions to
those problems, most of section~\ref{sec:good-simple-cuts} should
likewise be accessible.
Section~\ref{sec:rapidity-cuts}, on rapidity cuts and the interplay
between different kinds of cuts, becomes more technical from
section~\ref{sec:two-rap-cut} onwards.
However, the illustration of the key findings, in
section~\ref{sec:worked-example}, does not require a detailed
knowledge of the derivations that preceded it.
The cuts of section~\ref{sec:cbi-cuts}, which deliver
$\ptH$-independent acceptance, are likewise somewhat more complex than
the early sections of the paper.
We hope that we have provided sufficient explanations that it is
possible to comfortably work through the section, but some readers may
at first prefer to concentrate on the results for the performance of
the cuts,
figures~\ref{fig:acceptances-EBI} and \ref{fig:CBIHR-performance}.
Finally, section \ref{sec:Drell-Yan}, on Drell-Yan production can, to
a large extent, be read independently of
sections~\ref{sec:rapidity-cuts} and \ref{sec:cbi-cuts}.

\section{Existing cut strategies and their perturbative implications}
\label{sec:existing-cuts}

\subsection{Symmetric cuts}
\label{sec:symmetric-cuts}
A simple case in which to illustrate the issue of symmetric cuts is
that of the decay of a Higgs boson to two photons, where one wishes to
evaluate the fiducial cross section, i.e.\ the cross section as
measured after accounting for the cuts on the photons.
One can approach the problem in three steps:
(a) parameterise the kinematics of the Higgs boson decay photons in
terms of the Higgs transverse momentum and of the polar and azimuthal
decay angles for the photons in the Higgs rest frame;
(b) integrate over the decay angles, to work out how the cuts on the
photons affect the acceptance, i.e.\ evaluate the fraction
$f(\ptH)$ of Higgs bosons that pass the cut, as a function of the
Higgs boson transverse momentum, $\ptH$;
(c) determine how the $\ptH$-dependence of the acceptance,
specifically its small-$\ptH$ limit, affects the structure of the
perturbative series.

Let us place the Higgs, of mass $\mH$, at zero rapidity,
$\yH =\frac12 \ln \frac{E+p_z}{E-p_z}=0$.
When the Higgs boson has transverse momentum $\ptH$, we can
parameterise the momenta of the two photons (labelled $+$ and $-$) as
a function of polar and azimuthal angles $\theta$ and $\phi$,
\begin{multline}
  \label{eq:photon-momenta}
  p_{\pm}(\ptH,\theta,\phi) = 
  \frac{1}{2}
  \left\{
    \pm \sqrt{\mH^2+\ptH^2} \sin\theta \cos \phi + \ptH  \,,\,
    \pm \mH \sin\theta \sin \phi \,,\,
    \pm \mH \cos  \theta\,,\, \right.\\\left.
    \sqrt{\mH^2+\ptH^2} \pm \ptH \sin\theta \cos  \phi
  \right\},
\end{multline}
where the components are given in the order $x$, $y$, $z$, $E$, the
beams are along the $\pm z$ directions and, without loss of
generality, we have taken the Higgs boson transverse momentum to be
along the $x$ direction.
In this parametrisation, $\theta$ and $\phi$ are simply the usual
Collins--Soper angles~\cite{Collins:1977iv}.
When discussing $p_t$ cuts, it is sufficient to consider the domain
\begin{equation}
  \label{eq:theta-phi-regions}
  0 \le \theta \le \frac\pi2\,, \qquad\qquad
  -\frac\pi2 \le \phi \le \frac\pi2\,,
\end{equation}
where we have $p_{t,+} \ge p_{t,-}$.
We will refer to the higher (lower)-$p_t$ photon as the harder
(softer) one. 
In this domain, an identical (``symmetric'') transverse momentum cut
on both photons, $p_{t,+}, p_{t,-} \ge p_{t,\cut}$, reduces to a
requirement on the softer photon, $p_{t,-} \ge p_{t,\cut}$.
For other regions of $\theta$ and $\phi$, the argument 
would remain identical, simply taking care as to which of the two
photons has the smaller transverse momentum.

\begin{figure}
  \centering
  \includegraphics[width=0.8\textwidth]{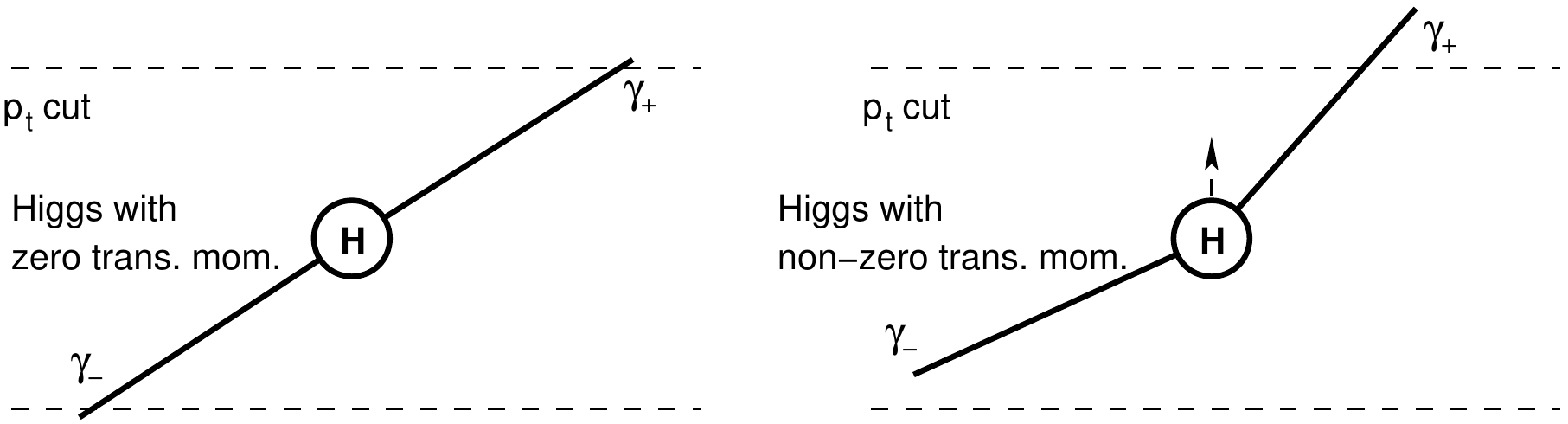}
  \caption{Illustration of an event where both Higgs decay photons
    just pass a symmetric $p_{t,\cut}$ when the Higgs boson is at rest
    (left), but where one of them no longer does when the Higgs boson
    is given a small transverse boost.
    This is the origin of the negative linear $\ptH$ dependence of the
    acceptance in Eq.~(\ref{eq:fsym-linear}).  }
  \label{fig:illustration}
\end{figure}

For a given $\ptH$, the fraction $f(\ptH)$ of Higgs boson decays
where both photons pass the cut is given by
\begin{equation}
  \label{eq:generic-acceptace}
  f(\ptH) =
  \int_{-\pi/2}^{\pi/2} \frac{d\phi}{\pi}
  \int_0^{\pi/2} \sin\theta d\theta\,
  \Theta(p_{t,-} > p_{t,\cut})\,.
\end{equation}
We can perform a simple integration over phase space,
independently of the Higgs production matrix element, because of the
spin-$0$ nature of the Higgs boson.
To evaluate $f(\ptH)$, it is convenient to work in the
small-$\ptH$ limit, where we have
\begin{equation}
\label{eq:pt-expansion}
  p_{t,\pm}(\ptH,\theta,\phi) =
  \frac{\mH}{2} \sin \theta 
  \pm\frac{1}{2}\ptH \cos \phi
  +\frac{\ptH^2}{4 \mH}  \left(\sin \theta \cos ^2\phi+\csc
      \theta \sin ^2\phi\right)
    + \ord3 
    \,,
\end{equation}
where the notation $\ord{n}$ is a shorthand that we introduce to
indicate that we neglect terms $\ptH^n$ and higher (and, later, the
$n^\text{th}$ power of any other factor in which we expand).
In Eq.~(\ref{eq:pt-expansion}), we have retained terms up to order
$\ptH^2/\mH^2$ because we will make use of the second-order term
later.
However, to keep the rest of this section as simple as possible, we
will now work with just the first two terms, and the requirement
$p_{t,-} > p_{t,\cut}$ translates to
\begin{equation}
  \label{eq:symcut-sintheta-expansion}
  \sin\theta >
  \frac{2 p_{t,\cut}}{\mH}
  + \cos\phi\, \frac{\ptH}{\mH}
  +\ord2
  \,,
\end{equation}
or equivalently
\begin{equation}
  \label{eq:symcut-costheta-expansion}
  \cos\theta <
  f_0
  - \frac{2}{f_0} \frac{p_{t,\cut}}{\mH} \cos\phi\, \frac{\ptH}{\mH}
  +\ord2
  \,,
  \qquad
  f_0 = \sqrt{1-\frac{4 p_{t,\cut}^2}{\mH^2}}\,.\quad
\end{equation}
Here $f_0$ is the acceptance for Born ($\ptH=0$) events.
The integral over $\theta$ in Eq.~(\ref{eq:generic-acceptace}) is
straightforwardly equal to the $\cos\theta$ limit of
Eq.~(\ref{eq:symcut-costheta-expansion}), while our use of a series
expansion makes it easy to carry out the $\phi$ integral.
Recall that the $\phi$ integral covers the range $-\pi/2<\phi<\pi/2$
so that $\cos\phi$ is always positive (extending to the full $\phi$
range one would need to replace $\cos\phi$ with $|\cos\phi|$ in the
equations given above).
The result for the acceptance with symmetric cuts is 
\begin{equation}
  \label{eq:fsym-linear}
  f^\text{sym}(\ptH) = f_0
  + f_1^\text{sym} \cdot \frac{\ptH}{\mH} +
  \ord2
  \,,\qquad
  f_1^\text{sym} = -\frac{4}{\pi f_0} \frac{p_{t,\cut}}{\mH}\,,
\end{equation}
with terms up to $(\ptH/\mH)^4$ given in
Appendix~\ref{sec:higher-order-expansions}.
The key feature of Eq.~(\ref{eq:fsym-linear}) is that for
small $\ptH$, the acceptance depends \textit{linearly} on $\ptH$,
with a negative slope.
This feature is easy to understand qualitatively, as illustrated in
Fig.~\ref{fig:illustration}: starting from a Higgs boson with zero
transverse momentum, where both photons have identical transverse
momenta, a small Higgs transverse boost increases the transverse
momentum of one of the photons by an amount of order $\ptH$, while the
other photon's transverse momentum is decreased by a corresponding
amount.
There is then a fraction of order $\ptH/\mH$ of decay orientations where both
photons would have passed the $p_{t,\cut}$ for a Higgs boson at rest,
but where one of the photons no longer passes the cut for the
transversely boosted Higgs boson.
For a cut $p_{t} > p_{t,\text{cut}} = 0.35 \mH$, one finds
$f_0 \simeq 0.714$ and $f_1^\text{sym} \simeq -0.624$,
while for a cut $p_{t} > p_{t,\text{cut}} = \mH/4$, one finds
$f_0 \simeq 0.866$ and $f_1^\text{sym} \simeq -0.368$.

One might expect that a linear dependence of the acceptance on the
Higgs boson transverse momentum should not be a major issue.
However it turns out to have drastic consequences for the behaviour of
calculations in perturbative QCD.
To understand why, consider the simplest possible approximation for
the distribution of $\ptH$, in which we take only the double
logarithmic (DL) terms $\as^n L^{2n-1}$, where $L = \ln \frac{\mH}{2\ptH}$,
\begin{subequations}
  \label{eq:2}
  \begin{align}
    \frac{d\sigma^\text{\sc dl}}{d\ptH}
    &=
      \frac{4 C_A \alpha_s L}{\pi \ptH} e^{-\frac{2C_A \alpha_s}{\pi} L^2} \sigma_\text{tot} \,,
    \\
    &=
      \sigma_\text{tot}\left[
      \delta(\ptH)
      +
      \sum_{n=1}^\infty \frac{(-1)^{n-1}}{(n-1)!}
      \left(\frac{2 C_A \alpha_s}{\pi}\right)^n
      \left(\frac{2 L^{2n-1}}{\ptH}\right)_{\!\!+}
      \right]
      \,,
  \label{eq:2b}
  \end{align}
\end{subequations}
with $\sigma_\text{tot}$ the total cross section.
In Eq.~(\ref{eq:2b}), we have included a plus prescription to account
for divergent virtual contributions at $\ptH=0$.

The fiducial cross section is given by the integral over the
differential cross section multiplied by the acceptance.
To understand how it behaves, we take the integration up to
$\ptH =\mH/2$, corresponding to $L=0$, 
\begin{subequations}
    \label{eq:sym-sigma-fid-explicit-series}
  \begin{align}
    \label{eq:sym-sigma-fid-explicit-series-integral}
    \sigma_{\text{fid,sym}}^\DL
    &= f^\text{sym}(0) \sigma_\text{tot}
    +  \int_{0}^{\mH/2} d\ptH \,
    \big[f^\text{sym}(\ptH) - f^\text{sym}(0)\big]
    \frac{d\sigma^\DL}{d\ptH},
    \\
    &=\left[f_0 + f_1^\text{sym} \sum_{n=1}^\infty (-1)^{n+1}
    \frac{(2n)!}{2(n!)}
    \left(\frac{2 C_A \alpha_s}{\pi}\right)^n
    + \ldots\right] \sigma_\text{tot}\,.
    \label{eq:sym-sigma-fid-explicit-series-result}
  \end{align}
\end{subequations}
On the first line, we have separated the result into the total cross
section multiplied by the Born ($\ptH=0$) acceptance, plus an integral
over $\ptH$ that accounts for the difference between the acceptance at
finite $\ptH$ and the Born acceptance (one could also view the first
line as the consequence of the plus prescription in Eq.~(\ref{eq:2b}),
however in general cases, such an interpretation involves ambiguities
that would complicate our subsequent interpretation of the contents of
the integral, cf.\ Appendix~\ref{sec:interpretations}).
On the second line, the sum multiplying $f_1$ can be evaluated in
closed form, but we have chosen to write it as a power series in the
coupling, as would arise in a fixed-order fiducial calculation.
For these terms, the underlying integrals, which can be expressed as
$\int_0^\infty L^{2n-1} e^{-L} dL = (2n-1)!$, generate a larger
factorial structure in the numerator than is inherited in the
denominator ($1/(n-1)!$) from the expansion of the exponential in
Eq.~(\ref{eq:2}).
Ultimately, the resulting $(2n)!/n!$ factor grows similarly to $n!$
for large $n$.

The appearance of factorial growth in QCD perturbative series is often
associated with infrared QCD renormalons~\cite{Beneke:1998ui} and a
fundamental ambiguity connected with the non-perturbative region.
Here, unlike those infrared renormalons, the terms alternate in sign
from one order to the next, and there is an unambiguous result for
their sum, which can be obtained by performing the integral in
Eq.~(\ref{eq:sym-sigma-fid-explicit-series}) with the resummed Higgs
$p_t$ distribution (technically, one would say that the series is
Borel resummable).
However, if one carries out an order-by-order calculation, the result
will start to diverge for
\logbook{}{$(2n-1)(2n)/n = 2(2n-1) = (2C_A\as/\pi^){-1}$, so $2n-1=\pi/(4C_A\as\pi)$. }
\begin{equation}
  \label{eq:n-smallest}
  n > n_\text{min} \simeq
  \frac{\pi}{8 \alpha_s C_A}+ \frac12\,.
\end{equation}
Using $C_A=3$, $\as(\mH/2) \simeq 0.125$, that translates to
$n\gtrsim 1.5$, as one can verify by examining the explicit series for
the terms proportional to $f_1$
\begin{subequations}
  \label{eq:illustrative-sym-expansion}
  \begin{align}
    \frac{\sigma_\text{fid,sym}^\DL}{f_0 \sigma_\text{tot}} -1 &= 
            \frac{f_1^\text{sym}}{f_0} \sum_{n=1}^\infty (-1)^{n+1} \frac{(2n)!}{2(n!)} 
             \left(\frac{2 C_A \alpha_s}{\pi}\right)^n
                                         + \cdots
    \\
  \label{eq:illustrative-sym-expansion-result}    
 &\simeq 
   \frac{f_1^\text{sym}}{f_0} \left(
   \underbrace{\,0.24\,}_{\as} 
   - \underbrace{\,0.34\,}_{\as^2}
   + \underbrace{\,0.82\,}_{\as^3}
   - \underbrace{\,2.73\,}_{\as^4}
   + \underbrace{\,11.72\,}_{\as^5}
   + \ldots
   \right)
   \simeq \frac{f_1^\text{sym}}{f_0} \times
   \underbrace{0.12}_\text{resummed}\,,
  \end{align}
\end{subequations}
where the right-hand answer is the resummed
result.
As expected, the series fails to converge from the first term (one
should keep in mind that the number for each term in the series
already includes the relevant power of $\as$).
Furthermore, no truncation of the series reproduces the all-order
result.

Even though the divergence that we see in
Eq.~(\ref{eq:sym-sigma-fid-explicit-series}) is of different origin
from infrared renormalons~\cite{Beneke:1998ui} (and with alternating
signs rather than the same-sign structure that appears for infrared
renormalons), one can express the size of the smallest term as a power
of $(\Lambda/\mH)$ and compare it to the $(\Lambda/\mH)^2$ infrared
renormalon expected for inclusive~\cite{Beneke:1995pq} and
rapidity-differential~\cite{Dasgupta:1999zm} cross sections for heavy
colour-singlets.%
\footnote{Recently, evidence has emerged suggesting that
  $(\Lambda/Q)^2$ scaling applies also for the $p_t$ distribution of a
  colour-singlet~\cite{FerrarioRavasio:2020guj,Caola:2021kzt} (where the singlet
  $p_t$ and mass are considered to be commensurate and of order $Q$).
  However, there still remain open questions in order to conclusively
  demonstrate the absence of $(\Lambda/\mH)$ corrections in general
  for Drell-Yan production~\cite{Beneke:1998ui}.
}
To do so, we use Stirling's approximation for the factorials,
Eq.~(\ref{eq:n-smallest}) to replace $(2C_A\as/\pi)$ with
$1/4n_{\min}$, and write the smallest term of the expansion in
Eq.~(\ref{eq:sym-sigma-fid-explicit-series}) as
\begin{equation}
  \label{eq:use-of-stirling}
  \left.  \frac{(2n)!}{n!}\left(\frac{2C_A\as}{\pi}\right)^n
  \right|_{n=n_{\min}}
  \sim
  \left.  2^{2n} \left(\frac{n}{e}\right)^n
    \left(\frac{1}{4n_{\min}}\right)^n
  \right|_{n=n_{\min}}
  \sim
  e^{-n_{\min}}.
\end{equation}
We have ignored numerical prefactors, as well as the additional
$+1/2$ in Eq.~(\ref{eq:n-smallest}), on the grounds that it only
contributes to the prefactor.
We then again invoke Eq.~(\ref{eq:n-smallest}), now to express
$n_{\min}$ in terms of $\as$ and use the 1-loop expression for $\as(Q)
= (2 b_0 \ln Q/\Lambda)^{-1}$ with $b_0 = (11C_A-2n_f)/12\pi$. 
This gives the following result for the size of the smallest term,
\logbook{dbcec96}{math-2021/power-of-min-term} \logbook{}{
  In general, we have terms of the form (ignoring square-root pre-factors)
  $f_n = (2n)!/2(n!) u^n
  \sim 1/2 (2n/e)^(2n) (e/n)^n u^n
  \sim 2^(2n-1) (n/e)^n u^n
  $
  The smallest term occurs for $4n u = 1$ or $n=1/(4u)$, i.e.\
  $f_{\min} = 2^(2n-1) (1/4u)^n e^{-n} u^n = e^{-n}/2 = e^{-1/4u} $.
  We have $u = 2C_A \as/\pi = 2C_A/\pi \cdot 1 / (2b_0 \ln Q/\Lambda)$
  i.e.\ $u = 12 C_A/(11 C_A - 2n_f) \cdot (1/\ln Q/\Lambda)$.
  
}
\begin{equation}
  \label{eq:effective-power-correction}
  \left(\frac{\Lambda}{Q}\right)^{\frac{(11 C_A - 2n_f)p^2}{48 C}}\,,
\end{equation}
where we have written the result generically for a process with a hard
scale $Q$, two
incoming legs having an average colour factor $C$ (e.g.\ $Q = m_H$, $C = C_A$ for
$gg\to H$ and $Q=m_Z$, $C=C_F = 4/3$ for $q\bar q \to Z$) and for acceptance
corrections that vanish as $p_t^p$.
For Higgs production with symmetric cuts, i.e.\ linear acceptance
corrections ($p=1$), using $n_f=5$, this goes as $(\Lambda/\mH)^{23/144}
\sim (\Lambda/\mH)^{0.160}$.
The specific powers that emerge from
Eq.~(\ref{eq:effective-power-correction}) are, in general, not the
same as those that will appear with the full QCD series.
This is because they can be altered by a variety of subleading
effects, a question that we return to briefly in
section~\ref{sec:further-pert-discussion} and
Appendix~\ref{sec:remarks-asysmptotics}. 
Still, despite this reservation, we believe that the general pattern
expressed by Eq.~(\ref{eq:effective-power-correction}) should be robust:
the problem caused by the factorial divergence is worse with a $C_A$
colour factor than $C_F$, and if one manages to replace linear $p_t$
dependence of the acceptance ($p=1$) with a higher power ($p=2$ or
higher), the situation will improve.
\logbook{}{See summary at the end of math-2021/bessel-games.nb}

\subsection{Asymmetric cuts}
\label{sec:asymmetric-cuts}

The concerns raised long ago about symmetric cuts have
led to the widespread adoption of so-called \emph{asymmetric} cuts,
where one places different cuts on the harder and softer of the two
decay products~\cite{Klasen:1995xe,Harris:1997hz,Frixione:1997ks}.
Let us require the harder photon to have $p_{t,+} > p_{t,\cut}$ and
the softer one to have $p_{t,-} > p_{t,\cut} - \Delta$.
We will work in an approximation where both $\ptH$ and $\Delta$ are
much smaller than $\mH$.
It is straightforward to show that $p_{t,+} - p_{t,-} \le \ptH$ (up
to second order in $\ptH/\mH$, this is evident from
Eq.~(\ref{eq:pt-expansion}), where $p_{t,+} - p_{t,-} =
\ptH \cos\phi + \ord3$).
Accordingly, for $\ptH < \Delta$, we know that
$p_{t,+} - p_{t,-} < \Delta$, and for any decay configuration where
the harder photon passes its cut $p_{t,+} > p_{t,\cut}$, the softer
photon automatically passes its cut too,
$p_{t,-} > p_{t,\cut} - \Delta$.
In this region, the evaluation of the $\ptH$ dependence of the
acceptance is a trivial repetition of the derivation in
section~\ref{sec:symmetric-cuts}, with the difference that we consider
$p_{t,+} > p_{t,\cut}$ instead of $p_{t,-} > p_{t,\cut}$.
At each stage, this simply changes the sign of terms linear in
$\ptH$, giving us
\begin{equation}
  \label{eq:fasym-linear-lt-Delta}
  f^\text{asym}(\ptH) = f_0
  + \left(\frac{\ptH}{\mH}\right)f_1^\text{asym} + \ord2
  \,,\qquad
  \ptH < \Delta \,,\quad
  f_1^\text{asym} = \frac{4}{\pi f_0} \frac{p_{t,\cut}}{\mH}\,,
\end{equation}
i.e.\ an acceptance that rises linearly with $\ptH$.
This linear rise occurs because, as $\ptH$ increases there is an
increasing range of decay orientations for which the transverse boost
of the Higgs boson can bring the harder photon above $p_{t,\cut}$,
while the cut on the softer photon is irrelevant.

\begin{figure}
  \centering
  \includegraphics[width=0.7\textwidth,page=1]{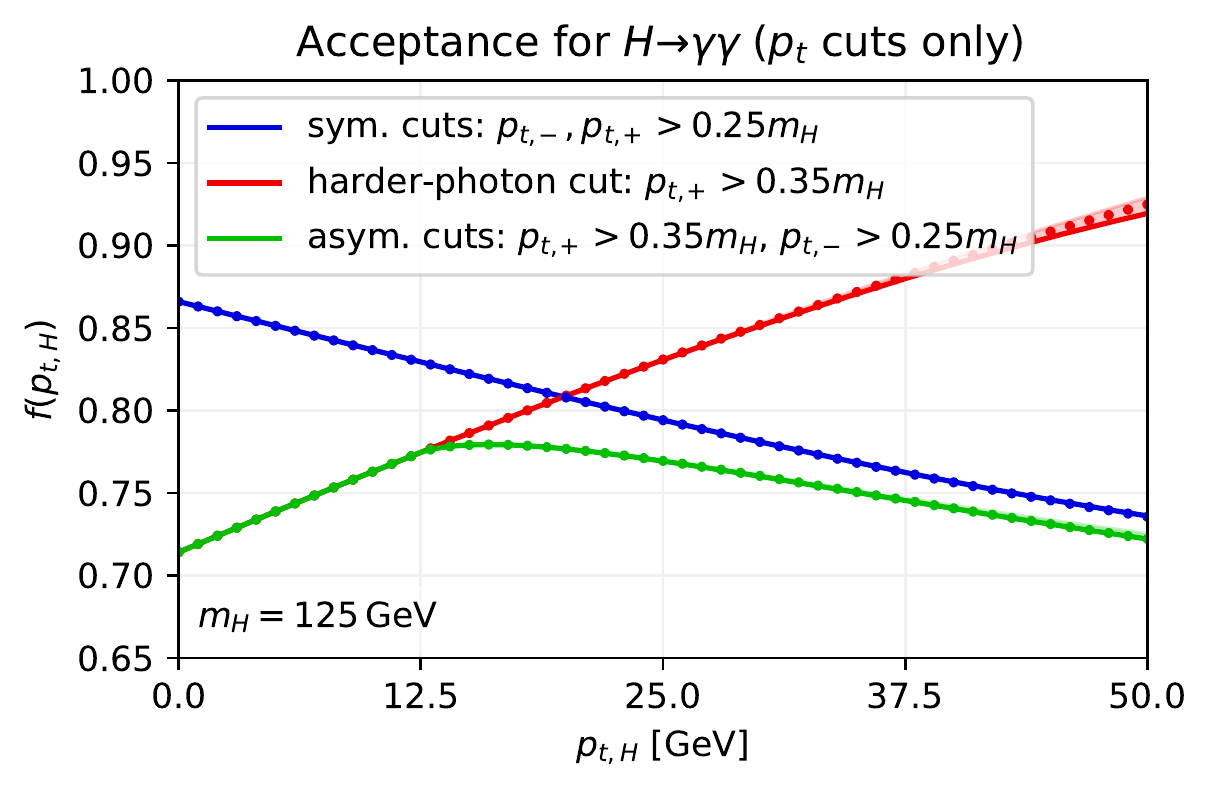}
  \caption{Acceptance for Higgs to di-photon decays, $f(\ptH)$, as a
    function of $\ptH$, for a symmetric cut on the photons ($p_{t,-},p_{t,+}>0.25 \mH$), a
    cut just on the harder photon ($p_{t,+}>0.35 \mH$) and an asymmetric
    cut, where both conditions are imposed.
    Points are Monte Carlo evaluations of the acceptance (whose value
    is independent of any perturbative order), while the
    lines use Eqs.~(\ref{eq:fsym-linear}),
    (\ref{eq:fasym-linear-lt-Delta}) and (\ref{eq:fasym-linear-full}),
    extended to fourth order in $\ptH/\mH$.
    Where a band is visible, its width corresponds to the difference
    between third and fourth order expansions.  }
  \label{fig:acceptances-sym-asym}
\end{figure}

For $\ptH>\Delta$, the cut on the softer photon starts to
matter.
It is instructive to explicitly write the constraint on $\cos\theta$
from each of the two cuts:
\begin{subequations}
  \label{eq:two-costheta-cuts}
  \begin{align}
    p_{t,+} &> p_{t,\cut} &
    &\to &\cos \theta &<  
                        f_0
                        + \frac{2}{f_0} \frac{p_{t,\cut}}{\mH^2}
                        \cdot \ptH \cos\phi\,
                       + \ord2\,,
    \\
    \label{eq:ptminus-cut-with-delta-costheta}
    p_{t,-} &> p_{t,\cut} - \Delta & 
    &\to & \cos \theta &< f_0 +
        \frac{2}{f_0}\frac{p_{t,\cut}}{\mH^2} \left(2\Delta
                         - \ptH \cos\phi\right)  + \ord2\,,
  \end{align}
\end{subequations}
where $\ord2$ includes contributions with two or more powers in total
of either $\ptH$ or $\Delta$.
For $\phi < \arccos (\Delta/\ptH)$ the $p_{t,-}$ cut replaces the
usual $p_{t,+}$ cut as being more constraining, resulting in an
overall acceptance of
\begin{equation}
  \label{eq:3}
  f^{\text{asym}} =
  f_0 + \left(\frac{\ptH}{\mH}\right)f_1^\text{asym}
  - \frac{2}{f_0}\frac{p_{t,\cut}}{\mH^2} \cdot
  \frac{2}{\pi}
  \int_0^{\arccos (\Delta/\ptH)} (2\ptH\cos\phi - 2\Delta) d\phi
  +\ord2\,,
\end{equation}
whose integrand involves the difference between the two $\cos\theta$
limits in Eqs.~(\ref{eq:two-costheta-cuts}).
Evaluating the $\phi$ integral, one obtains
\logbook{}{asym-cuts-new.nb}
\begin{equation}
  \label{eq:fasym-linear-full}
  f^\text{asym}(\ptH) = f_0
  + f_1^\text{asym}
  \left[\frac{\ptH}{\mH}
    - 2\frac{\chi\left(\ptH, \Delta\right)}{\mH}
  \right]
  + \ord2\,,
\end{equation}
where we have introduced the function
\begin{equation}
  \label{eq:chi}
  \chi(p_t,\Delta) = \left(\sqrt{p_t^2 - \Delta^2} - \Delta \arccos
    \frac{\Delta}{p_t} \right)
  \Theta\left(\frac{p_t}{\Delta}-1\right)\,,
\end{equation}
which will reappear below when discussing other combinations of cuts.
\logbook{}{See ``Properties of $\chi$'' section of rapidity-pt-cuts.nb}
It has the property that it is $0$ for $p_t \le \Delta$, it goes as
$\sqrt{8/9\Delta} (p_{t}-\Delta)^{3/2}$ for $p_{t}$ just above
$\Delta$ and as $p_{t} - \frac\pi2\Delta$ for $p_{t} \gg \Delta$.
The acceptance for the asymmetric cut is plotted as a function of
$\ptH$ in Fig.~\ref{fig:acceptances-sym-asym} (the green line), using
ATLAS values~\cite{Aaboud:2018xdt} for the photon thresholds.
The figure includes a comparison to a symmetric cut (in blue), as well
as a cut just on the harder photon (in red).
One sees that the asymmetric cut gives identical results to the
harder-photon cut up to $\ptH=\Delta = 0.1 \mH = 12.5\GeV$, while it
mostly tracks the symmetric cut beyond that point, a consequence of
the fact that for $p_t\gg \Delta$, $\chi(p_t,\Delta) \simeq p_t$,
effectively replacing $f_1^\text{asym} \ptH/\mH$ with
$-f_1^\text{asym} \ptH/\mH$ in Eq.~(\ref{eq:fasym-linear-full}).

The next step is to examine how
Eq.~(\ref{eq:illustrative-sym-expansion}) is modified with asymmetric
cuts.
With $\Delta = 0.1 \mH$, concentrating on the part of the acceptance
proportional to $f_1$, we obtain
\logbook{}{math-2021/cross-sec-impact.nb}
\begin{equation}
  \label{eq:illustrative-asym-expansion}
  \frac{\sigma_\text{asym}^\DL}{f_0 \sigma_\text{tot}} -1 
  \simeq 
  \frac{f_1^\text{asym}}{f_0} \left(
    \underbrace{\,0.16\,}_{\as} 
    - \underbrace{\,0.33\,}_{\as^2}
    + \underbrace{\,0.82\,}_{\as^3}
    - \underbrace{\,2.73\,}_{\as^4}
    + \underbrace{\,11.72\,}_{\as^5}
    + \ldots
  \right)
  \simeq \frac{f_1^\text{asym}}{f_0} \times \underbrace{0.05}_\text{resummed}\!\!.
\end{equation}
Comparing to Eq.~(\ref{eq:illustrative-sym-expansion}), there is an
overall replacement $f^\text{sym} \to f^{\text{asym}}$ (recall that
they have opposite signs).
The coefficient of the order $\alpha_s$ term is somewhat reduced, and
the resummed acceptance correction is also reduced (cf.\ the result of
$0.05f_1/f_0$ in in Eq.~(\ref{eq:illustrative-asym-expansion}) versus $0.12f_1/f_0$ in
Eq.~(\ref{eq:illustrative-sym-expansion-result})).
However, the underlying problem with the perturbative series, namely
the divergence of the series, is essentially identical to that in
Eq.~(\ref{eq:illustrative-sym-expansion-result}), with the terms from order
$\alpha_s^2$ onwards almost the same (aside from the overall
replacement of $f_1^\text{sym} \to f_1^{\text{asym}}$).
The conclusion is that relative to symmetric cuts, asymmetric cuts
bring essentially no improvement as regards one of the
fundamental issues of symmetric cuts, namely the poor
asymptotic behaviour of the perturbative series.\footnote{If anything,
  asymmetric cuts may even worsen it.
  So far we have factored out $f_1$ in all the series.
  But a symmetric cut with $p_{t,+},p_{t-} > 0.25 \mH$ has
  $f_1^\text{sym}/f_0 \simeq -0.42$
  while an asymmetric cut with $p_{t,+}>0.35 \mH$ and $p_{t,-}>0.25
  \mH$ has  $f_1^\text{asym}/f_0 \simeq 0.87$, i.e.\ roughly double
  the overall coefficient.
}

The reason why asymmetric cuts do not improve the perturbative
convergence is that the integral of a linearly dependent acceptance
with a perturbative term that goes as $\as^n L^N$ ($N=2n-1$) is
dominated by large values of $L$, or equivalently small values of
$\ptH$.
Specifically, half of the integral to a given perturbative term, in an
expression such as
Eq.~(\ref{eq:sym-sigma-fid-explicit-series-integral}), comes
from $L > N + 2/3 + \order{1/N}$.
Taking the example of the $\alpha_s^3$ term, where $N=5$, that means
that half of the
acceptance correction integral comes from $\ptH \lesssim 0.2 \GeV$.
It is not surprising, therefore, that the behaviour of the acceptance
beyond $\ptH > \Delta = 12.5\GeV$ (where the $p_{t,-}$ cut sets
in), should have little impact on the convergence of the perturbative
series.

The fact that so much of the perturbative series in
Eqs.~(\ref{eq:illustrative-sym-expansion}) and
(\ref{eq:illustrative-asym-expansion}) comes from low $\ptH$ values
is part of the reason why it has been found to be necessary to have very
small technical cuts in high-order perturbative calculations (cf.\
Fig.~2 of Ref.~\cite{Billis:2021ecs}; the direct N3LO calculation of
Ref.~\cite{Chen:2021isd} has also used extremely small technical cuts).
Indeed, within our DL approximation, one can show that to obtain
$95\%$ of the perturbative coefficient in integrals such as
Eq.~(\ref{eq:sym-sigma-fid-explicit-series-integral}), one needs to go
to
\begin{equation}
  \label{eq:95percent-result}
  L\gtrsim N + 1.64485\sqrt{N} + 1.56851 + \order{1/\sqrt{N}}.
\end{equation}
For the $\alpha_s^3$ term, this translates to $\ptH \lesssim 0.002
\GeV$!
\logbook{e01aec9}{For the derivation of the expansion, see
  math-2021/incomplete-gamma-games.nb;
  The true $95\%$ value is given by gammaincinv(1+5,0.95)=10.513
  instead of $10.247$.
  The expansion for the $50\%$ value is much better,
  gammaincinv(1+5,0.5)=5.67016 versus $5 + 2/3$.
}%
Aside from the technical difficulty associated with reliably
integrating over such small values of $\ptH$ in fixed-order codes,
one may legitimately worry that perturbation theory loses all meaning
if the perturbative coefficients receive substantial contributions
from the incomplete cancellation between real and virtual terms at 
transverse momentum scales that are well below the fundamental
non-perturbative scale $\Lambda_\text{QCD} \simeq 0.2
\GeV$.\footnote{%
  The identification of the contribution from low
  $\ptH$ values is unambiguous, as discussed in
  Appendix~\ref{sec:interpretations}.
  The condition for this statement to be true is that we should
  explicitly consider the difference between the fiducial cross
  section and the product of the Born acceptance and total cross
  section, as we do throughout this work.
  That appendix also discusses the connection with technical cuts in
  subtraction and slicing methods.
  We are grateful to the referee of this paper for encouraging us to
  make this connection more explicit.  }

Note that the sensitivity to small $\ptH$ values is relevant not just
to phase-space slicing perturbative calculations, but to any
perturbative calculation, including calculations that use local
subtraction methods.
\logbook{}{but MC stat. error may be different.}

\subsection{Further discussion of perturbative behaviour}
\label{sec:further-pert-discussion}

One issue with our discussion so far is that we have neglected the
impact of subleading logarithmic terms in the $\ptH$ distribution on
the high-order behaviour of the fiducial cross section.
The fundamental consideration to keep in mind is the following.
The DL term at order $\as^n$ has a structure $L^{2n-1}/(n-1)!$, which
after integration with the acceptance translates to
$(2n-1)!/(n-1)! = (2n)!/2(n!)$.
Subleading terms, while having fewer logarithms, may also have a
larger coefficient.
It is the interplay between the coefficient and the number of
logarithms that determines the ultimate contribution to the
perturbative series for the acceptance.
In general, a complete understanding of this question appears to be
somewhat delicate, in particular as regards the effects associated
with cancellations between transverse momenta of different emissions
and their treatment in Fourier-transform ($b$)
space~\cite{Parisi:1979se} or directly in transverse momentum
space~\cite{Monni:2016ktx}, as well as their interplay with
running-coupling effects.
Some of the subtleties are outlined briefly in
Appendix~\ref{sec:remarks-asysmptotics}.

To obtain a sense of the behaviour of the perturbative series at high
orders, we will consider a simplified approach, where we examine its
structure with four models for the series: one based on a DL
resummation, using Eq.~(\ref{eq:2}); one based on a
leading-logarithmic (LL) $p_t$-space resummation
\begin{equation}
  \label{eq:dsigmaLL}
  \frac{d\sigma^\LL}{d\ptH}
  =
  \frac{\sigma_\text{tot}}{\ptH} \frac{d}{dL} e^{-2 C_A L
    r_1(\alpha_s L b_0)}\,,\qquad
  r_1(\lambda) = -\frac{2\lambda + \ln(1-2\lambda)}{2b_0 \pi\lambda}\,,
  \qquad
  b_0 = \frac{11 C_A - 2n_f}{12\pi}\,,
\end{equation}
which supplements the DL result with running-coupling effects;
and two based, respectively, on the NNLL and N3LL calculations within the RadISH
approach~\cite{Bizon:2017rah,Bizon:2018foh}\footnote{We are grateful
  to the authors of those publications for supplying us with the
  numerical results for the resummations and their expansions.
  We use them with unmodified logarithms, a resummation scale set to
  $\mH$, default renormalisation and factorisation scales
  $\mu_R = \mu_F = \mH/2$, the PDF4LHC15\_nnlo parton distribution
  set~\cite{Butterworth:2015oua} and for a centre of mass energy of
  $\sqrt{s} = 13\TeV$.
} (other
N3LL $p_t$ resummations
include~\cite{Scimemi:2019cmh,Bacchetta:2019sam,Ebert:2020dfc,Billis:2021ecs,Becher:2020ugp,Camarda:2021ict,Re:2021con}).
The LL and DL acceptance results can be easily expanded to high
orders, and in many of our investigations of other possible
logarithmic effects, the asymptotic scaling of the terms (though not
their absolute values) is between that of the LL and DL results,
sometimes at one of the extremities.\logbook{}{bessel-games.nb}
The RadISH NNLO and N3LL expansions are available up to N3LO, and at
N3LL they account for all terms up to N3LO in $d\sigma/d\ptH$ that
have a $1/\ptH$ enhancement at small $\ptH$.
As before, we will integrate acceptances up to $\mH/2$, and we will
study
\begin{equation}
  \label{eq:generic-sigma-fid}
  \frac{\sigma_{\text{fid}} - f_0 \sigma_{\text{inc}}}{\sigma_0 f_0} =
  \int_\epsilon^{\frac{\mH}{2}} d\ptH \,
  \left(\frac{f^\text{fid}(\ptH)}{f_0} -1 \right)
  \frac1{\sigma_0} \frac{d\sigma}{d\ptH} \,,
\end{equation}
where $\sigma_\text{inc}$ is the inclusive cross section integrated up
to $\ptH=\mH/2$.\footnote{Were we considering results matched to fixed
  order, we would integrate up to the kinematic limit for $\ptH$ and
  then write $\sigma_\text{tot}$ instead of $\sigma_\text{inc}$.
  However, for the resummed approximation that we use here, it makes
  little sense to integrate beyond $\mH/2$.
  Note also that as compared to the standard resummation, the full cross
  section will include additional relative corrections suppressed by
  powers of $(\ptH/\mH)^2$~\cite{Ebert:2018gsn,Ebert:2020dfc}.
  We expect the additional contributions from such terms to be smaller
  than the leading power contributions discussed here.
}
Note that we include a cutoff $\epsilon$ for the lower limit of the
$\ptH$ integral.
Unless otherwise stated, when quoting numbers we will take
$\epsilon\to 0$, however we will also plot the $\epsilon$ dependence
of the result to gauge the effect of a $\ptH$ cutoff in a
projection-to-Born type~\cite{Cacciari:2015jma} subtraction approach
for perturbative calculations, as used in
Ref.~\cite{Chen:2021isd}. (In practice, such calculations impose a
cutoff $m_\text{min}^2$ on the invariant mass of parton pairs, and a cut
$m_\text{min}^2 \lesssim \epsilon^2$ is required to fully cover
transverse momenta down to a scale $\epsilon$.)

For asymmetric cuts with the ATLAS thresholds of $p_{t,+} > 0.35 \mH$
and $p_{t,-} > 0.25 \mH$ (using not just the $f_1$ part of the
acceptance, but its full structure), we obtain the following results
for the acceptances for each of the perturbative models,%
\logbook{}{These numbers were produced by n3-benchmarks.py}
\begin{align}
  \label{eq:asym-PT-series}
  \frac{\sigma_{\text{asym}} - f_0 \sigma_{\text{inc}}}{\sigma_0 f_0} 
&\simeq    0.15_{\as} - 0.29_{\as^{2}} +    0.71_{\as^{3}} - 2.39_{\as^{4}} +   10.31_{\as^{5}} + \ldots &\simeq    0.06 \;\;&@\text{DL},\nonumber\\[-6pt]
&\simeq    0.15_{\as} - 0.23_{\as^{2}} +    0.44_{\as^{3}} - 1.15_{\as^{4}} +    3.86_{\as^{5}} + \ldots &\simeq    0.06 \;\;&@\text{LL},\nonumber\\
&\simeq    0.18_{\as} - 0.15_{\as^{2}} +    0.29_{\as^{3}} + \ldots &\simeq    0.10 \;\;&@\text{NNLL},\nonumber\\
&\simeq    0.18_{\as} - 0.15_{\as^{2}} +    0.31_{\as^{3}} + \ldots &\simeq    0.12 \;\;&@\text{N3LL}.\nonumber\\
\end{align}
In these results, the $\as^n$ subscript indicates that the
corresponding term is the $\as^n$ contribution to the result, while
the right-hand side of the equality corresponds to the acceptance as
determined from the resummation (in the case of the LL result, we stop
the integration at the Landau pole).
The DL and LL results clearly show how the series start to diverge
towards higher orders.
In the LL case, the terms grow a little more slowly, and numerically
fitting the structure of the series to high orders leads to the
conclusion that (for $n_f=5$) the smallest term in the series scales
as $(\Lambda/Q)^{0.205}$ rather than the
$(\Lambda/Q)^{23/144} \simeq (\Lambda/Q)^{0.160}$ seen at DL level.
\logbook{}{look at the summary for bessel-games.nb}
The investigations reported in Appendix~\ref{sec:remarks-asysmptotics}
suggest that the $(\Lambda/Q)^{0.205}$ scaling may be robust with
respect to $b$-space versus $p_t$ space complications, as well as to
other subleading effects.

\begin{figure}
  \centering
  \includegraphics[page=1,scale=0.9]{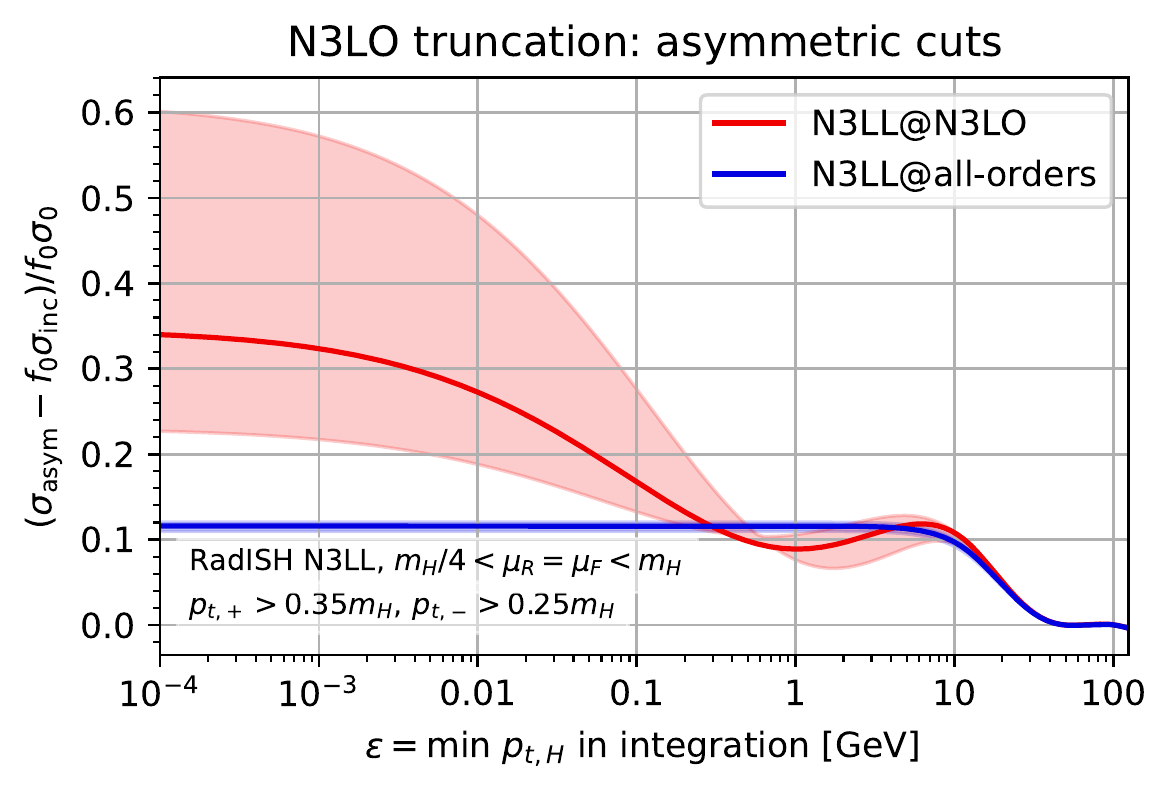}
  \caption{The N3LL resummed result and its truncation at N3LO for the
    fiducial corrections to the Higgs cross section,
    as defined in Eq.~(\ref{eq:generic-sigma-fid}), for asymmetric
    $p_{t,\gamma}$ cuts, $p_{t,+} > 0.35 \mH$ and $p_{t,-} > 0.25
    \mH$.
    The results are shown as a function of $\epsilon$, the minimum
    Higgs $p_t$ used in the integration  (conceptually
    analogous to a technical cutoff in a projection-to-Born
    fixed-order calculation, cf.\ Appendix~\ref{sec:interpretations}).
    The bands are the result of varying renormalisation and
    factorisation scales by a factor of two around $\mH/2$.
    The N3LL distribution and expansion used to obtain these results
    were kindly supplied by the authors of the RadISH
    framework~\cite{Bizon:2017rah}.  }
  \label{fig:minpth-impact}
\end{figure}

Next, we examine the NNLL and N3LL results in Eq.~(\ref{eq:asym-PT-series}).
The all-order results are twice as large in the NNLL and N3LL cases as compared
to the DL and LL cases, which is  a consequence of the fact that
the NNLL and N3LL results includes a substantial part of the $K$ factor for
inclusive Higgs production.
The NNLL and N3LL results are themselves close.
Examining the fixed-order results, the
main feature to note is that up to N3LO there is no truncation of the
series that agrees with the resummed result.

Fig.~\ref{fig:minpth-impact} illustrates the N3LO truncation
compared to the resummation, as a function of the cutoff $\epsilon$
in Eq.~(\ref{eq:generic-sigma-fid}).
First considering the small-$\epsilon$ limit, the difference of $0.22$
between the central N3LO result and the resummation corresponds to a
roughly $7\%$ relative effect on the full cross section (after
accounting for an overall $K$-factor of about $3$).
This is significantly larger than the perturbative scale uncertainty
on the inclusive N3LO cross section~\cite{Anastasiou:2016cez}.
The scale variation bands demonstrate a large scale sensitivity for
the fixed-order result, which does not overlap with the resummed result
(though contributions beyond the resummation could modify this
aspect, for example by increasing the width of the resummed scale
variation band).
The pattern of $\epsilon$-dependence in Fig.~\ref{fig:minpth-impact}
confirms the expectation from Eq.~(\ref{eq:95percent-result}) that the
fixed-order result is highly sensitive to unphysically low $\ptH$
values.%
\footnote{One intriguing feature is that setting $\epsilon$ in the
  range of a few hundred MeV to one GeV gives an N3LO truncated result
  that is much closer to the full N3LL result, and with a reduced
  scale uncertainty.
  We have yet to reach a conclusion as to the significance to
  attribute to this observation.
}

One may ask whether a badly divergent perturbative series for a
fiducial cross section is a problem: after all, there are various ways
of evaluating the fiducial cross section via the matching of
resummations and fixed order, including the $\ptH$ dependence
acceptance factor as part of the procedure.
This is the approach that has been taken in Refs.~\cite{Ebert:2020dfc,Billis:2021ecs}.
However there are reasons why it would be preferable to retain the ability
to make predictions with pure fixed-order perturbation theory.
For example pure fixed-order calculations are conceptually simple,
whereas matched resummed calculations tend to have additional scales
(e.g. a resummation scale), matching prescriptions, choices for
so-called modified logarithms, or equivalently, of profile functions,
etc.

A more general issue is that the $\ptH$ dependence of the acceptance
that we have seen implies a whole set of problems that one might
prefer to avoid associated with scales in the range of one to ten GeV.
For example it results in sensitivity to effects related to the interplay
between the $p_t$ scale and finite mass effects for the charm and
bottom quarks.
Even when using resummation, the coupling is effectively being
evaluated at a much lower scale than that of the hard process, which
may result in poorer convergence.
Additionally, regardless of the calculational approach being used,
non-perturbative uncertainties tend to be enhanced in the low-$p_{t}$
region.
It may be that none of these issues is major, but if one could avoid
them altogether, that would clearly be preferable.

Ultimately, fiducial cross sections are supposed to be conceptually
the cleanest of experimental measurements, but the effects discussed
here clearly interfere with that simplicity.

\section{Simple proposals for transverse momentum cuts}
\label{sec:good-simple-cuts}

To improve the convergence of the perturbative series, and reduce the
size of the minimal truncation ambiguity, the approach that we take in
this section is to modify the parametric dependence of the acceptance
on $\ptH$ in the small-$\ptH$ region.
In particular, we will see that it is straightforward to engineer cuts
such that the linear dependence of the acceptance on $\ptH$
disappears, leaving only a residual quadratic dependence.
To understand how we expect this to affect the perturbative series,
consider the generic form
\begin{equation}
  \label{eq:f-quadratic}
  f^\text{quadratic}(\ptH) = f_0 + f_2 \left(\frac{\ptH}{\mH}\right)^2 + \cdots\,.
\end{equation}
It is straightforward to show that the DL estimate of the fiducial
cross section is
\begin{subequations}
    \label{eq:quadratic-sigma-fid-explicit-series}
  \begin{align}
    \frac{\sigma_{\text{quadratic}}^\DL}{f_0\sigma_\text{tot}}-1
    &=
      \frac{f_2}{f_0} \sum_{n=1}^\infty (-1)^{n+1} \frac{1}{2^{2n+1}} \frac{(2n)!}{2(n!)} \left(\frac{2 C_A \alpha_s}{\pi}\right)^n
      + \ldots\,,
    \\
    &\simeq \frac{f_2}{f_0} \bigg(
      \underbrace{\,0.030\,}_{\as} 
      - \underbrace{\,0.011\,}_{\as^2}
      + \underbrace{\,0.006\,}_{\as^3}
      - \underbrace{\,0.005\,}_{\as^4}
      + \underbrace{\,0.006\,}_{\as^5}
      + \ldots
      \bigg),
      \\
    &\simeq \frac{f_2}{f_0} \times 0.023\,\quad \text{(DL resummed})\,.
  \end{align}
\end{subequations}
Relative to the series multiplying $f_1$ in
Eq.~(\ref{eq:sym-sigma-fid-explicit-series}), note the extra factor of
$1/2^{2n+1}$ for the $\alpha_s^n$ term.
This significantly reduces the size of the individual perturbative
terms.
Formally, the problem of the alternating sign factorial divergence
remains, but its impact is postponed to significantly higher orders,
parametrically $n \gtrsim
\pi/(2 \alpha_s C_A)$ instead of $n \gtrsim \pi/(8\alpha_s C_A)$ at
the DL level.
In practice, the sum of the first three orders of the series is now
quite close to the full (DL) resummed result in the last line, and the
ambiguity associated with the smallest term of the series is
$\sim 0.5\% \times f_2/f_0$.
Furthermore if one determines the region in $p_t$ that dominates the
integral, one finds that $95\%$ of it comes from
\begin{equation}
  \label{eq:1}
  L \lesssim \frac{N + 1.64485\sqrt{N} + 1.56851}{2} + \order{N^{-1/2}}, \qquad
  N = 2n-1
\end{equation}
which, for the N3LL term, translates to $\ptH \gtrsim 0.3 \GeV$.

The conclusion is that if we can engineer cuts where the acceptance
depends only quadratically on $\ptH$, we expect to see much smaller
perturbative corrections to the integrated acceptance, and those
corrections will come from a region where perturbation theory is more
likely to be under control.
Note that if we think about such ``quadratic cuts'' in terms of the
power scaling associated with the minimal truncation ambiguity in the
DL perturbative model used in
Eq.~(\ref{eq:effective-power-correction}), the resulting size
$(\Lambda/Q)^{23/36}$ will still be much worse than
$\Lambda^2/Q^2$.\footnote{It is quite likely that the
  $(\Lambda/Q)^{23/36}$ power will be modified by higher logarithmic
  effects, as discussed in Appendix~\ref{sec:remarks-asysmptotics},
  but a quantitative understanding remains a subject for further
  work.}

%

\subsection{Sum cuts}
\label{sec:sum-cuts}

One simple approach to obtaining a quadratic dependence of the
acceptance on $\ptH$ is to examine Eq.~(\ref{eq:pt-expansion}) and
note that the arithmetic average of the two photon momenta
\begin{equation}
\label{eq:pt-expansion-sum}
  p_{t,\text{sum}}(\ptH,\theta,\phi) = \frac12\left(p_{t,+}+p_{t,-}\right) =
  \frac{\mH}{2} \sin \theta 
  +\frac{\ptH^2}{4 \mH}  \left(\sin \theta \cos ^2\phi+\csc
      \theta \sin ^2\phi\right)
    + \ord4\,
    ,
\end{equation}
is free of any linear dependence on $\ptH$.
Rather than cutting on either of the photon momenta directly, one
obvious option is therefore to cut on $p_{t,\text{sum}}$ (a proposal
that was examined at HERA~\cite{Carli:1998zr,Adloff:2000tq} and was
seen numerically to reduce convergence problems in approximate NNLO
calculations for dijet cross sections in Ref.~\cite{Rubin:2010xp}).
We include a factor of a $1/2$ in the definition so that a given threshold on
$p_{t,\text{sum}}$ yields the same Born acceptance as the same threshold on either
$p_{t,+}$ or $p_{t,-}$.\footnote{The reader may ask themselves why we
  haven't called it an ``average'', and the answer is that below we
  will consider the product of transverse momenta, and if we need to
  start distinguishing between arithmetic and geometric averages, our
  labels will become too verbose.}
Repeating the analysis of section~\ref{sec:symmetric-cuts}, the
requirement $p_{t,\text{sum}} > p_{t,\cut}$ gives the condition
\begin{equation}
  \label{eq:sin-cut-sum}
  \sin\theta > \frac{2p_{t,\cut}}{\mH} -
  \left(\frac{p_{t,\cut}}{\mH}\cos^2\phi +
    \frac{\mH}{2p_{t,\cut}}\sin^2\phi\right) \frac{\ptH^2}{\mH^2}
  + \ord4\,,
\end{equation}
resulting in an acceptance,
\begin{equation}
  \label{eq:sum-cut-f2}
  f^\text{sum}(\ptH) = f_0 + f_2^\text{sum} \cdot \frac{\ptH^2}{\mH^2}
  +\ord4
  \,,
  \qquad
  f_2^\text{sum} = \frac{\mH^2 + 4p_{t,\cut}^2}{4 \mH^2 f_0}\,,
\end{equation}
which has the form of Eq.~(\ref{eq:f-quadratic}).
For $p_{t,\cut}= 0.35 \mH$ we have $f_2^\text{sum} / f_0 \simeq
0.73$. 

On its own, a sum cut naturally places a constraint on both photons,
but the constraint on the softer photon may be weaker than is
experimentally admissible.
In such a situation one may place an explicit cut on the softer
photon, $p_{t,-} > p_{t,\cut}-\Delta$.
The analysis is similar to that for the asymmetric cut in
section~\ref{sec:asymmetric-cuts} and with the condition
$\Delta \ll \mH$, the result that emerges is that
Eq.~(\ref{eq:sum-cut-f2}) is replaced by
\begin{equation}
  \label{eq:f-sum-Delta}
  f^\text{sum}(\ptH) = f_0 +
  f_2^\text{sum} \cdot \frac{\ptH^2}{\mH^2}
  + \mathcal{O}_4
  - \frac{4p_{t,\cut}}{\pi \mH f_0} \frac{\chi(\ptH,2\Delta)}{\mH}
  \left(1 + \mathcal{O}_2\right)\,.
\end{equation}
Note the same function $\chi$ that appeared in
Eq.~(\ref{eq:fasym-linear-full}), but now as $\chi(\ptH,2\Delta)$
instead of $\chi(\ptH,\Delta)$, so that the transition intervenes
for $\ptH>2\Delta$, i.e.\ at twice the value that occurred with
standard asymmetric cuts.
It is simple to understand why: to have $p_{t,-} = p_{t,\cut}-\Delta$
and $p_{t,+}+p_{t,-} = 2p_{t,\cut}$, then it is necessary to have
$p_{t,+} -p_{t,-}= 2\Delta$, which implies $\ptH \ge 2\Delta$.

In general, when we refer to sum cuts, we will always understand them
to involve an additional requirement on the $p_t$ of the softer decay
product, 
and similarly for all the other cuts that we discuss below.

\subsection{Product cuts}
\label{sec:product-cuts}

Another simple solution to engineering an acceptance with a quadratic
dependence on $\ptH$ is to consider the (square-root) of the
product of the two photon transverse momenta
\begin{equation}
\label{eq:pt-expansion-prod}
  p_{t,\text{prod}}(\ptH,\theta,\phi) = \sqrt{p_{t,+} p_{t,-}} =
  \frac{\mH}{2} \sin \theta 
  +\frac{\ptH^2}{4 \mH}
  \frac{\sin^2\phi - \cos^2\theta \cos^2\phi}{\sin\theta}
  + \ord4\,.
\end{equation}
Again, the fact that $p_{t,\text{prod}}$ has no linear dependence on
$\ptH$ will have the consequence that a cut $p_{t,\text{prod}} >
p_{t,\cut}$ will have an acceptance with only quadratic dependence on
$\ptH$. 
Specifically, the acceptance is given by
\begin{equation}
  \label{eq:prod-cut-f2}
  f^\text{prod}(\ptH) = f_0 + f_2^\text{prod} \left(\frac{\ptH}{\mH}\right)^2
  + \ord4\,,
  \qquad
  f_2^\text{prod} = \frac{p_{t,\cut}^2}{\mH^2 f_0}\,.
\end{equation}
The coefficient of the quadratic dependence, $f_2^\text{prod}$, is
somewhat smaller than with sum cuts: for example, for
$p_{t,\cut}= 0.35 \mH$, we have $f_2^\text{prod}/f_0\simeq 0.24$,
i.e.\ about $3$ times smaller than $f_2^\text{sum} / f_0$. 

As in the previous subsection, a cut just on $p_{t,\text{prod}}$ may
not be sufficient experimentally, since the constraint it places on
the softer photon is rather weak.
However, it is once again possible to combine a $p_{t,\text{prod}}$
cut with a cut on the softer photon, $p_{t,-} > p_{t,\cut} - \Delta$,
and for small $\Delta$ one obtains a result structurally very similar
to Eq.~(\ref{eq:f-sum-Delta}):
\begin{equation}
  \label{eq:f-prod-Delta}
  f^\text{prod}(\ptH) = f_0 +
  f_2^\text{prod} \left(\frac{\ptH}{\mH}\right)^2
  + \ord4
  - \frac{4p_{t,\cut}}{\pi \mH f_0} \frac{\chi(\ptH,2\Delta + \ord2)}{\mH}
  \left(1 + \mathcal{O}_2\right).
\end{equation}
In particular, for small $\ptH$ one obtains the same acceptance as
without the $p_{t,-}$ cut, and the transition for
$\ptH \gtrsim 2\Delta$ has the same form at first order in
$\ptH/\mH$.
One small difference is that the transition is not exactly at
$2\Delta$, but rather slightly higher, at
$\Delta(1 + 1/(1 - \Delta/p_{t,\cut}))$.

\subsection{Staggered cuts}
\label{sec:staggered-cuts}

The linear $\ptH$ dependence on the acceptance in
section~\ref{sec:existing-cuts} came about because a non-zero Higgs
$p_{t}$ breaks the symmetry between the transverse momenta of the two decay
photons, and any cut that specifically targets the harder or softer
photon is directly sensitive to that broken symmetry.
As emphasised recently in Ref.~\cite{Alekhin:2021xcu} in the context
of $W$ and $Z$ decays, there are ways of applying separate cuts to the
two leptons where the distinction between the two leptons is made not
based on their transverse momentum, but on their charge.
For example in $W$ decay, one may place different $p_t$ cuts on the
charged lepton and the neutrino~\cite{D0:2014kma}, while in $Z$ decay one may place
different $p_t$ cuts on the positively and negatively charged
leptons.
Staggered cuts can also be adapted for $H\to \gamma\gamma$ decays, or
other processes with two apparently identical objects (e.g.\ jets):
examining Eq.~(\ref{eq:photon-momenta}), one notes that rapidity
ordering of the two photons is unaffected by $\ptH$, and so one may
place staggered cuts on the higher and lower rapidity photons.
This may be considered somewhat inelegant in a $H\to\gamma\gamma$
context, because it breaks the intrinsic symmetry between the two
photons, but let us still examine the consequences.
Carrying out the usual analysis, but with the $\phi$ integral in
Eq.~(\ref{eq:generic-acceptace}) extended to cover the range
$-\pi<\phi<\pi$, we find
\begin{equation}
  \label{eq:f-stag-Delta}
  f^\text{stag}(\ptH) = f_0 +
  f_2^\text{stag} \left(\frac{\ptH}{\mH}\right)^2
  + \ord4
    - \frac{4p_{t,\cut}}{\pi \mH f_0} \frac{\chi(\ptH,\Delta)}{\mH}(1+\ord2),
\end{equation}
with
\begin{equation}
  \label{eq:f2-stag}
  f_2^\text{stag} = -\frac{4p_{t,\cut}^4}{\mH^4 f_0^3}.
\end{equation}
For concreteness, taking the larger of two cuts to be on the higher rapidity photon,
which we denote $p_{t,y_+} > 0.35 \mH$, yields
$f_2^\text{stag}/f_0 \simeq -0.23$, which is similar in magnitude to
corresponding ratio for product cuts, but with the opposite sign.

Whereas the quadratic behaviour for sum and product cuts came about
because they cut on a quantity that is free of linear $\ptH$
dependence, in the case of staggered cuts the quadratic dependence
arises because the linear $\ptH$ dependence averages to zero after
integrating over the azimuthal angle $\phi$ in
Eq.~(\ref{eq:photon-momenta}).
The departure from quadratic behaviour induced by the lower cut sets
in earlier than for sum and product cuts, for $\ptH > \Delta$,
associated with a $\phi = 0$ configuration, for which we have
$p_{t,y_+}-p_{t,y_-} = \ptH$.
The form of the departure comes with the usual structure.
Note that the expansion in Eq.~(\ref{eq:f2-stag}) breaks down
relatively early, i.e.\ beyond
$\ptH = \frac{\mH^2 - 4p_{t,\cut}^2}{4p_{t,\cut}^2}$, which is
about $45.5\GeV$ for our standard cut
values.\logbook{}{staggered-cuts.nb, ``Work out where the pTH value
  from which at $\phi=\pi$, there is no longer a solution''.}

\subsection{Comparative discussion of quadratic cuts}
\label{sec:quadratic-comparison}

\begin{figure}
  \centering
  \includegraphics[width=0.7\textwidth,page=2]{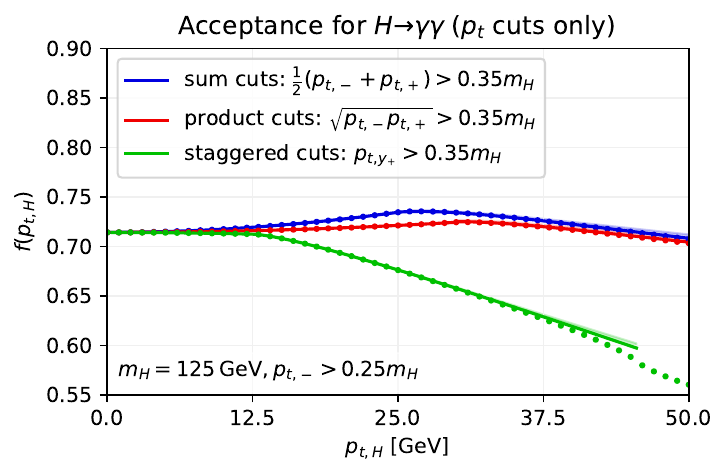}
  \caption{Comparison of the $\ptH$-dependent acceptances for the
    sum, product and staggered cuts.
    For the staggered cuts, $p_{t,y_+}$ corresponds to the transverse
    momentum of the photon at higher rapidity.
    As in Fig.~\ref{fig:acceptances-sym-asym}, the points corresponds
    to Monte Carlo evaluations of the acceptances.
    Lines use series expansions to fourth order and bands (where
    visible) show the size of the fourth order term.
  }
  \label{fig:acceptances-sum-prd}
\end{figure}

The acceptances for the three sets of ``quadratic'' cuts are shown in
Fig.~\ref{fig:acceptances-sum-prd}, and can be compared to the
acceptances for standard symmetric and asymmetric cuts (with linear
$\ptH$ dependence) in Fig.~\ref{fig:acceptances-sym-asym}.
The vertical scales have the same extent, but are shifted by $0.1$.
All three quadratic cuts lead to considerably flatter $\ptH$
dependence, as was expected, and represent good, simple alternatives
to standard symmetric and asymmetric cuts.
One clearly sees the transition to linear $\ptH$ dependence for
$\ptH\gtrsim 2\Delta$ in the case of the sum and product cuts and for
$\ptH > \Delta$ for the staggered cuts.

\begin{figure}
  \centering
  \includegraphics[page=2,scale=0.9]{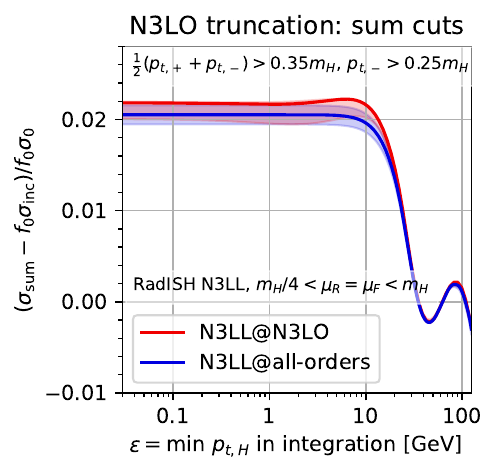}%
  \includegraphics[page=3,scale=0.9]{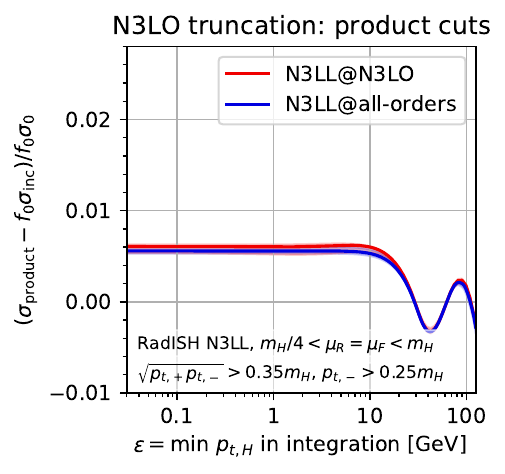}
  \caption{The N3LL resummed result and its N3LO truncation, for sum
    cuts (left)
    and product cuts (right), as a function of $\epsilon$, the minimum
    $\ptH$ in Eq.~(\ref{eq:generic-sigma-fid}).
    Note the different scale relative to Fig.~\ref{fig:minpth-impact}.
  }
  \label{fig:minpth-impact-product}
\end{figure}

The perturbative convergence of the acceptance with sum and product
cuts is illustrated with the following results for the perturbative
series, first for the sum cuts,
\begin{align}
  \label{eq:sum-PT-series}
  \frac{\sigma_{\text{sum}} - f_0 \sigma_{\text{inc}}}{\sigma_0 f_0} 
&\simeq   0.013_{\as} - 0.007_{\as^{2}} +   0.005_{\as^{3}} - 0.004_{\as^{4}} +   0.004_{\as^{5}} + \ldots &\simeq   0.009 \;\;&@\text{DL},\nonumber\\[-6pt]
&\simeq   0.013_{\as} - 0.005_{\as^{2}} +   0.001_{\as^{3}} - 0.001_{\as^{4}} +   0.000_{\as^{5}} + \ldots &\simeq   0.010 \;\;&@\text{LL},\nonumber\\
&\simeq   0.016_{\as} +   0.007_{\as^{2}} - 0.004_{\as^{3}} + \ldots &\simeq   0.019 \;\;&@\text{NNLL}\nonumber,\\
&\simeq   0.016_{\as} +   0.007_{\as^{2}} - 0.001_{\as^{3}} + \ldots &\simeq   0.021 \;\;&@\text{N3LL}\nonumber,\\
\end{align}
and next for product cuts,
\begin{align}
  \label{eq:prod-PT-series}
  \frac{\sigma_{\text{prod}} - f_0 \sigma_{\text{inc}}}{\sigma_0 f_0} 
&\simeq   0.005_{\as} - 0.002_{\as^{2}} +   0.002_{\as^{3}} - 0.001_{\as^{4}} +   0.001_{\as^{5}} + \ldots &\simeq   0.003 \;\;&@\text{DL}\nonumber,\\[-6pt]
&\simeq   0.005_{\as} - 0.002_{\as^{2}} +   0.000_{\as^{3}} - 0.000_{\as^{4}} +   0.000_{\as^{5}} + \ldots &\simeq   0.003 \;\;&@\text{LL}\nonumber,\\
&\simeq   0.005_{\as} +   0.002_{\as^{2}} - 0.001_{\as^{3}} + \ldots &\simeq   0.005 \;\;&@\text{NNLL}\nonumber,\\
&\simeq   0.005_{\as} +   0.002_{\as^{2}} - 0.001_{\as^{3}} + \ldots &\simeq   0.006 \;\;&@\text{N3LL}\nonumber.\\
\end{align}
The improvement in convergence relative to the corresponding results
for asymmetric cuts, Eq.~(\ref{eq:asym-PT-series}) is
striking.

Fig.~\ref{fig:minpth-impact-product}, which is to be compared to its
analogue for asymmetric cuts, i.e.\ Fig.~\ref{fig:minpth-impact}, shows the
sensitivity to the infrared cutoff in
Eq.~(\ref{eq:generic-sigma-fid}), as well as the impact of scale
variation.
N3LO (from N3LL) and the full N3LL resummation now agree
well and the N3LO result is much less sensitive to the minimum
$\ptH$ in the integration, converging at a few GeV, rather than at
MeV scales for asymmetric cuts.
These are precisely the features that we had anticipated in the
introduction to this section.
Note also that the residual scale uncertainty is now essentially
negligible (at least by today's standards for Higgs physics), and that
the overall size of the fiducial acceptance correction is much
smaller than for asymmetric cuts.
Note that in Eqs.~(\ref{eq:sum-PT-series}) and
(\ref{eq:prod-PT-series}), at N3LL the coefficient of the $\as^2$
term is now positive, whereas it is negative at DL and LL.
The most likely explanation for the change of sign is that it is
related to the interplay between the acceptance cuts and the large
(positive) NLO $K$-factor for Higgs production.

Of the three quadratic cuts discussed so far, overall the best choice
appears to be the product cuts, for several reasons:
\begin{enumerate}
\item The coefficient of the quadratic dependence is small (though
  staggered cuts give a smaller coefficient for $2p_{t\cut}/\mH <
  1/\sqrt{2}$).
\item The transition point to quasi-linear $\ptH$ dependence, at
  $\ptH \simeq 2\Delta$, is the highest of the three (sum cuts
  transition at a similar, though slightly lower value of $\ptH$).
  Having a high transition point is of value because it means that the
  substantial $\ptH$ dependence occurs in a region where the
  perturbative prediction for the $\ptH$ spectrum is more likely to
  be reliable, providing confidence in the use of pure fixed-order
  perturbation theory to calculate acceptances.
\item The acceptance remains high at the highest values of $\ptH$
  shown in Fig.~\ref{fig:acceptances-sum-prd}, significantly higher
  than with staggered cuts, albeit slightly lower than with sum cuts.
\end{enumerate}
One context in which one might prefer staggered cuts (or a cut just on
the charged lepton~\cite{Sirunyan:2020oum}) is for the study
of $W$ bosons, where significantly different experimental resolutions
and background contributions for neutrinos versus charged leptons
may favour the application of distinct cuts.
One may also wish to use staggered cuts for $Z/\gamma^*$ decays in
situations where the $Z/\gamma^*$ decays are effectively being used as
a calibration for the $W$'s (the cuts on the $Z$ decay products should
then perhaps then be scaled by a factor $m_Z/m_W$ relative to the $W$
cuts).
For $W$ and $Z$ production, having the quadratic regime extend only to
$p_t=\Delta$ rather than $p_t\simeq 2\Delta$ should be less of an issue than
in Higgs production: the presence of a $C_F=4/3$ versus
$C_A=3$ coefficient in the $p_t$ resummation means that the region of
good perturbative control at fixed order will extend to lower $p_t$
than for Higgs production.
Further discussion about $Z$ production is to be found in
section~\ref{sec:Drell-Yan}. 

Our final remarks concern the values of the thresholds.
If one adopts product or sum cuts with the same values for the
thresholds as used currently for asymmetric cuts (e.g.\ $0.35\mH$ and
$0.25\mH$) any event that passes existing asymmetric cuts will also
pass the sum and product cuts.
Thus the events collected with asymmetric cuts could, in principle, be
straightforwardly reanalysed with the sum or product cuts. 
One should be aware that because the resummed fiducial correction is
smaller with sum and product cuts than with the asymmetric cuts (cf.\
Eqs.~(\ref{eq:sum-PT-series}) and (\ref{eq:prod-PT-series}) versus Eq.~(\ref{eq:asym-PT-series})), there will be a slight loss in Higgs
statistics with sum or product cuts (taking into account a Higgs
$K$-factor of about $3$ relative to the Born cross section, the
numbers suggest roughly $3{-}4\%$, with the background also being
reduced%
\footnote{Further investigation goes beyond the scope of this article,
  notably because of the difficulty of simulating the fake photon
  spectrum.
  However a preliminary study with a simulated genuine di-photon
  sample from the Sherpa~\cite{Bothmann:2019yzt} program suggests that
  when switching from asymmetric to product cuts while retaining the
  same thresholds, the impact of the changed acceptance on the Higgs
  statistical significance is indeed modest.
  We are grateful to Marek Sch\"onherr and Frank Siegert for
  assistance with the event generation.  }).
However, in future data collection (and depending on triggers, perhaps
also with already recorded data), there may be freedom to adjust cut
thresholds.
For example, lowering the $0.35\mH$ threshold to $0.30\mH$ (while
retaining a fixed $p_{t,-}$ threshold of $0.25\mH$) would still
leave a significant region where the quadratic $\ptH$ dependence holds
(up to $\ptH \simeq 12.5\GeV$), while raising the acceptance by more
than enough to compensate for the loss in switching from asymmetric to
product cuts.
Optimisation of the choice of thresholds probably needs experimental
input, e.g.\ as concerns the behaviour of the continuum di-photon
background (including its fake-photon contribution), as well as
theoretical input, e.g.\ in terms of verifying the perturbative behaviour of
each given set of cuts.

\subsection{Boost-invariant cuts via Collins-Soper decay transverse momentum}
\label{sec:cuts-collins-soper}

The quadratic cuts that we have seen so far are already a significant
improvement on widely used symmetric and asymmetric cuts and
Fig.~\ref{fig:minpth-impact-product} suggests that any
residual non-convergence issues of the perturbative series are not
likely to be of imminent practical importance.
Still, the perturbative series in
Eq.~(\ref{eq:quadratic-sigma-fid-explicit-series}) reaches a breakdown
in convergence earlier than one might hope for, even if the
coefficients of the breakdown are small.
Furthermore the effective power scaling of the minimal term of the
perturbative series in the DL approximation, $(\Lambda/Q)^{23/36}$
(Eq.~(\ref{eq:effective-power-correction})), is not entirely
reassuring, even if subleading logarithmic effects are likely to modify the
power.

In this subsection we start our investigation of cuts for which the
acceptance is entirely independent of $\ptH$ at small values of
$\ptH$ (with the remainder of our analysis to be given in
section~\ref{sec:cbi-cuts}, after we have considered the issue of
rapidity cuts).
We will work within the constraint that a direct cut of
$p_{t,\cut}-\Delta$ on the $p_t$ on the softer of the two photons is
experimentally unavoidable.
Additionally, we will aim to maintain the same $\ptH=0$ acceptance
as for the asymmetric, product and sum cuts.

At $\ptH=0$ a cut on $p_{t,\pm}$ is equivalent to a cut on
$\sin\theta$ in the Collins-Soper parametrisation of the decay phase
space in Eq.~(\ref{eq:photon-momenta}).
Our core idea here is to cut on $\sin \theta$ as defined in that
parametrisation also for non-zero $\ptH$.
In practice we express this condition by introducing a
``Collins-Soper'' decay transverse momentum in terms of the kinematics
of the two decay products
\begin{equation}
  \label{eq:ptCS}
  \vec p_{t,\CS} = \frac{1}{2} \left[\vec \delta_t + 
    \frac{\vec p_{t,12}.\vec \delta_t}{p_{t,12}^2}
    \left(\frac{m_{12}}
      {\sqrt{\smash[b]{m_{12}^2+p_{t,12}^2}}}-1\right)\vec p_{t,12}
  \right]
  \,,\qquad
  \vec \delta_t = \vec p_{t,1} - \vec p_{t,2}\,,
\end{equation}
where $\vec p_t$ is the (two-dimensional) vector transverse component
of a momentum $p$ and the dot product is a two-dimensional
scalar product.
It is irrelevant which of the decay products is labelled $1$ and $2$.
We have written the definition in terms $m_{12}$, the invariant mass
of the two-body system and $\vec p_{t,12}$, the net transverse momentum of
the two-body system, which in the Higgs decay case are simply $\mH$
and ${\vec p}_{t,\Higgs}$.
At $p_{t,12}\equiv \ptH=0$, the second term in the square brackets
vanishes and since the two decay products are back to back,
$p_{t,\CS} = p_{t,1} = p_{t,2}$.
For general $\ptH$ it is straightforward to verify that
Eq.~(\ref{eq:ptCS}) yields $2p_{t,\CS}/m_{12} \equiv \sin\theta$.
Our cuts are then $p_{t,\CS} > p_{t,\cut}$ and
$p_{t,-} > p_{t,\cut} - \Delta$.

For values of $\ptH$ that are not too large,%
\logbook{}{xpTHboundary in boostinv-cuts.nb}
\begin{subequations}
  \begin{align}
    \label{eq:ptcs-asym-transition}
    \ptH < \ptH^{\CS\text{-threshold}}
    &= \frac{2 \mH \left(p_{t,\cut} \sqrt{\mH^2+4 \Delta  (\Delta -2 p_{t,\cut})}+\mH
      (\Delta -p_{t,\cut})\right)}{\mH^2-4 p_{t,\cut}^2},
    \\
    &= 2\Delta + \frac{4p_{t,\cut}\Delta^2}{\mH^2} + \ord3\,,
  \end{align}
\end{subequations}
the acceptance is simply $f(\ptH) = f_0$.
Including the region beyond the transition at first order in
$\ptH/\mH$ we have
\begin{equation}
  \label{eq:f-prod-CS}
  f^\text{\CS}(\ptH) = f_0
  - \frac{4p_{t,\cut}}{\pi \mH f_0} \frac{\chi(\ptH,2\Delta + \ord2)}{\mH}
  \left(1 + \mathcal{O}_2\right),
\end{equation}
with, once again, the usual transition beyond $\ptH \simeq2\Delta$.

\begin{figure}
  \centering
  \includegraphics[width=0.7\textwidth,page=3]{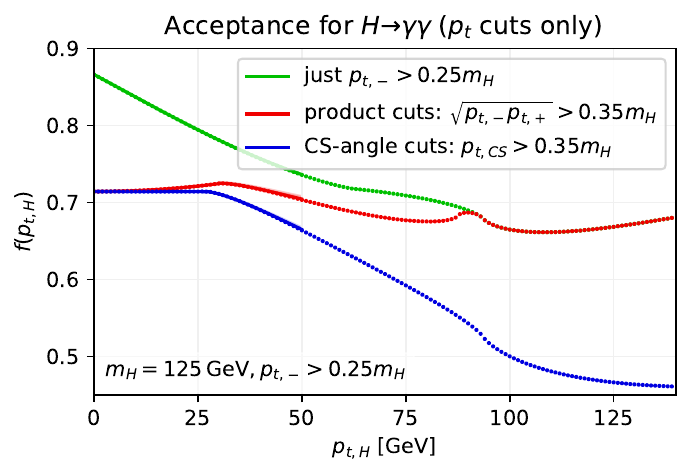}
  \caption{Comparison of the $\ptH$-dependent acceptance of the
    Collins-Soper 
    $p_{t,\CS}$ cut with that of the product cuts (both sets of cuts
    also include the constraint $p_{t,-}>0.25\mH$), as well as a cut
    just on the softer photon $p_t$ (i.e.\ a symmetric cut).
    As in Figs.~\ref{fig:acceptances-sym-asym} and
    \ref{fig:acceptances-sum-prd}, the points corresponds 
    to Monte Carlo evaluations of the acceptances.
    Note the different scale used here.
  }
  \label{fig:acceptances-ptCS}
\end{figure}

Fig.~\ref{fig:acceptances-ptCS} compares the $p_{t,\CS}$ cut
acceptance with that of the product cut (both additionally include the
requirement $p_{t,-}>0.25\mH$).
Up to $\ptH \simeq 2\Delta = 0.2 \mH$, the $p_{t,\CS}$ cut
acceptance is independent of $\ptH$, as desired.
However at larger $\ptH$ values, it has a considerably lower
acceptance than the product cut.
This is less than ideal phenomenologically, because in that region
events are relatively rare and one may wish to maximise
acceptance within the constraints posed by the photon reconstruction
requirement, $p_{t,-}>0.25\mH$ (shown in green on the plot).
For $\ptH \gtrsim 2\Delta$, the product cut comes much closer to
achieving this.\footnote{This discussion neglects the question of
  the relative impact of the cuts on signals and backgrounds.}
It turns out that it is possible to design techniques to recover the
high-$\ptH$ acceptance that is lost with the $p_{t,\CS}$ cuts.
The approach that we take will be useful not just for addressing
$\ptH$-acceptance dependence with hardness cuts, but also with
rapidity cuts.
Accordingly we introduce it in section~\ref{sec:cbi-cuts}, after our
discussion of rapidity cuts and their interplay with hardness cuts.

One feature that the reader may have noticed is that the coefficient
of the $\chi$ function is identical for all of the cuts in this
section, yet even those that share transitions at $\ptH \simeq
2\Delta$ show rather different behaviours beyond that point (for
example product and $p_{t,\CS}$ cuts in Fig.~\ref{fig:acceptances-ptCS}).
Recall, however, that the $\chi$ function in
Eqs.~(\ref{eq:f-sum-Delta},\ref{eq:f-prod-Delta},\ref{eq:f-prod-CS})
is accompanied by higher powers of $\ptH/\mH$ and $\Delta/\mH$, which
we have not explicitly written down analytically (but which are included in
the lines in
Figs.~\ref{fig:acceptances-sum-prd} and \ref{fig:acceptances-ptCS}). 
It is those higher-power terms that are responsible for the different
behaviours beyond $2\Delta$.

\section{Rapidity cuts}
\label{sec:rapidity-cuts}

Our focus so far has been on hardness cuts, which are essential for
eliminating the high event rates associated with low-$p_t$ photons,
leptons, etc.\ and for
avoiding momentum regions where objects may be poorly measured.
Real experiments also have limitations on the range of
(pseudo)rapidities over which they can measure particles, which also
induce $\ptH$ dependence of the acceptance.

Since we are working with an assumption of massless (or
quasi-massless) decay products, rapidities and pseudorapidities are
identical.
For brevity, here, we just use the term rapidity.

The key results that we will obtain in this section are that a single
rapidity cut leads to quadratic acceptance,
Eq.~(\ref{eq:onerap-acceptance}); that a linear $\ptH$ dependence
reappears when the Higgs rapidity is midway between two photon
rapidity cuts; and, for a rapidity cut in combination with quadratic
(or better) hardness cuts, linear dependence arises also at the Higgs
rapidity where the photon hardness and rapidity cuts are equivalent
($\delta_y = 0$ in Eq.~(\ref{eq:deltay-ptrap})).
Those linear behaviours are eliminated once one takes finite rapidity
bins around the critical rapidity points, though the residual
quadratic behaviours for the corresponding bins have a coefficient of
$\ptH^2$ that scales as the inverse bin width, suggesting that the
bins around those points should be reasonably wide.

On a first pass, some readers may wish to skip the technical details
that we provide in this section and go straight to the practical
example in section~\ref{sec:worked-example}.

\subsection{A single rapidity cut}
\label{sec:one-rap-cut}

Suppose that we have a Higgs boson at rapidity $\yH = 0$, decaying to
positive and negative-rapidity photons, $\gamma_{y_+}$ and
$\gamma_{y_{-}}$, with rapidities $y_{y_+}$ and $y_{y_-}$
respectively.
We start our study of rapidity cuts by examining what happens with a
single rapidity cut, $y_{y_+} < y_\cut$.
We will again follow the methods of section~\ref{sec:existing-cuts},
in particular keeping the parametrisation
Eq.~(\ref{eq:photon-momenta}).
The one critical difference is that while we will maintain
$0 \le \theta \le \pi/2$, we will allow the full azimuthal range
$-\pi \le \phi \le \pi$, because that full range yields $y_{+} \ge 0$
and $y_{-} \le 0$.
For a given value of $\phi$, the rapidity cut can straightforwardly be
translated to a limit on $\cos\theta$,
\begin{equation}
  \label{eq:onerap-costheta}
  \cos\theta < \tanh y_\cut \left[
  1 + \frac{\cos\phi}{\cosh y_\cut} \cdot \frac{\ptH}{\mH}
  + \frac12 \left(\csch^2 y_\cut - \cos 2\phi\right)\tanh^2 y_\cut \cdot
  \frac{\ptH^2}{\mH^2}
  + \ord3\right].
\end{equation}
Integrating over the full $\phi$ range, the term linear in $\ptH$
averages to zero and we are left with an acceptance with a quadratic
dependence on $\ptH$, which for general Higgs boson rapidity $\yH$
reads
\begin{equation}
  \label{eq:onerap-acceptance}
  f^{<y_\cut}(\ptH,\yH) = \Theta(y_\cut - \yH) \tanh (y_\cut-\yH)\left[1 
    + \frac12 \sech^2 (y_\cut-\yH) \cdot \frac{\ptH^2}{\mH^2}
    + \ord4
  \right].
\end{equation}
The mechanism that causes the linear term to vanish is fundamentally
different from that in the sum and product $p_t$ cuts in
section~\ref{sec:good-simple-cuts}: there the product and sum
kinematic variables were free of any linear $\ptH$ term,
independently of $\phi$; here the linear term is present for almost
all $\phi$ values and it is only after the $\phi$ integral that it
drops out, as for the staggered cuts of
section~\ref{sec:staggered-cuts}. 
Anything that breaks the $\phi \to \pi + \phi$ cancellation will
result in the linear dependence reappearing.

\subsection{Two rapidity cuts}
\label{sec:two-rap-cut}

To understand the behaviour of the acceptance in situations with two
rapidity cuts, we situate the Higgs boson at rapidity $\yH$ and apply
a requirement that both decay photons should satisfy $|y| < y_\cut$.
Our first observation is that if $\yH = 0$, we have
Eq.~(\ref{eq:onerap-costheta}) for the positive-rapidity photon and a
similar relation for the negative-rapidity photon with the replacement
$\phi \to \pi + \phi$, i.e.\ a change of sign for the $\cos \phi$
term.
Both conditions must be satisfied simultaneously: in the region
$|\phi| < \pi/2$ it is the condition on the negative-rapidity photon
that will be more constraining, while for $\pi/2<|\phi|<\pi$ it is the
condition on the positive-rapidity photon that will be more
constraining.
The acceptance can be obtained by integrating just the condition for
the negative-rapidity photon in the region $|\phi| < \pi/2$, giving
\begin{equation}
  \label{eq:f-symrapcut}
  f^{\lessgtr\pm y_\cut}(\ptH,\yH=0) =
  \tanh y_\cut\left[1
    - \frac{2}{\pi} \sech y_\cut \cdot \frac{\ptH}{\mH}
    + \frac12 \sech^2 y_\cut \cdot \frac{\ptH^2}{\mH^2}
    + \ord3
    \right].
\end{equation}
The presence of a linear $\ptH$ term is a consequence of the loss
of the azimuthal cancellation between $\phi$ and $\phi + \pi$.
The linear $\ptH$ dependence extends down to $\ptH=0$ only if the
Higgs boson is exactly mid-way between the rapidity cuts.
If we retain the cuts at $\pm y_\cut$, but give a small non-zero
rapidity to the Higgs boson, we obtain
\begin{equation}
  \label{eq:f-symrapcut-offset}
  f^{\lessgtr\pm y_\cut}(\ptH,\yH) = f^{y_\cut}(\ptH,|\yH|)
  - \frac{2 \tanh y_\cut }{\pi \mH \cosh y_\cut} \,
  \chi\!\left(\ptH, \frac{|\yH|\, \mH}{\sinh y_\cut} +\ord2\right)
  (1 + \ord2)\,,
\end{equation}
i.e.\ we retain the quadratic behaviour of
Eq.~(\ref{eq:onerap-acceptance}) up to
$\ptH = |\yH|\, \mH/\sinh y_\cut$ and then observe a transition
with the usual $\chi$ function.

In practice, experimental results that are differential in $\yH$ are
presented in finite bins of $\yH$.
The rapidity bin that is most critical is the one that contains a
Higgs rapidity midway between two photon rapidity cuts (i.e.\ the
$\yH = 0$ point in our simple example here).
If we consider a rapidity bin of half-width $\delta$ that covers
the region $|\yH| < \delta$, assuming that we can ignore the dependence of the
Higgs differential cross section on $\yH$,\footnote{This is valid
  to first order in $\delta$, since first order effects cancel between
  positive and negative $\yH$ values.} we obtain
\begin{multline}
  \label{eq:f-symrapcut-rapbin-integral}
  \left\langle f^{\lessgtr\pm y_\cut}(\ptH, \yH) \right\rangle_{|\yH| < \delta}
  = \frac{1}{\delta}\int_{0}^{\delta}
  f^{y_\cut}(\ptH,|\yH|) d\yH\,
  +
  \\
  - \frac{1}{\delta}
  \cdot  \frac{2 \tanh y_\cut }{\pi \mH \cosh y_\cut}
  \int_{0}^{\min\left(\delta,\frac{\ptH \sinh y_\cut}{\mH}\right)}
  \chi\!\left(\ptH, \frac{\yH\, \mH}{\sinh y_\cut}\right) d\yH
  + \ord2\,.
\end{multline}
At this point, it is useful to define
\begin{equation}
  \label{eq:chibar-def}
  \bar\chi(x,\delta) = \int_0^\delta d\delta_y \, \chi(x,\delta_y)\,,
\end{equation}
which is zero for negative values of $\delta$ and otherwise evaluates
to
\begin{equation}
  \label{eq:chibar-result}
  \bar\chi(x,\delta) \equiv
  \left\{
    \begin{array}{lll}
      \frac{\pi}{8}x^2\,, && x \le \delta\,,
         \\
      \frac{3 \delta}{4} \sqrt{x^2-\delta^2}
      +\frac{x^2}{4} \left(2 \arctan \frac{\delta}{\sqrt{x^2-\delta^2}}
                           - \arcsin\frac{\delta}{x}\right)
      -\frac{\delta^2}{2} \arccos\frac{\delta}{x}\,,
      && x > \delta\,.
    \end{array}
  \right.
\end{equation}
The critical features to observe are the quadratic behaviour in $x$
for small values of $x$, while for $x\gg \delta$, $\bar\chi(x,\delta)$
is approximately equal to $\delta \cdot x$.
The result for Eq.~(\ref{eq:f-symrapcut-rapbin-integral}) can now be
written
\begin{equation}
  \label{eq:f-symrapcut-rapbin-result}
  \left\langle f^{\lessgtr \pm y_\cut}(\ptH, \yH) \right\rangle_{|\yH| < \delta}
  =
  \tanh y_\cut - \left(\frac{\delta}{2\cosh^2y_\cut} +
    \frac{2\tanh^2 y_\cut}{\delta \cdot  \pi \mH^2}
    \bar\chi\!\left(\ptH, \frac{\yH\, \mH}{\sinh y_\cut}\right) \right) + \ord2\,,
\end{equation}
where we count powers of $\ptH/\mH$ and $\delta$ on the same
footing from the point of view of the series expansion.
When $\ptH < \delta \cdot \mH  / \sinh y_\cut$, this becomes
\begin{multline}
  \label{eq:f-symrapcut-rapbin-integral-full-result}
  \left\langle f^{\lessgtr \pm y_\cut}(\ptH, \yH) \right\rangle_{|\yH| < \delta}
  =
    \tanh y_\cut - \left(\frac{\delta}{2\cosh^2y_\cut} + \frac{\tanh^2
            y_\cut}{4\delta} \cdot \frac{\ptH^2}{\mH^2} \right) + \ord2\,.
\end{multline}
We see that for small $\ptH$, the linear $\ptH$ dependence that
was present in Eq.~(\ref{eq:f-symrapcut}) vanishes, to be replaced by
a quadratic dependence on $\ptH$ that is enhanced by $1/\delta$.
This leads to an important practical consideration: in the vicinity of
a rapidity value that is midway between two rapidity cuts, one should
ensure that the rapidity bin has a half-width $\delta$ that is not too
small, so as to ensure that the coefficient of the quadratic
$\ptH^2/\mH^2$ dependence, $\tanh^2 (y_\cut-\yH) / 4\delta$, is not
too large.

Our final comments here concern other combinations of rapidity cuts,
which are relevant when considering rapidity ranges with excluded
bands, e.g.\ $|\eta_\gamma| < 2.37$ but excluding
$1.37 < |\eta_\gamma| < 1.52$ for ATLAS~\cite{Aaboud:2018xdt} and
similar cuts for CMS~\cite{Sirunyan:2018kta}.
Considering a pair of cuts at a time, there are two generic situations
beyond that discussed above.
One can be cast as a requirement $y_{y_-} < -y_\cut$ and
$y_{y_+} < y_\cut$, in which case the result is given by
\begin{equation}
  \label{eq:rapcuts-lt-lt}
  f^{<\pm y_\cut}(\ptH,\yH) = f^{<+y_\cut}(\ptH,\yH) - f^{\lessgtr \pm y_\cut}(\ptH,\yH)\,,
\end{equation}
while the other can be cast as the requirement $y_{y_-} < -y_\cut$ and
$y_{y_+} > y_\cut$, yielding 
\begin{equation}
  \label{eq:rapcuts-gt-gt}
  f^{\gtrless \pm y_\cut}(\ptH,\yH) =
  1  - \left(f^{<+y_\cut}(\ptH,\yH) + f^{>-y_\cut}(\ptH,\yH)
             - f^{\lessgtr \pm y_\cut}(\ptH,\yH) \right).
\end{equation}
From these results, one sees that the midpoint between any pair of
rapidity cuts will involve the same kinds of structures.
\logbook{}{See
  GavinsCode/c++/2021-rapcut/noptcut-excl-checks.pdf for checks at two
  $\yH$ values}

\subsection{Combination of rapidity and $p_t$ cuts}
\label{sec:ptrap-cut}

\logbook{}{All results here are from math-2021/rapidity-pt-cuts.nb}

The final situation that needs to be considered is that when the
rapidity and transverse momentum cuts for the decay products cover
similar phase space.
In terms of our understanding of the small-$\ptH$ behaviour of the
acceptance, the relevant region is when the rapidity and leading
photon $p_t$ cut lead to similar acceptances for $\ptH = 0$, i.e.\
when
\begin{equation}
  \label{eq:deltay-ptrap}
  \delta_y \equiv \yH - \left(y_\cut - \operatorname{arccosh}
    \frac{\mH}{2p_{t,\cut}} \right)
   \ll 1\,,
\end{equation}
working in the regime $\yH < y_\cut$ and where the rapidity cut vetoes
photons with $y > y_\cut$.
Many of our hardness cuts involve two scales: a main hardness cut,
$p_{t,\cut}$, and a subsidiary condition on the softer decay product,
$p_{t,-} > p_{t,\cut} - \Delta$.
To simplify our analysis here, we will work with the assumption that
we can neglect that softer cut, or correspondingly,
$\delta_y \ll \Delta/\mH$.

To keep our results compact it will be helpful to define a ``base''
result for a hardness cut H
\begin{equation}
  \label{eq:fbase}
  f_\text{base}^\text{H}(\ptH, \delta_y) =
  \left\{
  \begin{array}{lll}
    f^\text{H}(\ptH) && \delta_y < 0\,,
    \\
    f^{<y_\cut}(\ptH, \delta_y) && \delta_y > 0 \,,
  \end{array}
  \right.
\end{equation}
which selects the correct small $\ptH$ behaviour for the efficiency
depending on whether $\delta_y< 0$ (the hardness cut is the only
relevant one) or $\delta_y> 0$ (the rapidity cut is the only relevant
one).

It will be convenient to use the shorthands
\begin{equation}
  \label{eq:shorthands}
  s_0 = \frac{2p_{t,\cut}}{\mH}\,,\qquad f_0 = \sqrt{1-s_0^2}\,.
\end{equation}
For a symmetric $p_t$ cut we obtain
\begin{equation}
  \label{eq:sym+rap}
  f^\text{sym}(\ptH, \delta_y) \simeq
  f_\text{base}^\text{sym}(\ptH, \delta_y)
  - \frac{\left(1+f_0^2\right) s_0 }{\pi  f_0 }
  \chi \!\left(\frac{\ptH}{\mH},\frac{f_0 s_0 \delta _y}{1+f_0^2}\right)
  - \frac{s_0^3 }{\pi  f_0} \chi\! \left(\frac{\ptH}{\mH},\frac{f_0 \delta _y}{s_0}\right).
\end{equation}
Throughout this section, when we write $\simeq$, it means that
we drop $\ord2$ terms in the structure associated with the $\chi$
functions, both for the shape of the function and the location of its
turn-on. 
Recall that our definition of the $\chi$ function, Eq.~(\ref{eq:chi}),
includes a $\Theta$ function such that $\chi$ is non-zero only when
the first argument (which is always positive) is larger than the
second one and the second one is positive.
Thus Eq.~(\ref{eq:sym+rap}) shows that the combination of a $p_t$ and
rapidity cut induces changes in the acceptance relative to
Eq.~(\ref{eq:fbase}) only when $\delta_y > 0$.\footnote{
  This is when we work at first order in $\delta$ and $\ptH$.
  Working to second order, there is an additional transition for
  $\delta_y < 0$ at
  $\ptH = \mH\sqrt{-2f_0 \delta_y} + \order{\mH \delta_y}$, which is
  parametrically larger than the other transitions that we see, which
  are all at $\ptH$ of order $\delta_y \mH$.  }
The presence of two $\chi$ functions tells us that there are two
values of $\ptH$ at which there will be a transition in the
$\ptH$ dependence.
They are associated with the intersections of the $p_{t}$ and rapidity
cuts at $\phi = \pi$ and $\phi = 0$ respectively.

\begin{figure}
  \centering
  \includegraphics[width=\textwidth,page=5]{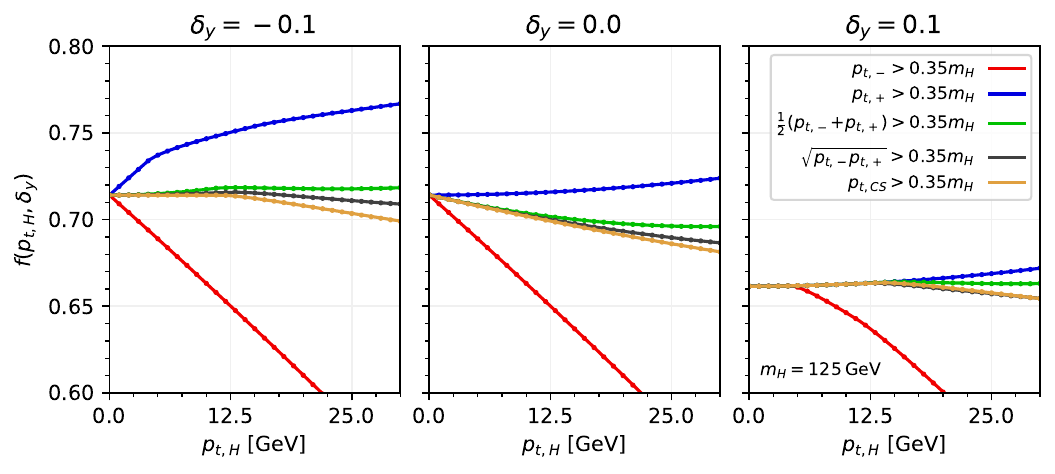}
  \caption{%
    Acceptance for a combination of a hardness cut (as specified in
    the legend) and a single photon rapidity cut.  %
    The quantity $\delta_y$ is defined in
    Eq.~(\ref{eq:deltay-ptrap}).  %
    For the hardness cuts that would normally have two separate
    hardness
    scales (i.e.\ all but the symmetric cut) we have set the lower cut to
    zero, so as to concentrate on the interplay between the harder cut
    and rapidity cut, as is appropriate in the region
    $\delta_y \ll \Delta/\mH$ that we set out to study.
    The points, lines and bands follow the usual convention of
    Fig.~\ref{fig:acceptances-sym-asym}. 
  }
  \label{fig:acceptances-ptrap}
\end{figure}

The resulting acceptance is shown as the red lines
($p_{t,-}>0.35 \mH$) in Fig.~\ref{fig:acceptances-ptrap}, with each
panel corresponding to a different value of $\delta y$.
With the cuts shown, $s_0 = 0.7$ and $f_0 \simeq 0.714$ are almost
equal.
For $\delta_y \le 0$, the rapidity cut has no impact over the range
shown, while for $\delta_y =0.1$, one sees a first kink at
$\ptH\simeq 0.33\, \delta_y \mH \simeq 4.1\GeV$ and a second, weaker
kink at $\ptH \simeq 1.02\,\delta_y \mH \simeq 12.8\GeV$.
Note that for $\ptH$ significantly larger than the position of the
kinks, the normalisations of the $\chi$-terms in
Eq.~(\ref{eq:sym+rap}) add up to give $-\frac{2s_0}{\pi f_0}$ multiplying
$\ptH/\mH$, which is precisely $f_1^\text{sym}$ in
Eq.~(\ref{eq:fsym-linear}).

\logbook{}{$1+f_0^2 + s_0^2 = 2$, and one can then verify that for
  $\ptH$ above both cuts, one obtains the slope in
  Eq.~(\ref{eq:fsym-linear}), which is $2s_0/\pi f_0$.
  This is what we expect.
}

For a standard asymmetric cut, which in our limit of
$\delta_y \ll \Delta/\mH$ is equivalent to a cut on the $p_t$ just of the
harder photon, we obtain 
\begin{equation}
  \label{eq:max+rap}
  f^\text{asym}(\ptH, \delta_y) \simeq
  f_\text{base}^\text{asym}(\ptH, \delta_y)
  - \frac{\left(1+f_0^2\right)s_0 }{\pi  f_0 }\chi \left(\frac{\ptH}{\mH},-\frac{f_0 s_0 \delta _y}{f_0^2+1}\right)
  -\frac{s_0^3 }{\pi  f_0} \chi \left(\frac{\ptH}{\mH},-\frac{f_0 \delta  _y}{s_0}\right).
\end{equation}
Now the two transitions are present only for negative values of
$\delta_y$.
The resulting acceptance is again shown in
Fig.~\ref{fig:acceptances-ptrap}, as the blue lines
($p_{t,+}>0.35\mH$).
For $\delta_y \ge 0$,  the result corresponds to just the
rapidity cut, i.e.\ quadratic $\ptH$ dependence, while for
$\delta_y = -0.1$ one sees a clear first kink around
$0.33 |\delta_y| \mH$ and a weaker second kink around
$1.02 |\delta_y| \mH$.
The normalisation is such that sufficiently far beyond the second kink
(but before $\ptH = \Delta$ asymmetric cut transition would be
reached), the positive linear $\ptH$ dependence from $f_1^\text{asym}$
of Eq.~(\ref{eq:fasym-linear-lt-Delta}) is cancelled out.

Finally, using H to denote any of the sum, product and Collins--Soper boost-invariant
hardness cuts, we have
\begin{equation}
  f^\text{H}(\ptH, \delta_y) \simeq
  f^\text{H}_\text{base}(\ptH, \delta_y) 
  -\frac{f_0 s_0 }{\pi } \chi \left(\frac{\ptH}{\mH},\frac{s_0 |\delta _y|}{f_0}\right),
\end{equation}
where there is just a single transition, which is present for both
positive and negative values of $\delta_y$.
Again, the results are shown in Fig.~\ref{fig:acceptances-ptrap}.
For $\delta_y = -0.1$, one sees the three quadratic (or flat)
low-$\ptH$ acceptances for each of the sum, product or $p_{t,\CS}$
cuts, with a kink at
$\ptH\simeq 0.98 |\delta_y| \mH \simeq 12.3\GeV$, transitioning to a
linear $\ptH$ dependence.
For $\delta_y = 0$ that linear $\ptH$ dependence is present from
$\ptH = 0$ and the small differences between the three cuts reflect
their differing quadratic terms.
For $\delta_y = 0.1$, we initially have the quadratic $\ptH$
dependence that is characteristic of just the rapidity cut, and
following the kink a hint of linear dependence arising.
Note that the coefficient of the linear dependence, $f_0s_0/\pi$, is
about four times smaller than that for the symmetric or asymmetric
cuts.

At first sight, it is concerning that hardness cuts that led to
quadratic or flat $\ptH$ dependence, when combined with a rapidity
cut, now reacquire linear dependence at $\delta_y = 0$.
However, as with the discussion of pairs of rapidity cuts in
section~\ref{sec:two-rap-cut}, the question that is ultimately
relevant for practical perturbative calculations is not what happens
at a given point in rapidity, but what happens after integration over
a given rapidity bin.
For $H$ denoting any of the sum, product, and $p_{t,\CS}$ cuts,
considering a bin of half width $\delta$ centred at $\delta_y = 0$, we
have
\begin{equation}
  \label{eq:f-ptrapcut-rapbin-result}
  \left\langle f^{H,y_+ < y_\cut}(\ptH, \delta_y) \right\rangle_{|\delta_y| < \delta}
  \simeq
  f^{H}(\ptH) - \left(\frac{s_0^2}{4}\delta
    +
    \frac{f_0^2}{\pi  \delta} \bar{\chi
    }\left(\frac{\ptH}{\mH},\frac{ s_0 \delta}{f_0}\right)
  \right),
\end{equation}
with $\bar\chi$ as given in Eq.~(\ref{eq:chibar-result}).
For $\ptH < \frac{s_0 \delta}{f_0} \mH $\, the result reduces to
\begin{equation}
  \label{eq:f-ptrapcut-rapbin-result-lowpt}
  \left\langle f^{H,y_+ < y_\cut}(\ptH, \delta_y) \right\rangle_{|\delta_y| < \delta}
  \simeq
  f^{H}(\ptH) - \left(\frac{s_0^2}{4}\delta
    +
    \frac{f_0^2}{8 \delta} \frac{\ptH^2}{\mH^2}
  \right).
\end{equation}
As with the rapidity integral for the pair of rapidity cuts, we see
quadratic $\ptH$ dependence, multiplied by $1/\delta$, i.e.\ it is
once again important to ensure that the bin in Higgs rapidity around
$\delta_y = 0$ is not too small.
In practice, the coefficient $f_{0}^2 / 8$ is quite small, and taking
$\delta$ in the range $0.1{-}0.2$ is probably adequate.

A final comment concerns the case where we combine a hardness cut with
a rapidity cut $y_{+} > y_{\cut}$.
The acceptance can be straightforwardly deduced from our existing
results
\begin{equation}
  \label{eq:fX-yplus-gt-ycut}
  f^{H,y_+ > y_\cut}(\ptH, \delta_y)
  =
  f^{H}(\ptH) - 
  f^{H,y_+ < y_\cut}(\ptH, \delta_y)\,.
\end{equation}
%

\subsection{A worked example}
\label{sec:worked-example}

\begin{figure}
  \centering
  \includegraphics[width=0.7\textwidth,page=1]{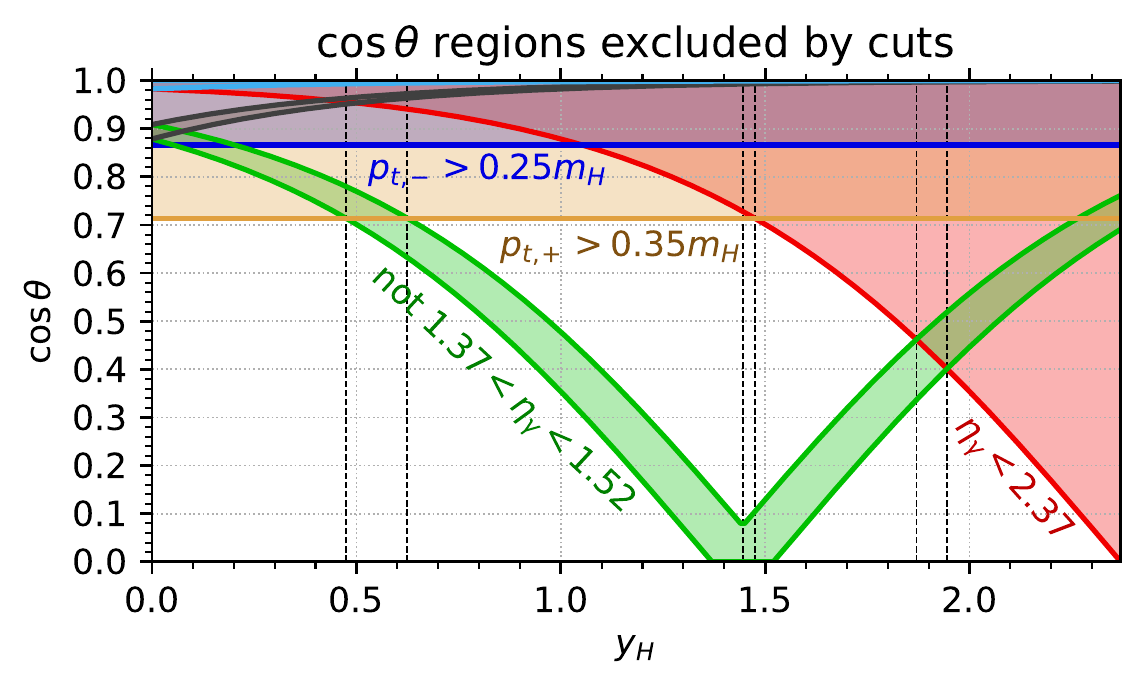}
  \includegraphics[width=0.7\textwidth,page=2]{2021-figs/acceptances-v-rapidity.pdf}
  \caption{Upper plot: the $\cos\theta$ regions that are excluded by
    the main relevant photon $p_t$ and rapidity cuts, as a function of
    the Higgs boson rapidity $\yH$, for $\ptH = 0$.
    The vertical lines indicate the $\yH$ values where pairs of cuts
    have equivalent actions, i.e.\ corresponding to the special cases
    outlined in sections~\ref{sec:two-rap-cut} and \ref{sec:ptrap-cut}.
    The lower plot shows the Born ($\ptH=0$) acceptance with these
    cuts, as a function of $\yH$. }
  \label{fig:all-ATLAS-cuts}
\end{figure}

To help make this section's discussion a bit more concrete, we
conclude it with a worked example using a concrete set of cuts.
We take the cuts used by the ATLAS
collaboration~\cite{Aaboud:2018xdt,ATLAS:2019jst},
$p_{t,+} > p_{t,\cut} = 0.35 m_{\gamma\gamma}$,
$p_{t,-} > 0.25 m_{\gamma\gamma}$,
$|\eta_\gamma|\equiv|y_\gamma| < y_{\max}$ and excluding photons in
the region $y_1 < |\eta_\gamma| < y_2$, with $y_{1}=1.37$, $y_2=1.52$
and $y_{\max} =2.37$.
The CMS collaboration uses a similar structure of
cuts~\cite{Sirunyan:2018kta}, with the same value of $\mH/4$ for the $p_{t,-}$ cut,
$p_{t,\cut} = m_{\gamma\gamma}/3$ and $y_{1}=1.4442$, $y_2=1.566$ and
$y_{\max} =2.50$.\footnote{We do not discuss the impact of photon
  isolation, which has been considered from a perturbative point of
  view in Refs.~\cite{Ebert:2019zkb,Becher:2020ugp}.
  The ATLAS and CMS fiducial isolation procedures differ
  substantially.
  The ATLAS fiducial isolation~\cite{Aaboud:2018xdt,ATLAS:2019jst}
  requires the scalar sum of transverse momenta of charged particles
  with $p_t > 1\GeV$ within a radius of 0.2 around the photon to be
  less than $5\%$ of the photon transverse momentum.
  This is an intrinsically non-perturbative definition, since it
  involves charged particles with a momentum cut, and it is also
  likely to be quite sensitive to multi-parton
  interactions.\logbook{c9432ab}{It actually is quite sensitive, see
    mc-analysis/2021-runs/README.md at 14 TeV with Monash 13; it costs
    $2.6\%$ of events with UE, $1.6\%$ without. }
  The CMS fiducial isolation criterion~\cite{Sirunyan:2018kta},\logbook{}{see
    p.8 of that ref}
  in contrast, is perturbative, simply requiring less than $10\GeV$ of
  transverse energy within $\Delta R = 0.3$ around each photon
  candidate.
  \logbook{7cde020}{It costs about $0.8
    $\pm 25\%$ seems insensitive to UE.
  }
  In discussing isolation perturbatively, one element to keep in mind
  is that fragmentation photon contributions will contribute to the
  continuum (e.g.\ combining one fragmentation and one direct photon),
  while direct photons from Higgs decay will contribute to the
  resonance peak.
  It is then conceptually important (though practically probably less
  so) to understand whether a quoted Higgs fiducial cross section
  includes just the resonant $\gamma\gamma$ contribution.}
Since we are considering just Higgs decays in this article, and not
the background of continuum $\gamma\gamma$ production, we will work
with the assumption $m_{\gamma\gamma}=\mH$.
The structures that we will identify at various different Higgs
rapidities will remain the same for a general $m_{\gamma\gamma}$ as
long as hardness cuts remain expressed as a fraction of
$m_{\gamma\gamma}$.

Fig.~\ref{fig:all-ATLAS-cuts} (upper plot) shows the regions of
$\cos\theta$ that are excluded by the ATLAS cuts, as a function of the
Higgs rapidity, for $\ptH = 0$.
The lower plot shows the resulting efficiency.
Each value of $\yH$ where two cuts intersect (so long as the
intersection borders the allowed region) leads to one of the special
configurations discussed in sections~\ref{sec:two-rap-cut} and
\ref{sec:ptrap-cut}.
Those $\yH$ values are indicated with dashed vertical lines and they
correspond to kinks in the $\yH$-dependence of the acceptance.
They arise at the midpoints between any pair of the (same-sign) rapidity cut
values,
\begin{subequations}
  \label{eq:critical-yH}
  \begin{equation}
    \label{eq:critical-yH-ycut-pair}
    |\yH| = \left\{\frac{y_1+y_2}{2} = 1.445,\quad
    \frac{y_1+y_{\max}}{2} = 1.87, \quad
    \frac{y_2+y_{\max}}{2} = 1.945\right\}, \quad
  \end{equation}
  and at each of the points where the rapidity cut and the main
  hardness cut are equivalent
  \begin{equation}
    \label{eq:critical-yH-ycut-ptcut}
    |\yH| = y_{\cut} - \operatorname{arccosh}
    \frac{\mH}{2p_{t,\cut}} \simeq \left\{ 0.474, 0.624, 1.474\right\},
  \end{equation}
\end{subequations}
where $y_{\cut}$ is any of $y_1$, $y_2$ and $y_{\max}$.

\begin{figure}
  \centering
  \includegraphics[scale=0.80,page=3]{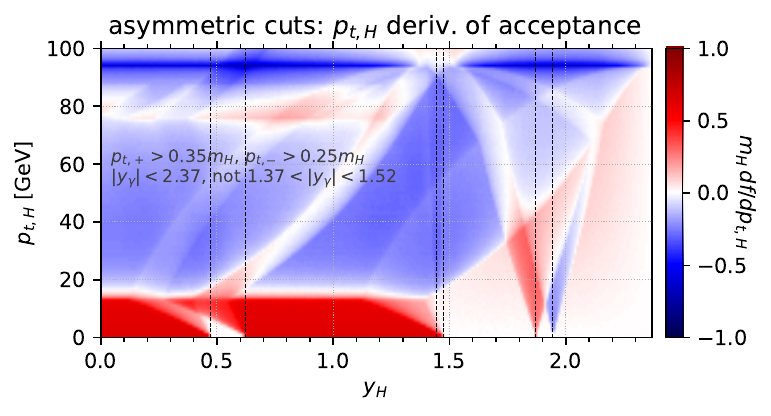}
  \includegraphics[scale=0.80,page=5]{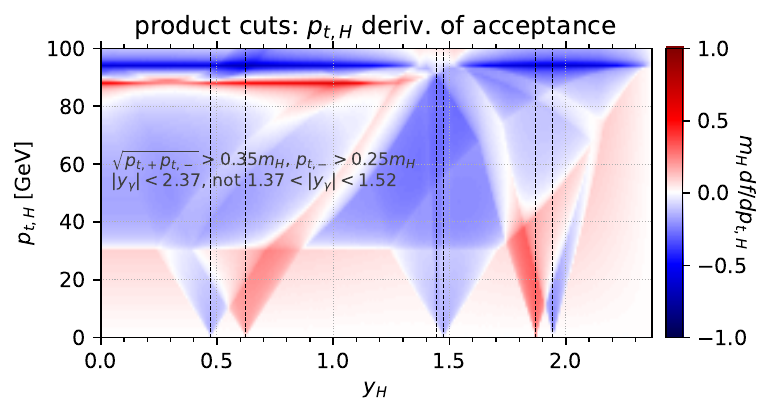}
  \includegraphics[scale=0.80,page=6]{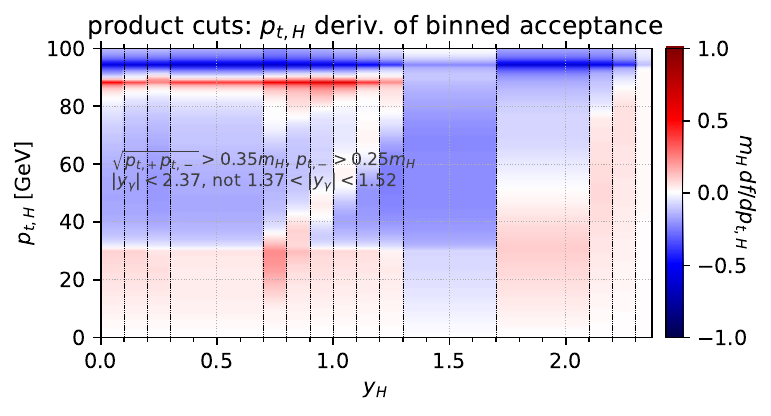}
  \caption{The $\ptH$ derivative of the acceptance, as a function
    of the Higgs boson rapidity (horizontal axis) and transverse
    momentum (vertical axis).
    Linear $\ptH$ dependence of the acceptance at small $\ptH$
    appears as a blue or red colour that persists in the $\ptH \to
    0$ limit.
    The upper plot shows the results for the standard ATLAS asymmetric
    cuts.
    The middle plot shows the results obtained replacing the
    $p_{t,+} > 0.35 \mH$ cut with a product cut,
    $\sqrt{p_{t,-}p_{t,+}} > 0.35\mH$, while retaining the
    $p_{t,-} > 0.25 \mH$ cut and all photon rapidity cuts.
    The lower plot shows the same results averaged over rapidity
    bins.  }
  \label{fig:all-ATLAS-cuts-eff-derivatives}
\end{figure}

To help understand the $\ptH$ dependence of the cuts in different
regions, Fig.~\ref{fig:all-ATLAS-cuts-eff-derivatives} shows the
$\ptH$ derivative of the acceptance,
\begin{equation}
  \label{eq:deriv-figure}
  \mH \frac{df(\ptH,\yH)}{d\ptH}\,,
\end{equation}
as a function of $\yH$ (horizontal axis) and $\ptH$ (vertical
axis).
The value of the derivative is encoded in the colour of the points.
The top panel is for standard asymmetric cuts used by the ATLAS
collaboration.
The deep red colour at low $\ptH$, over a wide range of $\yH$
values is the signal of linear dependence of the acceptance on the
cuts.
One also sees a complex structure of bands at higher $\ptH$ values,
reflecting the interplay between the many rapidity and $p_{t}$ cuts.

The middle panel of Fig.~\ref{fig:all-ATLAS-cuts-eff-derivatives} shows
the result obtained if one replaces the asymmetric cuts with product
cuts, $\sqrt{p_{t,-}p_{t,+}} > 0.35 \mH$ and $p_{t,-} > 0.25 \mH$,
maintaining the ATLAS rapidity cuts.
In the low $\ptH$ region, over most $\yH$ values, one sees that the
$\ptH$ derivative of the acceptance vanishes at $\ptH = 0$,
consistent with an overall quadratic $\ptH$ dependence of the
acceptance.
This pattern breaks down at each of the special rapidities highlighted
with a vertical dashed line, i.e.\ the rapidities of
Eq.~(\ref{eq:critical-yH}), where one sees that the acceptance
derivative remains non-zero all the way to $\ptH=0$, precisely as
expected from our discussion in sections~\ref{sec:two-rap-cut} and
\ref{sec:ptrap-cut} (the $\yH=1.445$ transition is too narrow to see
with the plot's resolution).%
\logbook{}{But I checked with higher resolution that it does seem to
  go down to zero --- the various raw data files should include the
  $\yH=1.445$ point.}
The transition from quadratic to linear $\ptH$ dependence is at progressively
higher $\ptH$ as one moves away from those $\yH$ values.
This is why, if we integrate over Higgs rapidity bins that are
sufficiently large in those regions, e.g.\ $0.4{-}0.7$, $1.3{-}1.7$
and $1.7{-}2.1$, we expect to recover mild quadratic $\ptH$ dependence
at low $\ptH$, with linear quadratic dependence setting in only for
higher $\ptH$ values (e.g.\ $\ptH > \delta s_0/f_0 \mH$ in the case of
the combination of hardness and rapidity cuts, for a rapidity bin of
half-width $\delta$).
This is illustrated in the bottom panel of
Fig.~\ref{fig:all-ATLAS-cuts-eff-derivatives} (the pattern of colours
alone perhaps does not give full confidence that the linear term is
consistent with zero, however an explicit inspection of the results in
each rapidity bin confirms that this is indeed the case).

\section{Compensating Boost Invariant  cuts}
\label{sec:cbi-cuts}

We saw in section~\ref{sec:cuts-collins-soper} that if one uses a
hardness cut on the photon transverse momentum in the Collins-Soper
frame, Eq.~(\ref{eq:ptCS}), the acceptance is independent of $\ptH$
at low $\ptH$.
As it stands, that approach faces two problems.
Firstly, at larger $\ptH$ values, the acceptance is noticeably
lower than with other cuts.
Secondly, as we saw in section~\ref{sec:ptrap-cut}, when combining a
$p_{t,\CS}$ cut with rapidity cuts, the latter bring back $\ptH$
dependence of the acceptance at low $\ptH$.
In this section, we will examine how to alleviate both of these
problems, with techniques that balance loss and gain of acceptance
from different decay phase space regions.\footnote{An alternative,
  valid in the scalar decay case, is to take the approach of
  defiducialisation~\cite{Glazov:2020gza}.
  In some respects this is simpler than the approach that we explore
  here, though it is rigorously applicable only to the case of scalar
  decays and requires some form of rapid evaluation of the acceptance.  
  The approach that we explore here can, to some extent, be applied
  also to the vector case, as we shall see in
  section~\ref{sec:Drell-Yan}, and the underlying methods can also be
  of direct help with defiducialisation.
  Further discussion of defiducialisation is given in
  Appendix~\ref{sec:defid-remark}.  }

Achieving an acceptance that has no $\ptH$ dependence at low $\ptH$
has interesting implications.
A first observation is that for experimental measurements that correct
to a total cross section or an STXS cross
section~\cite{Berger:2019wnu}, having an experimental acceptance that
is independent of $\ptH$ removes one source of systematic (knowledge
of the $\ptH$ distribution) in the extrapolation to the more inclusive
cross section.
The second observation concerns the perturbative structure of fiducial
cross sections in the approximation that one can neglect the
perturbative impact of isolation cuts.\footnote{
  Where isolation cuts are non-perturbative, such as those imposed by
  ATLAS, they are in any case perhaps best modelled separately from a
  perturbative calculation, and one might even consider a data-driven
  approach to remove their impact from measurements.
  Where the isolation cuts are amenable to perturbative treatment, as
  with the CMS cuts, further thought would be warranted regarding
  their integration into the discussion here. }
Consider an acceptance $f(\ptH, \yH)$ that is independent of
$\ptH$ and equal to $f_0(\yH)$ up to some threshold transverse
momentum scale $p_t^\text{thresh}(\yH)$.
Suppose that we know the Higgs cross section differentially in
rapidity $d\sigma/d\yH$ (integrated over $\ptH$) and differentially
in both $\ptH$ and $ \yH$, $d\sigma/d\ptH d\yH$.
We can then write the fiducial cross section as
\begin{equation}
  \label{eq:sigma-fid-CBI}
  \sigma_\text{fid} = \int_{-\infty}^{\infty} d\yH \frac{d\sigma}{d\yH} f_0(\yH)
  - \int_{-\infty}^{\infty} d\yH
  \int_{p_t^\text{thresh}(\yH)}^{\infty}
  \!\!\!\!\!\!\!\!
  d\ptH 
  \frac{d\sigma}{d\ptH d\yH}
  \left[f_0(\yH) - f_0(\ptH, \yH)\right]
  \,,
\end{equation}
where infinite integration limits are to be understood as extending to
the kinematic limit.
In this way of writing the fiducial cross section, there is no
dependence at all on the details of the differential cross section at
any $\ptH$ below $p_t^\text{thresh}(\yH)$.

Eq.~(\ref{eq:sigma-fid-CBI}) means that all problems of
low-$\ptH$ acceptance-induced factorial divergences in the perturbative series
disappear, and the only issues that remain in the fiducial cross
section will be those intrinsic to hard cross sections.
Based on the experience with rapidity differential Drell-Yan cross
sections~\cite{Dasgupta:1999zm}, and assuming similar conclusions to
be valid for Higgs production, one expects the first term in
Eq.~(\ref{eq:sigma-fid-CBI}) to have renormalon power corrections of
the form $(\Lambda/\mH)^2$ (within the caveats mentioned in
Ref.~\cite{Beneke:1998ui}).
The work of Ref.~\cite{FerrarioRavasio:2020guj,Caola:2021kzt}, on corrections to
Drell-Yan production at finite $p_t$ was consistent with
$(\Lambda/p_t)^2$ corrections and if the conclusions apply also in the
case of Higgs production, would imply power corrections to the second
term that are no larger than $(\Lambda/p_t^\text{thresh})^2$.
If $p_t^\text{thresh}$ is sufficiently large, the fiducial cross
section should then have a perturbative description that is as
reliable as that of a normal (rapidity-differential) total cross
section.\footnote{One potential concern is that the cross section for an
  electroweak boson to be in some high-$p_t$ bin (inclusive over the
  boson decay orientations) could conceivably be subject to the same
  kinds of quadratic ``acceptance'' corrections as arise for a cut on
  a single photon in Higgs decay (cf.\
  section~\ref{sec:staggered-cuts}), but now the quadratic dependence
  is on the net $p_t$ of the boson plus recoiling jet system, rather
  than the $p_t$ of just the boson.
  This question perhaps warrants further study.  }

A more general comment about Eq.~(\ref{eq:sigma-fid-CBI}) is that
the dominant contribution will often come from the first term.
The second term is suppressed for two reasons: firstly, for
sufficiently large $p_t^\text{thresh}$, only a small fraction of the
cross section is above $p_t^\text{thresh}$;  secondly, in practice
$f_0(\ptH, \yH)$ is often numerically quite close to $f_0(\yH)$.

\subsection{The case with just hardness cuts}
\label{sec:extended-cs-cuts}
\logbook{54994de}{Results from this section are in 2021-math/boostinv-cuts-extra.nb}

As before when considering just hardness cuts, we work within a
framework where the only experimental $p_t$ cut that is essential is
that on $p_{t,-}$, i.e.\ there is some minimal $p_t$ below which the
experiments cannot reliably reconstruct photons (or leptons, etc.\ as
appropriate), but that apart from that we have complete flexibility
with other cuts.
We will start from the $p_{t,\CS}$ cut of
section~\ref{sec:cuts-collins-soper} and use the flexibility so as to
enhance the size of the region at low and moderate $\ptH$ where the
acceptance is exactly $\ptH$ independent, while at large values of
$\ptH$ we will seek to make the acceptance as close as possible to
that obtained with just a $p_{t,-}$ cut.

\begin{figure}
  \centering
  \includegraphics[width=\textwidth,page=1]{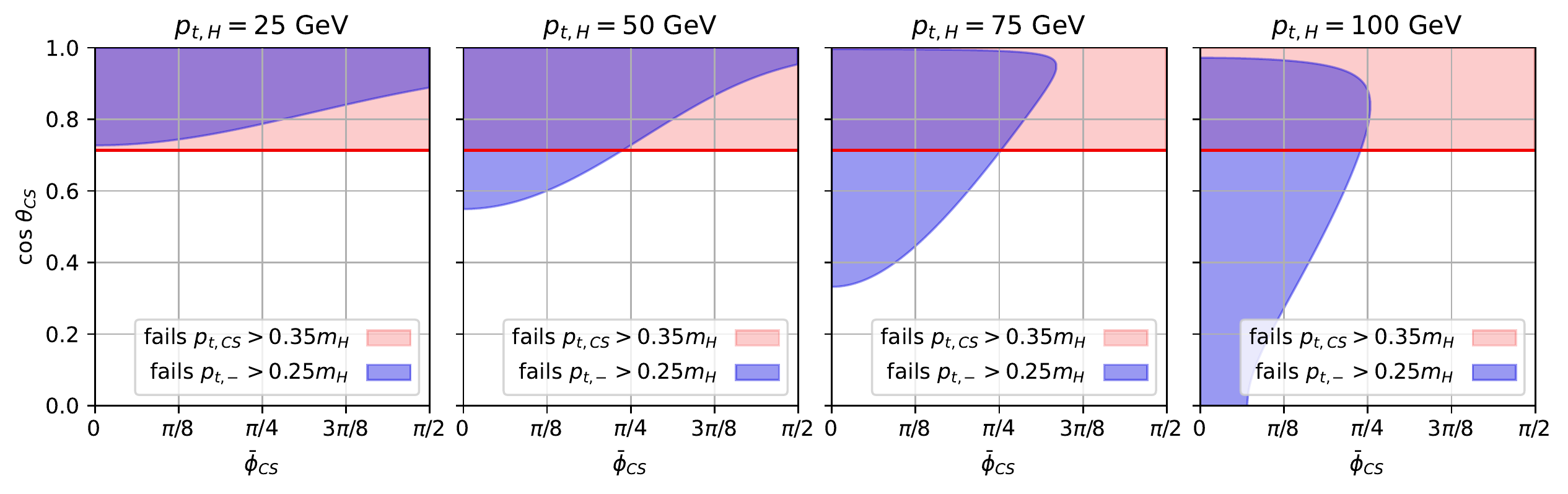}%
  \caption{Action of the $p_{t,\CS}$ and $p_{t,-}$ cuts of
    section~\ref{sec:cuts-collins-soper} in the
    $\bar \phi_{\CS}$--$\theta_{\CS}$ plane, as defined in
    Eq.~(\ref{eq:theta_phi_CS}), for four values of $\ptH$.  }
  \label{fig:ptCS-ptcut2-phasespace}
\end{figure}

Given $\vec p_{t,\CS}$ from Eq.~(\ref{eq:ptCS}), it is helpful to
define
\begin{equation}
  \label{eq:theta_phi_CS}
  \theta_{\CS} = \sin^{-1} \frac{2p_t,\CS}{m_{12}}\,,
  \qquad
  \phi_{\CS} = \cos^{-1}
  \frac{\vec \delta_{t,12} \cdot \vec p_{t,12}}{\delta_{t,12}  p_{t,12}}\,,
  \qquad
  \bar \phi_{\CS} = \min(\phi_\CS, \pi - \phi_{\CS})\,,
\end{equation}
such that $0 < \theta_\CS < \pi/2$, $0 < \phi_\CS < \pi$ and $0 < \bar \phi_\CS < \pi/2$.
Fig.~\ref{fig:ptCS-ptcut2-phasespace} shows the
$\bar \phi_{\CS}$--$\cos\theta_{\CS}$ plane for Higgs decay for
several values of $\ptH$.
Recall from the discussion of section~\ref{sec:existing-cuts} that the
scalar nature of the Higgs boson results in uniform coverage of the
plane.
The figures shows the action of the cuts used in
section~\ref{sec:cuts-collins-soper}, i.e.\ a $p_{t,\CS} \ge 0.35\mH$
cut, which excludes the pink region, and a $p_{t,-}\ge 0.25 \mH$ cut,
which excludes the blue region.
The $p_{t,\CS}$ cut is, by construction, a horizontal line independent
of $\ptH$.
To understand the behaviour of the $p_{t,-}$ cut, it is useful to
identify, as a function of $\phi_\CS$, the $\sin \theta_\CS$ values where
$p_{t,-} = p_{t,-,\cut} \equiv p_{t,\cut} - \Delta$.
There are two solutions for $\sin\theta_\CS$ and at low $\ptH$, only one of them is physical,
\begin{subequations}
    \label{eq:sintheta-limit-minus}
  \begin{align}
    \label{eq:sintheta-limit-minus-main}
    \sin \theta_\CS^{-}(\phi) &= \frac{
                  +2 \ptH \sqrt{\mH^2+\ptH^2} |\cos \phi| + \sqrt{U(\phi)}}{2
                  \mH^2+\ptH^2(1+ \cos 2 \phi )}\,,
    \\
    \label{eq:sintheta-limit-minus-U}
    U(\phi) &= 2 \left(\ptH^2 \left(\mH^2+4 p_{t,-,\cut}^2\right) \cos
        2 \phi - \ptH^2 \left(\mH^2-4 p_{t,-,\cut}^2\right)+8 \mH^2
        p_{t,-,\cut}^2\right)\,,
  \end{align}
\end{subequations}
where for compactness, we have dropped the $\CS$ subscript on the
$\phi$ and we write $\ptH$ and $\mH$ rather than $p_{t,12}$ and
$m_{12}$. 
For $\ptH=0$, $\sqrt{U}=4\mH p_{t,-,\cut}$ is the only contribution
to the numerator, the physical solution is that with $+\sqrt{U}$ and
we get $\sin \theta_\CS^-= 2p_{t,-,\cut}/\mH$, as expected.
Eq.~(\ref{eq:sintheta-limit-minus}) is the generalisation of the
small-$\ptH$ expansion given in
Eq.~(\ref{eq:symcut-sintheta-expansion}).

For $\ptH \ge \ptH^{\CS\text{-threshold}}\simeq 27\GeV$, cf.\
Eq.~(\ref{eq:ptcs-asym-transition}),
Fig.~\ref{fig:ptCS-ptcut2-phasespace} illustrates how the $p_{t,-}$
cut starts to extend beyond the $p_{t,\CS}$ cut in the region around
$\bar\phi_\CS=0$, which leads to the loss of efficiency that is
visible in Fig.~\ref{fig:acceptances-ptCS}.
Examining the low-$\bar\phi_\CS$ region in  the $\ptH=50\GeV$ panel of
Fig.~\ref{fig:ptCS-ptcut2-phasespace}, we notice that the phase space that
has been lost for $\bar\phi_\CS <\pi/4$ can potentially be
recovered by relaxing the $p_{t,\CS}$ cut for $\bar\phi_\CS >\pi/4$.
Specifically if $\bar\phi_\CS > \pi/4$, we can determine the value of
$\theta_\CS^{-}$ that would be obtained if one mirrored the $\phi$
value around $\pi/4$.
Let us refer to that as
\begin{equation}
  \label{eq:theta-CS-m}
  \theta_\CS^m = \theta_\CS^{-}(\pi/2-\bar\phi_\CS)\,.
\end{equation}
When $\cos \theta_\CS^m < f_0$, we can recover the phase space that was
lost for $\bar\phi_\CS <\pi/4$ by allowing $\cos\theta_\CS$ values up
to $2f_0 - \cos \theta_\CS^m$ rather than the usual $f_0$.
For typical cut values, the Born acceptance, $f_0$, is then retained
as long as there is enough phase space at a given
$\bar\phi_\CS >\pi/4$ to compensate for the phase space lost to the
$p_{t,-}$ cut at the mirrored $\phi$ value.
One requirement for this to be true is that $\cos \theta_\CS^{-}(\pi/4)
> f_0$.
For small values of $\Delta$, this is a sufficient requirement and
the Born acceptance can be retained up to
\begin{equation}
  \label{eq:pTH-max-flat-CS-extended}
  \ptH
  =
            2\sqrt{2} \Delta + \sqrt{2}\frac{\mH^2 + 4
    p_{t,\cut}^2}{p_{t,\cut}\mH} \frac{\Delta^2}{\mH} + \ord3\,,
\end{equation}
rather than $2\Delta$ for the $p_{t,\CS}$ cut.
(One can write the full expression for the transition point in
closed form, but it is not especially illuminating).
\logbook{See midpointSeriesSoln and midpointSoln or midpointSolnDelta in boostinv-cuts-extra.nb}

Fig.~\ref{fig:ptCS-ptcut2-phasespace} is instructive also for thinking
about how to maximise the acceptance for yet larger $\ptH$.
In the $\ptH = 75\GeV$ panel, one sees a region of $\bar \phi_\CS
\gtrsim 5\pi/16$
where the $p_{t,-}$ cut is inactive, corresponding to negative values
for $U(\bar \phi_\CS)$ in Eq.~(\ref{eq:sintheta-limit-minus}).
Such a region starts to appear for $\ptH > 2p_{t,-,\cut}$, as can be
understood by substituting $\phi=\pi/2$ into
Eq.~(\ref{eq:photon-momenta}) and observing that
$p_{t,-} \ge \ptH/2$ for all $\theta$ values.
This suggests a strategy whereby one ignores the $p_{t,\CS}$ cut
altogether for $\phi_{\CS}$ values where $U(\phi_{\CS})$ is negative.
Additionally, Fig.~\ref{fig:ptCS-ptcut2-phasespace} shows that for
large $\ptH$ values, the region excluded by the $p_{t,-}$ cut no
longer extends to $\cos\theta_{CS} = 1$.
This occurs when the second solution for $\sin\theta_\CS$ that yields
$p_{t,-} = p_{t,-,\cut}$,
\begin{equation}
  \label{eq:sintheta-limit-plus}
  \sin \theta_\CS^{+}(\phi) = \frac{
    +2 \ptH \sqrt{\mH^2+\ptH^2} |\cos \phi| - \sqrt{U(\phi)}}{2
    \mH^2+\ptH^2(1+ \cos 2 \phi )}\,,
\end{equation}
is in the physical range $0<\sin \theta_\CS^{+}(\bar\phi_\CS)<1$.
This suggests a strategy whereby one accepts events with
$\cos\theta_\CS > \cos \theta_\CS^{+}(\bar\phi_\CS)$ whenever the
latter is physical.

\begin{algorithm}[th]
  \caption{Hardness Compensating Boost-Invariant (\CBIH) cut algorithm to
    determine whether an event with a two-body decay should be accepted.
    It takes a Born transverse-momentum threshold $p_{t\cut}$ and a
    minimum $p_t$ requirement, $p_{t,-,\cut} \equiv p_{t\cut} - \Delta$ on
    both decay products.  }
  \label{alg:EBI}
  \begin{algorithmic}[1]
    \STATE If $p_{t,-} < p_{t,-,\cut}$ discard the
    event.
    \STATE If $p_{t,\CS} \ge p_{t,\cut}$, with $p_{t,\CS}$ is defined
    in Eq.~(\ref{eq:ptCS}), accept the event. 
    \STATE If either of $U(\bar\phi_{\CS})$ and
    $U(\pi/2-\bar\phi_{\CS})$ is negative, as obtained using
    Eqs.~(\ref{eq:theta_phi_CS}) and (\ref{eq:sintheta-limit-minus}),
    accept the event.
    \STATE If $\sin\theta_\CS^{+}(\bar\phi_\CS)$ in
    Eq.~(\ref{eq:sintheta-limit-plus}) is between $0$ and $1$ and
    $\cos\theta_\CS > \cos\theta_\CS^{+}(\bar\phi_\CS)$, accept the
    event.
    \STATE If $\bar\phi_\CS > \pi/4$, determine the value of
    $\theta_\CS^{-}$, that would be obtained if one mirrored the $\phi$
    value around $\pi/4$.
    We refer to it as
    $\theta_\CS^m = \theta_\CS^{-}(\pi/2-\bar\phi_\CS)$.
    If $\cos \theta_\CS^m < f_0$, accept the event if
    $\cos\theta_\CS < 2f_0 - \cos \theta_\CS^m$.
    \STATE Reject the event.
\end{algorithmic}
\end{algorithm}

Assembling together these different elements, we obtain a hardness
compensating boost-invariant (\CBIH) cut procedure for selecting
$H\to \gamma \gamma$ events, given as Algorithm~\ref{alg:EBI}.
\begin{figure}
  \centering
  \includegraphics[width=\textwidth,page=2]{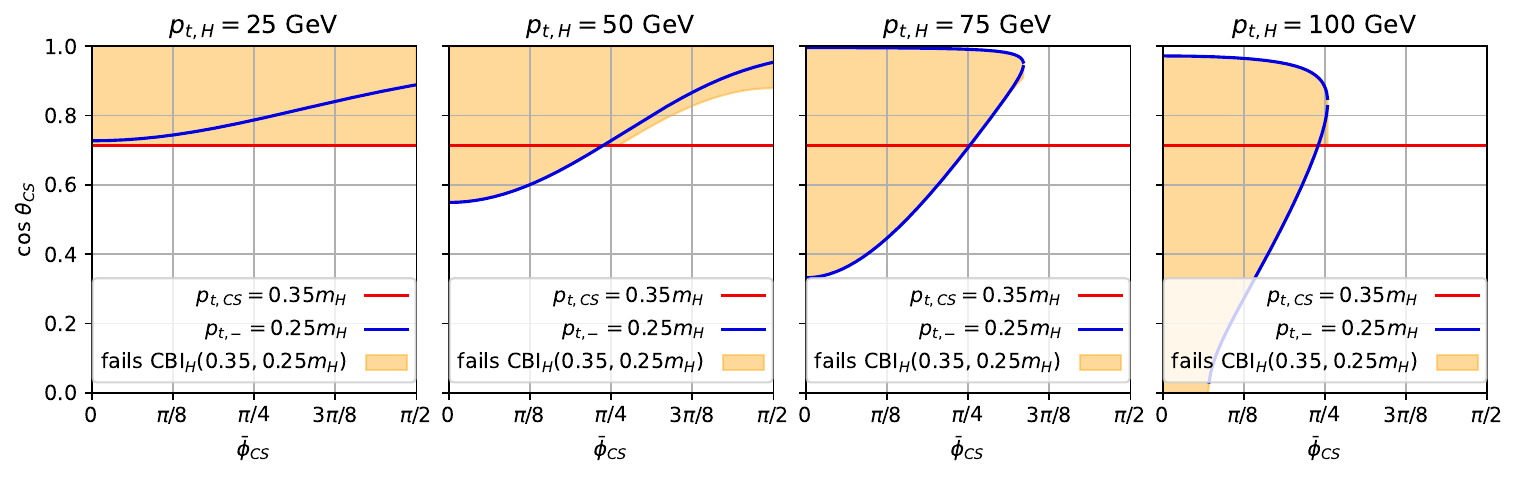}%
  \caption{Action of the hardness-compensating  boost-invariant (\CBIH) cuts in the
    $\bar \phi_{\CS}$--$\theta_{\CS}$ plane, as compared to the
    $p_{t,\CS}$ and $p_{t,-}$ cuts of
    section~\ref{sec:cuts-collins-soper}, whose action was shown in
    Fig.~\ref{fig:ptCS-ptcut2-phasespace}. }
  \label{fig:EBI-phasespace}
\end{figure}
It is generally speaking sensible to apply this algorithm if
$2\sqrt2(p_{t,\cut}-p_{t,-,\cut}) \equiv 2\sqrt{2}\Delta \lesssim
2p_{t,-,\cut}$.
The action of the \CBIH cut procedure on the decay phase space is shown
in Fig.~\ref{fig:EBI-phasespace}.
In the $\ptH = 25 \GeV$ panel, one sees that the $p_{t,\CS}$ cut
(red line) controls the acceptance.
In the $\ptH = 50 \GeV$ panel, the right hand part of the plot
illustrates the use of the region of $\phi > \pi/4$ between the red
line and the orange band (a region that is accepted) that compensates
for the loss of the region below the red line due to the $p_{t,-}$
cut.
In the remaining two panels, one sees how the \CBIH cut almost fully
tracks the $p_{t,-}$ cut, allowing for maximisation of the acceptance.

\begin{figure}
  \centering
  \includegraphics[width=0.7\textwidth,page=4]{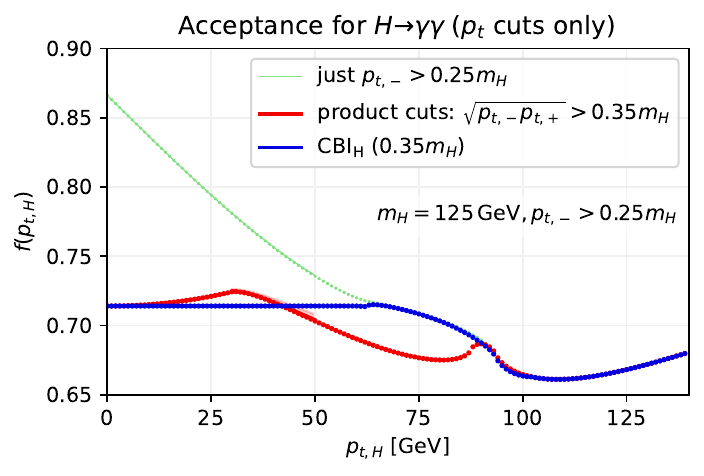}
  \caption{Acceptance of the hardness compensating  boost-invariant (\CBIH) cut, as a
    function of $\ptH$, 
    compared to the product cuts of section~\ref{sec:product-cuts} and
    the maximum possible acceptance that can be achieved with the underlying
    $p_{t,-} > 0.25 \mH$ requirement.
  }
  \label{fig:acceptances-EBI}
\end{figure}

The resulting acceptance as a function of $\ptH$ is shown in
Fig.~\ref{fig:acceptances-EBI} for our usual pair of cut thresholds.
The acceptance is exactly independent of $\ptH$ up to
$\ptH \simeq 60 \GeV$ (which, numerically, is substantially larger
than the naive expectation of $2\sqrt2{\Delta} \simeq 35 \GeV$).
This exact independence ensures that the perturbative series for the
fiducial cross section should be independent of the acceptance-induced
alternating-sign factorial divergence discussed in
sections~\ref{sec:existing-cuts} and \ref{sec:good-simple-cuts}.
At higher $\ptH$ values, the acceptance then closely tracks the
maximum possible acceptance that can be obtained with the $p_{t,-}$
cut.
Thus the \CBIH cut is near optimal.\footnote{Careful inspection of
  Fig.~\ref{fig:acceptances-EBI} reveals an efficiency around
  $100\GeV$ that is very slightly lower than that with just a
  $p_{t,-}$ cut (or its combination with a product cut).
  The origin is just barely visible in Fig.~\ref{fig:EBI-phasespace}
  where the $75$ and $100\GeV$ panels show a \CBIH exclusion region
  just below the rightmost edge of the $p_{t,-}$ curve.
  One could recover this region, by accepting an event whenever the
  $p_{t,-}$ cut is satisfied and the
  $\cos\theta_\CS^{+}(\bar\phi_\CS)$ solution is physical.
  This naive approach leads to a sharp feature in the acceptance at
  $\ptH = 2p_{t,-,\cut}$, which is the reason why we do not adopt
  it as our default.}

For all of the other hardness cuts considered so far, we have included
equations such as Eq.~(\ref{eq:asym-PT-series}) and plots such as
Fig.~\ref{fig:minpth-impact} to illustrate the perturbative behaviour
of the cuts.
For \CBIH cuts with the standard thresholds, the terms in the
perturbative series are essentially zero and the N3LL and N3LO
acceptance corrections are zero for $\epsilon \lesssim \mH/2$: the
fact that $f(\ptH) = f_0$ for $\ptH \lesssim \mH/2$ ensures that the
integrand of Eq.~(\ref{eq:generic-sigma-fid}) is zero.
Were we to extend the integral to the kinematic limit in $\ptH$, using
a matched fixed-order plus resummation $\ptH$ distribution, the
coefficients of the perturbative expansion would be non zero, but they
would show the convergence properties of a high-$\ptH$ cross section,
as expected from Eq.~(\ref{eq:sigma-fid-CBI}). 

\subsection{The case with hardness and rapidity cuts}
\label{sec:cbi-hardness-rap}

In section~\ref{sec:extended-cs-cuts} we worked with the assumption
that there is a non-negotiable minimum $p_t$ cut on the softer decay
product and adjusted the $p_{t,\CS}$ cut to retain a boost-invariant
acceptance over a wide range of $\ptH$ values.
Here we extend this approach, working with the additional constraint
that there are non-negotiable rapidity cuts on the decay products, but
still using adjustments of the $p_{t,\CS}$ cut (or, more directly, of
the $\cos \theta_\CS$ cut) to attempt to retain a $\ptH$-independent
acceptance.

The starting point is to establish the $\theta_\CS$ values for
which a decay product can be at the boundary of a cut $y_\cut$.
For any given decay $\phi$ there are up to two solutions,
\logbook{64c4200}{See the ``Try to understand what a single rap cuts
  look like'' section of math-2021/rapidity-pt-cuts.nb}
\begin{subequations}
    \label{eq:thetaCS-from-ycut}
  \begin{equation}
    \label{eq:thetaCS-from-ycut-main}
    \sin\theta_\CS^{y_\cut}(\phi) =
    \frac{2 s\, \ptH\sqrt{\mH^2+\ptH^2} \cos\phi \pm \sqrt{V(\phi)}}{
      \ptH^2 + \ptH^2 \cos2\phi + 2\mH^2(1 + c^2)\,,
    }\,,
    \qquad
    s = \left\{
    \begin{array}{c}
      +1\text{ for $\yH < y_\cut$}\,,
      \\[5pt]
      -1\text{ for $\yH > y_\cut$}\,,
    \end{array}
    \right.
  \end{equation}
  where
  \begin{equation}
    \label{eq:thetaCS-from-ycut-extras}
    V(\phi) = 2\mH^2(1 + c^2)\left[\ptH^2(\cos2\phi - 1) + 2\mH^2 c^2\right]\,,
    \,\qquad c = \frac{1}{\sinh (y_\cut - \yH)}\,.
  \end{equation}
\end{subequations}
A solution that gives $\sin\theta_\CS^{y_\cut}(\phi)>1$ ($<0$) is
considered to be at $\sin\theta_\CS = 1$ ($0$).\footnote{Recall that
  we consider the two decay products to be interchangeable, and define
  $\theta_\CS$ such that it is in the range $0<\theta_\CS<\pi/2$. }
Taking the case $\yH < y_\cut$, a requirement $y_\gamma < y_\cut$
implies a veto on $\theta_\CS$ values that lie between the two
solutions.
If $V(\phi)$ is negative, all $\theta_\CS$ values are allowed. 

One immediate difference relative to the case of just hardness cuts
(Eqs.~(\ref{eq:sintheta-limit-minus}, \ref{eq:sintheta-limit-plus})) is
that Eq.~(\ref{eq:thetaCS-from-ycut}) contains dependence on
$\cos\phi$ rather than $|\!\cos\phi|$, with the result that
configurations at $\phi$ and $\pi-\phi$ are not equivalent.
Consequently, it will be useful for our compensation algorithm to
consider four points in $\phi$ simultaneously rather than just two.
For any given $\phi$, the four points will be
\begin{equation}
  \label{eq:phi-mirror-points}
  \phi_{\{1,2,3,4\}} = \{\phi,\, \pi - \phi,\, |\pi/2 - \phi|,\, \pi - |\pi/2 - \phi|\}\,,
\end{equation}
recalling that from Eq.~(\ref{eq:theta_phi_CS}) we have $0<\phi<\pi$.
Eq.~(\ref{eq:phi-mirror-points}) ensures that we consider points with
opposite signs but equal values of both $\cos\phi$ and $\cos2\phi$.

Another difference relative to the case of just hardness cuts is the
extra degree of complexity brought in by the presence of multiple
rapidity cuts and the interplay between different rapidity cuts and
the hardness cuts (cf.\ Fig.~\ref{fig:all-ATLAS-cuts}, even just for
$\ptH=0$).
This leads us to formulate a balancing procedure where the adjustments
can largely be automated.

For this purpose, we use $\mathcal{R}_i(\phi,\ptH)$ to denote the
region(s) of $\cos\theta_\CS$ values allowed by cut $i$ for a given
$\phi$ and $\ptH$ (for example, an allowed region might consist of a
segment from $\cos\theta_\CS=0$ to $0.5$ and another segment from
$0.8$ to $1.0$).
We also introduce the notation $\mathcal{E}(\mathcal{R})$ to denote
the total extent of allowed region $\mathcal{R}$
($\mathcal{E}(\mathcal{R})=0.7$ in the example just given).
For a scalar decay, that extent is equal to the average acceptance
given the boson kinematic variables and the decay $\phi$.
Finally, regions can be combined logically, for example for cuts $a$
and $b$, we write $\mathcal{R}_{a,b}(\phi,\ptH)$ to indicate the
region(s) in $\cos\theta_\CS$ where a decay passes both sets of cuts.
In practice when combining multiple cuts, the final allowed region may
contain multiple non-contiguous allowed segments in
$\cos\theta_\CS=0$, which would be analytically tedious to deal with,
but can easily be encapsulated in computer code. 
We will use $\mathcal{R}_{-,\forall y}(\phi,\ptH)$ to denote the
region that is allowed after applying the $p_{t,-}$ cut and all photon
rapidity cuts, and $\mathcal{R}_{\CS,-,\forall y}(\phi,\ptH)$ to denote
the region that remains when additionally applying the $p_{t,\CS}$
cut.
As part of our high-$p_t$ enhancement approach, for the purpose
of the compensation calculations, $\mathcal{R}_{-}(\phi)$ will be
evaluated with the replacement $\cos\theta_\CS^{+}(\phi) \to 1$.
With the notation established, we can present our hardness and
rapidity compensating boost-invariant algorithm (\CBIHR),
algorithm~\ref{alg:CBI-hardness-rap} on
p.~\pageref{alg:CBI-hardness-rap}.

\begin{algorithm}[thp]
  \caption{Hardness and Rapidity Compensating boost-invariant (\CBIHR) cut algorithm.
    It takes a primary transverse-momentum threshold $p_{t,\cut}$,
    applied by default to $p_{t,\CS}$, a
    minimum $p_t$ requirement, $p_{t,-,\cut} \equiv p_{t\cut} - \Delta$ on
    both decay products, and a set of rapidity cuts. 
  }
  \label{alg:CBI-hardness-rap}
  \begin{algorithmic}[1]
    \STATE If $p_{t,-} < p_{t,-,\cut}$ or either of the decay products
    fails the rapidity cuts, discard the event.
    \STATE Use Eq.~(\ref{eq:theta_phi_CS}) to determine the
    Collins-Soper decay angles (we refer to $\phi_\CS$ as just
    $\phi$). 
    \STATE Apply high-$p_t$ enhancement: using
    Eq.~(\ref{eq:sintheta-limit-minus}), if $U(\phi_i) < 0$ for any
    of the $\phi_i$ in Eq.~(\ref{eq:phi-mirror-points}), accept the event; using
    Eq.~(\ref{eq:sintheta-limit-plus}), if
    $\cos\theta_\CS > \cos\theta_\CS^{+}(\phi_1)$, accept the
    event.
    \STATE Evaluate what the $\ptH=0$ acceptance extent would be
    for this $\phi$, 
    $f_0 = \mathcal{E}(\mathcal{R}_{\CS,-,\forall y}(\phi,\ptH=0))$. 
    %
    %
    \STATE For each of the original and mirrored $\phi_i$ values in
      Eq.~(\ref{eq:phi-mirror-points}), evaluate the acceptance
    extents both with and without the $p_{t,\CS}$ cut
    \begin{equation}
      \label{eq:fiCS}
      f_{i,\CS} = \mathcal{E}(\mathcal{R}_{\CS,-,\forall
        y}(\phi_i,\ptH))\,,
      \qquad
      f_{i} = \mathcal{E}(\mathcal{R}_{-,\forall y}(\phi_i,\ptH))\,.
    \end{equation}
    \STATE Determine the sums of the acceptance extents
    $f_{\Sigma,\CS} = \sum_{i=1}^4 f_{i,\CS}$ and
    $f_\Sigma = \sum_{i=1}^4 f_{i}$.
    \STATE If $f_{\Sigma,\CS} > 4f_0$, additional vetoes are needed,
    which should total $f_v = f_{\Sigma,\CS} - 4f_0$ when summed across the
    four $\phi_i$ values.
    The extent of the additional veto to be applied to this $\phi$
    value is,
    \begin{equation}
      \label{eq:delta-f1-v}
      \delta f_1^{(v)} = f_v
      \frac{\max(0, f_{1,\CS} - f_0)}{\sum_{i=1}^4 \max(0, f_{i,\CS} - f_0)}\,.
    \end{equation}
    We choose to apply it from the uppermost part of the allowed
    $\mathcal{R}_{\CS,-,\forall y}(\phi_1, \ptH)$ region, working
    downwards.
    Accept the event iff $\theta_\CS$ is in the allowed region
    $\mathcal{R}_{\CS,-,\forall y}(\phi_1, \ptH)$ but not in the
    additionally vetoed region.
    
    \STATE If $f_{\Sigma,\CS} < 4f_0$, one should attempt to find
    additional acceptance. 
    The total additional acceptance needed across all four $\phi_i$
    values is $f_a = 4f_0 - f_{\Sigma,\CS}$.
    The maximum recoverable acceptance, through elimination of the
    $p_{t,\CS}$ cut, is $f_{\Sigma} - f_{\Sigma,\CS}$.
    If $f_a$ is larger than this maximum recoverable acceptance,
    discard the $p_{t,\CS}$ cut, and accept the event iff $\theta_\CS$
    is in the $\mathcal{R}_{-,\forall y}(\phi_1, \ptH)$ region.
    Otherwise recover an additional acceptance 
    \begin{equation}
      \label{eq:delta-f1-a}
      \delta f_1^{(a)} = f_a
      \frac{f_1 - f_{1,\CS}}{
        \sum_{i=1}^4 (f_i - f_{i,\CS})}\,,
    \end{equation}
    for this $\phi$ value.
    We choose to do so working upwards from the lowermost
    $\cos\theta_{\CS}$ part of the difference between the
    $\mathcal{R}_{-,\forall y}(\phi_1, \ptH)$ and
    $\mathcal{R}_{\CS,-,\forall y}(\phi_1, \ptH)$ regions.
    Accept the event iff  $\theta_{\CS}$ is in
    $\mathcal{R}_{\CS,-,\forall y}(\phi_1, \ptH)$ or the additional
    allowed region.
\end{algorithmic}
\end{algorithm}

While the algorithm may appear to quite lengthy at first sight, its
two underlying principles are simple:
(1) within each group of mirror
$\phi$ values, adjust the $\theta_{\CS}$ cuts so as to retain the same
acceptance summed across those $\phi$ values as is obtained at $\ptH=0$;
and
(2) for $\ptH$ values above $2p_{t,-,\cut}$, which is when $U(\phi)$
in Eq.~(\ref{eq:sintheta-limit-minus-U}) can start to be negative,
start relaxing the $\theta_{\CS}$ cut so as to bring the acceptance
close to its maximal value without any $\theta_{\CS}$ cut.\footnote{In
  the code that accompanies this article,
  \url{https://github.com/gavinsalam/two-body-cuts},
  which internally relies on FastJet~\cite{Cacciari:2011ma},
  there are further options
  for controlling the exact behaviour in this region.}

There is one context where Algorithm~\ref{alg:CBI-hardness-rap} cannot
maintain constant acceptance away from $\ptH=0$, specifically when the
acceptance is limited by a pair of rapidity cuts, $y_\gamma > \yH - Y$
and $y_\gamma < \yH + Y$ for some $Y>0$.
Considering, as before, the ATLAS $H\to \gamma\gamma$ cuts, cf.\
Fig.~\ref{fig:all-ATLAS-cuts}, this occurs for $|y_H| = 1.945$, i.e.\
midway between the $|y_{\gamma}|>1.52$ and $|y_{\gamma}|<2.37$ cuts.
In such a situation, examining just the term linear in $\ptH$ in
Eq.~(\ref{eq:thetaCS-from-ycut-main}), one can see that for
$\cos\phi >0$ the $y_\gamma < \yH + Y$ condition raises the effective
$\sin\theta_{\CS}$ cut, while for $\cos\phi <0$ the
$y_\gamma > \yH - Y$ cut raises the effective $\sin\theta_{\CS}$
cut.
This means that taking $\ptH$ slightly away from zero always raises
the effective $\sin\theta_{\CS}$ cut and so reduces the acceptance.
With the default $p_{t,\CS}$ cut imposing a less stringent condition
on $\theta_{\CS}$ in this region than the rapidity cuts, adjusting the
$p_{t,\CS}$ cut as a function of $\ptH$ cannot recover the acceptance
that is being lost.

\begin{figure}
  \centering
  \includegraphics[scale=0.85,page=1]{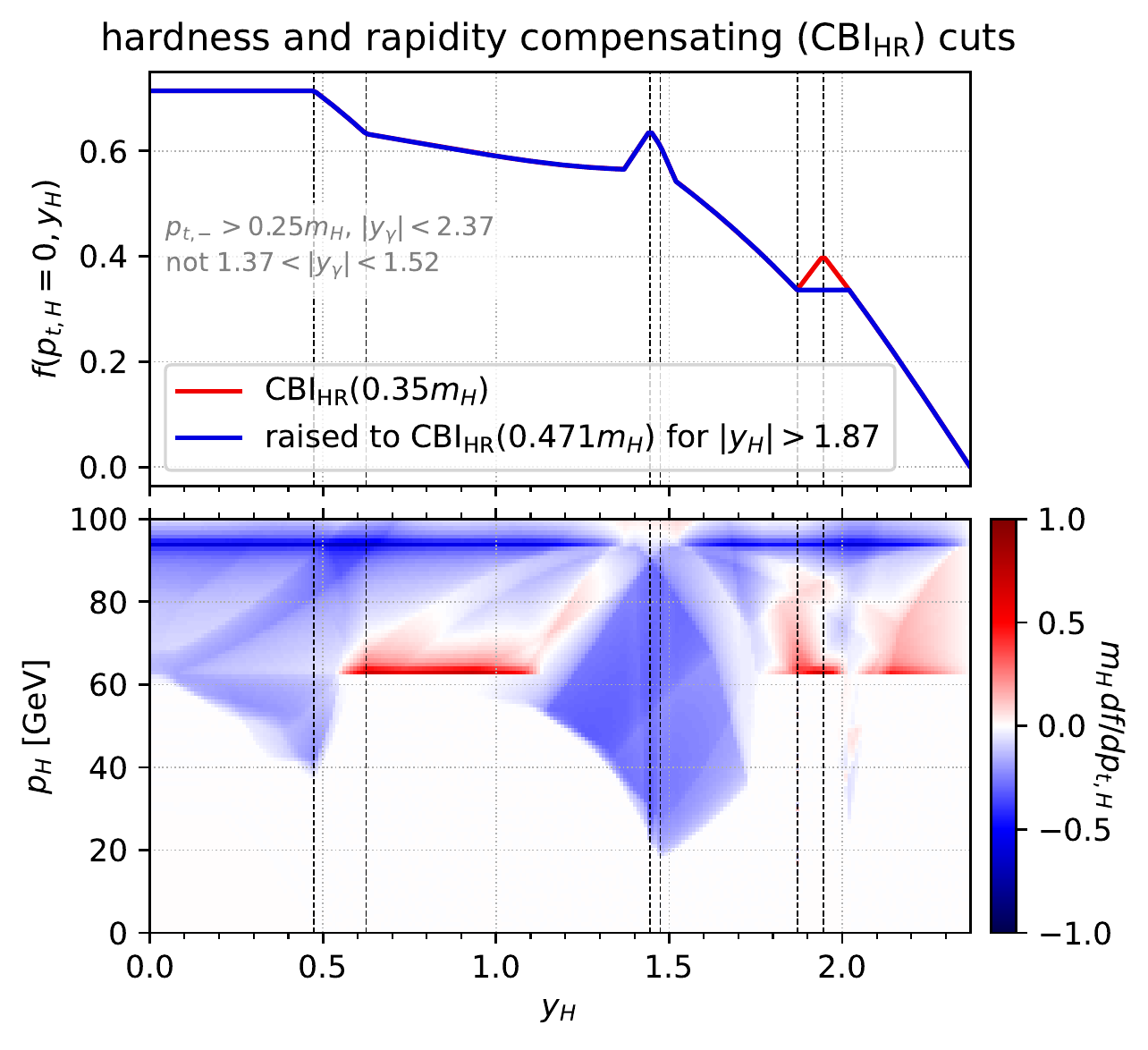}%
  \caption{Analogue of Figs.~\ref{fig:all-ATLAS-cuts} and
    \ref{fig:all-ATLAS-cuts-eff-derivatives},
    for the hardness and rapidity compensating boost-invariant
    (\CBIHR) cuts.
    The upper panel shows the impact on the $\ptH=0$ acceptance of the
    raised high-rapidity 
    $p_{t,\CS}$ cut that we impose, Eq.~(\ref{eq:extrapt-cut}).
    The lower panel shows the $\ptH$ derivative of the acceptance
    (including the raised high-rapidity $p_{t,\CS}$ cut), illustrating
    that up to $\ptH \simeq 20 \GeV$ the acceptance is independent of
    $\ptH$ for all Higgs rapidities.
    Absolute values of the acceptances as a function of $\ptH$ are to
    be found in Fig.~\ref{fig:CBIHR-many-acceptances}.  }
  \label{fig:CBIHR-performance} 
\end{figure}

A workaround for this issue is to raise the Born $p_{t,\CS}$ cut in a
suitable rapidity region.
One might worry about the resulting loss of acceptance, but as we
shall see, this is minimal.
For a generic situation with an overall rapidity
cut $|y_\gamma| < y_{\max}$ and an exclusion band $y_1 < |y_\gamma|
< y_2$ ($y_{1,2,\max} = 1.37,1.52,,2.37$ for ATLAS), we adopt the
following procedure:
starting from the midpoint between $y_1$ and $y_{\max}$ (where it is
the $y_2$ cut that sets the Born acceptance) we replace the usual Born
$p_{t,\CS}$ cut with the requirement (still supplemented with
$\ptH$-dependent compensation)
\begin{equation}
  \label{eq:extrapt-cut}
  p_{t,\CS} > \frac{\mH}{2 \cosh (y_{m} - y_2)}  \simeq 0.471
  m_H\,,
  \qquad
  \text{for } |\yH| > y_m = \frac{y_1 + y_{\max}}{2}\,.
\end{equation}
This corresponds to a constraint on $\cos\theta$ that is identical to
that of the $y_2$ cut at $y_H = y_m$.
The impact of this modification is illustrated for $\ptH=0$ in the
upper panel of Fig.~\ref{fig:CBIHR-performance} and it
corresponds to the difference between the red and blue lines for
$\yH \simeq 1.9$, confirming that it is a small overall effect.

\begin{figure}
  \centering
  \includegraphics[width=0.9\textwidth]{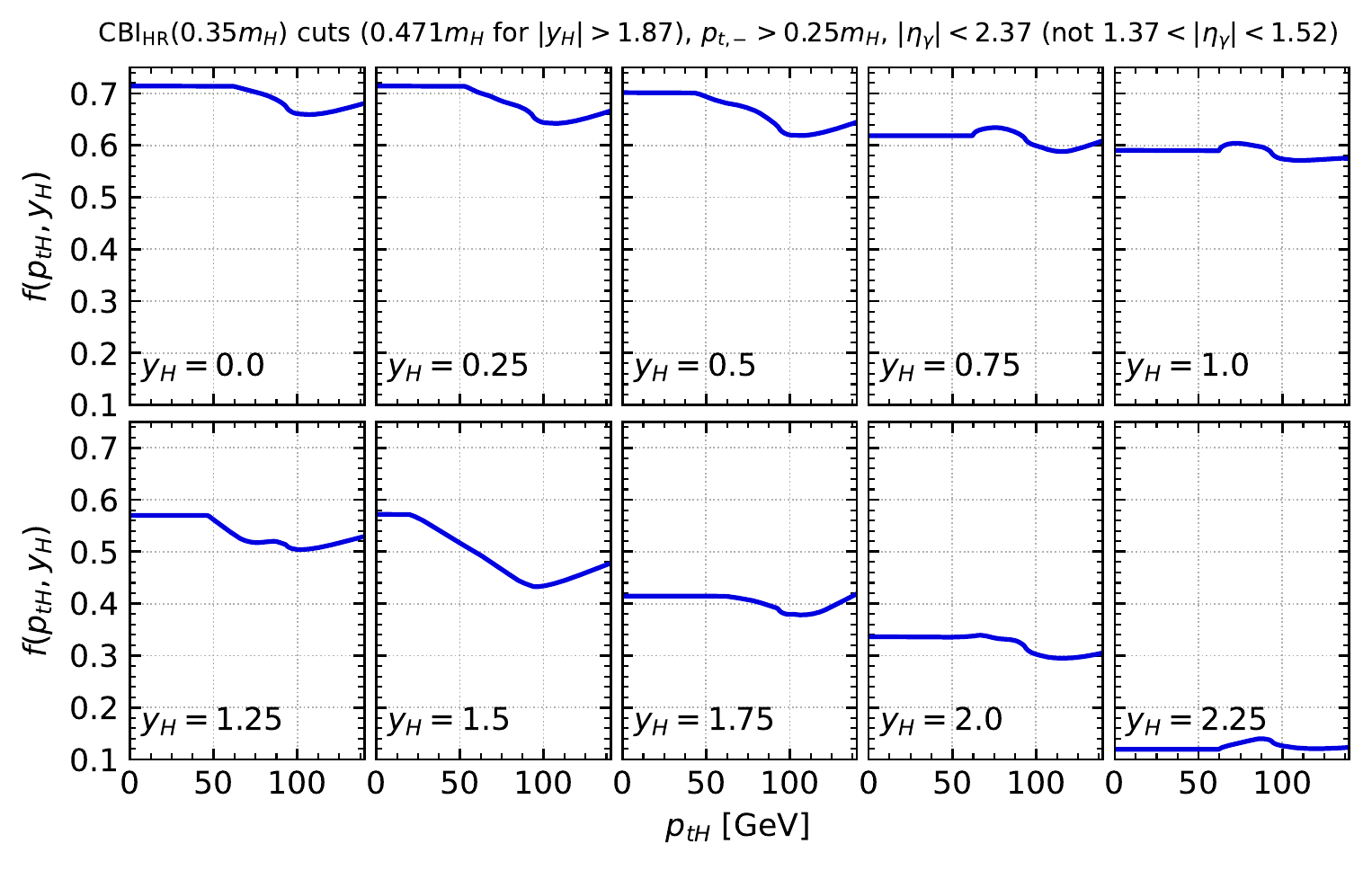}
  \caption{Acceptance for the hardness and rapidity compensating cuts,
    as a function of $\ptH$, for several $\yH$ values (\CBIHR, raised
    at high rapidity, as for the blue curve and the heat-map plot in
    the upper and lower panels respectively of
    Fig.~\ref{fig:CBIHR-performance}).  }
  \label{fig:CBIHR-many-acceptances}
\end{figure}

The lower panel of Fig.~\ref{fig:CBIHR-performance} shows the
$\ptH$ derivative of the acceptance, while the absolute values of the
acceptances as a function of $\ptH$ are shown for a representative set
of $\yH$ values in Fig.~\ref{fig:CBIHR-many-acceptances}.
One of the key objectives of this section was to eliminate all $\ptH$
dependence at low-$\ptH$.
Regions with no $\ptH$ dependence appear in white in
Fig.~\ref{fig:CBIHR-performance}, and the figure clearly demonstrates
that the acceptance in the low and moderate $\ptH$ regions is
independent of $\ptH$.
The lowest value of $\ptH$ where this independence breaks down,
corresponding to the $p_t^\text{thresh}$ threshold that enters into
our expression, Eq.~(\ref{eq:sigma-fid-CBI}), for the fiducial cross
section, is about $20\GeV$.
It arises at the $\yH$ value where the $p_{t,\CS}$ and the upper
rapidity cuts have equivalent $\ptH=0$ actions, i.e.\ for $y_H =
y_{\max} - \operatorname{arccosh}(\mH/2p_{t,\text{cut}}) \simeq 1.474$.
For low values of $\ptH$, the compensation mechanism balances a loss
of acceptance from the rapidity cut at $\phi > \pi/2$ (cf.\ the linear
term of Eq.~(\ref{eq:onerap-costheta})), with a corresponding gain of
acceptance from a loosening of the $p_{t,\CS}$ cut for $\phi < \pi/2$.
\logbook{}{
  See 2021-figs/cut_acceptances.ipynb to fiddle with the cuts and
 pt- cut: cos(theta) < f0 + s0/f0 (2Delta/mH - ptH/mH cos(phi))
 rap cut: cos(theta) < f0 + s0*f0 cos(phi) . ptH/mH
 So they coincide when (2Delta/mH = (1+f0^2) ptH/mH)
}
However that compensation mechanism becomes compromised for 
\begin{equation}
  \label{eq:pt-thresh}
  \ptH \gtrsim \frac{2\Delta}{1 + f_0^2} + \ord{2}\,,
\end{equation}
where, for $\phi=0$, the $\cos\theta$ limit from the $p_{t,-}$ cut,
Eq.~(\ref{eq:ptminus-cut-with-delta-costheta}), becomes lower than the
$\cos\theta$ limit that is needed to balance the loss of acceptance
from the rapidity cut for $\phi=\pi$.
%

\section{Comments on Drell--Yan ($Z$) production}
\label{sec:Drell-Yan}

\logbook{}{See DY-A0-term.nb for some first thoughts here}

A complete study of the Drell--Yan process, i.e.\ the production of a
charged-lepton pair, is beyond the scope of this article.
Still it may be useful to briefly outline some of the similarities and
differences relative to the Higgs production case.
A first consideration is that the Drell-Yan di-lepton mass
($m_{\elll}$) spectrum covers a continuum.
We will work within the assumption that one picks a narrow part of the
continuum (e.g.\ concentrating on resonant $Z$ production and/or
imposing a narrow $m_\elll$ window), or that hardness cuts are
formulated as fractions of $m_\elll$.
This serves to avoid the additional complications that come from the
interplay between a steeply-falling spectrum and fixed lepton hardness
cuts.
We will also ignore contributions from
$\gamma \gamma \to \ell^+ \ell^-$ (see e.g.\
Ref.~\cite{Harland-Lang:2020veo}) and other electroweak
contributions~\cite{Denner:2011vu,Frederix:2020nyw}.
This allows us to
adopt the widespread parametrisation of the cross section as a
function of the Drell--Yan exchanged 4-momentum $q$ and the
Collins-Soper~\cite{Collins:1977iv} angles $\theta$ and
$\phi$\footnote{Which coincide with the decay parametrisation in
  Eq.~(\ref{eq:photon-momenta}), where the $+$ ($-$) momentum
  corresponds to the (anti)lepton.}
\begin{equation}
  \label{eq:DY-decomposition}
  \frac{d\sigma}{d^4q d\cos\theta d\phi} =
  \frac{3}{16\pi} \frac{d\sigma^\text{unpol.}}{d^4q}
  \left(h_u(\theta,\phi) + \sum_{i=0}^7 A_i(q)\, h_i(\theta,\phi)\right),
\end{equation}
in terms of the unpolarised cross section and the spherical harmonic
functions $h_\text{\sc x}$
\begin{subequations}
  \label{eq:spherical-harmonics}
  \begin{align}
    h_u &= 1 + \cos^2\theta\,,
    & h_0 &= {\textstyle \frac12}(1-3\cos^2\theta)\,,
    & h_1 &= \sin2\theta \cos\phi\,,
    \\
    h_2 &={\textstyle \frac12}\sin^2\theta \cos2\phi\,,
    & h_3 &=\sin\theta \cos\phi\,,
    & h_4 &=\cos\theta,
    \\
    h_5 &=\sin^2\theta \sin2\phi\,,
    & h_6 &=\sin2\theta \sin\phi\,,
    & h_7 &=\sin\theta \sin\phi\,.
  \end{align}
\end{subequations}
Each of the $h_i$ is multiplied by a coefficient $A_i(q)$, where we
have made explicit that it depends on the Drell-Yan pair 4-momentum,
while the normalisation of $h_u$ is fixed by the requirement for the
expression to integrate to the unpolarised cross section.
The $A_i(q)$ coefficients have been calculated to NNLO for non-zero
$q_t$ in Ref.~\cite{Gauld:2017tww}.
We will work within the assumption that the Drell-Yan lepton-pair
transverse momentum, $p_{t,\elll}$, is identical to $q_t$, which
corresponds to an assumption that collinear photon radiation has been
clustered with the leptons (we ignore lepton isolation).
The Drell--Yan fiducial cross section with cuts on the final-state leptons
is then given by
\begin{equation}
  \label{eq:DY-sigma-f-fid}
  \frac{d\sigma_\text{fid}}{d^4q} =
  \frac{d\sigma^\text{unpol.}}{d^4q}\left[
    f^{(u)}(q)
    +
    \sum_{i=0\ldots 7} A_i(q) f^{(i)}(q)
    \right],
\end{equation}
with 
\begin{equation}
  \label{eq:DY-f-fid-def}
  f^{(\text{\textsc{x}})}(q) =
  \frac{3}{16\pi}
  \int_{-1}^1 \!d\!\cos\theta
  \int_{-\pi}^{\pi} \!d\phi \,\,
  h_\text{\textsc{x}}(\theta,\phi)\,\, \Theta_\text{cuts}(\theta,\phi,q)\,,
\end{equation}
where $\Theta_\text{cuts}(\theta,\phi,q)$ is $1$ ($0$) if the leptons
pass (do not pass) the fiducial cuts.
We refer to the $f^{(\text{\textsc{x}})}$ as harmonic acceptances.

To understand the overall behaviour of the cross section, one needs to
put together the behaviour of both the $A_i(q)$ and the
$f^{(\text{\textsc{x}})}$ functions, and there are some general
features that are worth keeping in mind.
Firstly, for $i=3,\ldots,7$, the $f^{(i)}$ are zero, which can be
seen by observing that the cuts have an effect that is unchanged under
symmetry operations that swap the two leptons ($\theta \to \pi-\theta$
and $\phi \to \pi+\phi$) or that correspond to a change of the sign of
$\phi$.
Under one or other of these symmetry operations the corresponding
$h_i$ functions flip sign.
Secondly, the discussions of
Refs.~\cite{Collins:1978yt,Boer:2006eq,Berger:2007si,Bodek:2010qg,Lindfors:1979rc}
indicate that the $A_{0\ldots2}$ are zero for $p_{t,\ell\ell}=0$ and
that for small $p_{t,\ell\ell}$, the $A_{0}$ and $A_{2}$ coefficients
scale quadratically with $p_{t,\ell\ell}$, while $A_{1}$ scales
linearly.

\begin{figure}
  \centering
  \includegraphics[scale=0.8,page=2]{2021-figs/n3-benchmarks-twoZ.pdf}
  \includegraphics[scale=0.8,page=1]{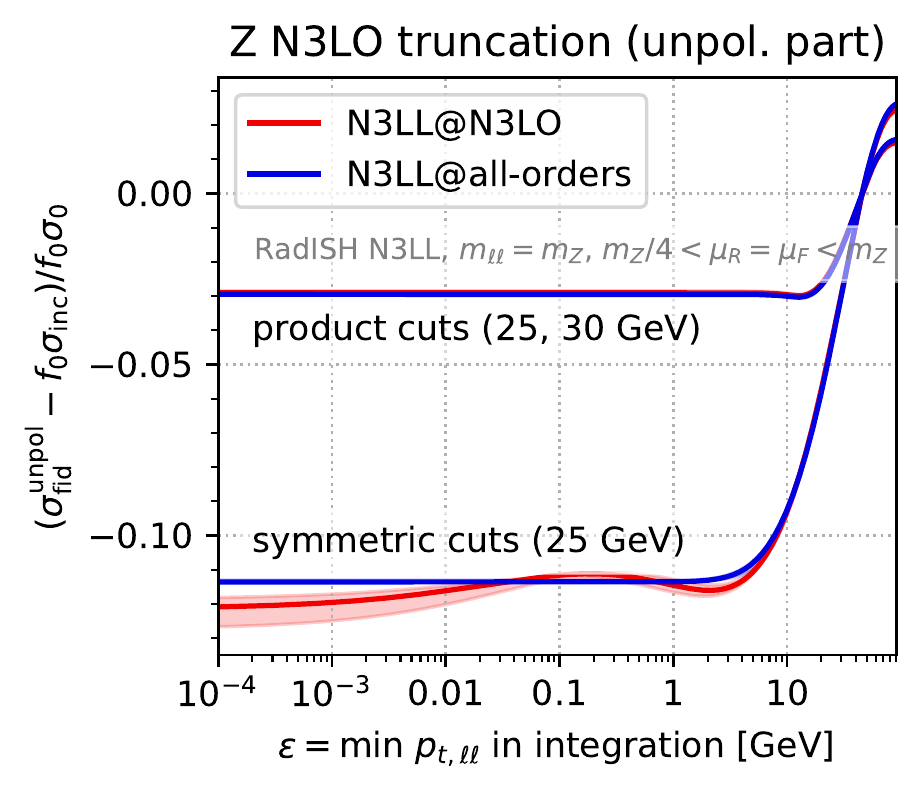}
  \caption{Left: acceptances for the non-zero spherical harmonics, as
    defined in Eq.~(\ref{eq:DY-f-fid-def}), for symmetric cuts
    ($p_{t,\ell} > 25\GeV$, in red), and product
    cuts ($\sqrt{p_{t,+}p_{t,-}} > 30\GeV$, supplemented with a
    minimum cut $p_{t,-} > 25\GeV$, in black).
    Right: the unpolarised part of the cross section (i.e.\
    corresponding to $f^{(u)}$) within an all-order N3LL calculation
    (in blue)
    and its truncation at N3LO (in red), as a function of the minimum
    $p_{t,\ell\ell}$ that is allowed in the integration.
    The results are shown for the same symmetric and product cuts as
    in the left-hand plot.
  }
  \label{fig:DY-basics}
\end{figure}

As in our discussion of Higgs boson cuts, we start by considering just
hardness cuts, using symmetric and product cuts for our illustration.
Fig.~\ref{fig:DY-basics} (left) shows the harmonic acceptances for
symmetric cuts ($p_{t,\ell} > 25\GeV$, as used by
CMS~\cite{Sirunyan:2019bzr}) and product cuts
($\sqrt{p_{t,+} p_{t,-}} > 30\GeV$ and $p_{t,-} > 25\GeV$).
One clearly sees a linear $p_{t,\ell\ell}$ dependence for $f^{(u)}$
and $f^{(0)}$ for the symmetric cuts, which appears for the same
reasons as discussed in the Higgs boson case (though with different
coefficients).
With product cuts, the linear dependence is absent for
$p_{t,\ell\ell} \lesssim 2\Delta = 10 \GeV$, and it is only quadratic
dependence that remains there.

The impact of the cuts on a perturbative calculation of the fiducial
cross section is illustrated with the following series for the
symmetric cuts (the spectrum and expansion used to obtain the N3LL
results were kindly provided by the authors of
Ref.~\cite{Bizon:2018foh})
\logbook{b54f05ba75}{see Higgs_N3LO_expansion/n3-benchmarks-2021-09-12.tex}
\begin{align}
  \label{eq:Zsym-PT-series}
  \frac{\sigma_{\text{sym}}^\text{(u)} - f_0 \sigma_{\text{inc}}}{\sigma_0 f_0} 
&\simeq  -0.074_{\as} +   0.051_{\as^{2}} - 0.057_{\as^{3}} +   0.090_{\as^{4}} - 0.181_{\as^{5}} + \ldots &\simeq  -0.047 \;\;&@\text{DL}\nonumber,\\[-6pt]
&\simeq  -0.074_{\as} +   0.027_{\as^{2}} - 0.014_{\as^{3}} +   0.010_{\as^{4}} - 0.010_{\as^{5}} + \ldots &\simeq  -0.055 \;\;&@\text{LL}\nonumber,\\
&\simeq  -0.118_{\as} +   0.012_{\as^{2}} - 0.016_{\as^{3}} + \ldots &\simeq  -0.114 \;\;&@\text{NNLL},\nonumber\\
&\simeq  -0.118_{\as} +   0.012_{\as^{2}} - 0.016_{\as^{3}} + \ldots &\simeq  -0.114 \;\;&@\text{N3LL}.\nonumber\\
\end{align}
It is the linear dependence of $f^{(u)}$ that will be critical, so the
above equations show just the contribution to the cross section from
$f^{(u)}$.
The DL and LL results both show a breakdown in the convergence of the
series, though at somewhat different orders and with fairly different
normalisations for the smallest term.\footnote{
  In the LL case, the smallest term in the series scales as
  $(\Lambda/Q)^{0.76}$ rather than the
  $(\Lambda/Q)^{23/64} \simeq (\Lambda/Q)^{0.36}$ seen at DL level in
  Eq.~(\ref{eq:effective-power-correction}), cf.\ Appendix~\ref{sec:remarks-asysmptotics}.
  As in the Higgs case, the investigations of
  Appendix~\ref{sec:remarks-asysmptotics} suggest that for linear
  cuts, the power scaling seen at LL may well hold beyond, while for
  quadratic cuts we have not conclusively established the power.  }
Considering the N3LL series, the all-order N3LL result and its N3LO
truncation disagree at the order of a percent relative to the Born
cross section.

The dependence of the unpolarised part of the fiducial cross section
on a $\ptll$ cutoff and the impact of scale variation are illustrated
in Fig.~\ref{fig:DY-basics} (right).
The N3LO
truncation is noticeably sensitive to the minimum $p_{t,\ell\ell}$ allowed
in the integration, converging only when including unphysically low
values down to $1\MeV$ and below.
The pattern can be compared to that in Fig.~\ref{fig:minpth-impact}
(Higgs production with asymmetric cuts).
The normalisation of the discrepancy here is much reduced because of
the $C_F$ factor instead of a $C_A$ factor in the resummation (keeping
in mind that the missing DL terms, i.e.\ those from N4LO onwards,
scale as the colour factor to the fourth power).
However, the accuracy of the data is also
much higher for $Z$ production than for Higgs production, and so
percent-level issues are conceivably relevant for the $Z$ case.
The minimum $p_{t,\ell\ell}$ value that is needed for a reliable
estimate of the N3LO
coefficient is similar to that in the Higgs case,
Eq.~(\ref{eq:95percent-result}), because it is determined by the
number of logarithms in the fixed-order expansion, which is the same.

The corresponding results for product cuts are
\begin{align}
  \label{eq:Zprod-PT-series}
  \frac{\sigma_{\text{prod}}^\text{(u)} - f_0 \sigma_{\text{inc}}}{\sigma_0 f_0}
&\simeq  -0.006_{\as} - 0.000_{\as^{2}} +   0.000_{\as^{3}} - 0.000_{\as^{4}} - 0.000_{\as^{5}} + \ldots &\simeq  -0.006 \;\;&@\text{DL}\nonumber,\\[-6pt]
&\simeq  -0.006_{\as} - 0.000_{\as^{2}} - 0.000_{\as^{3}} +   0.000_{\as^{4}} - 0.000_{\as^{5}} + \ldots &\simeq  -0.007 \;\;&@\text{LL}\nonumber,\\
&\simeq  -0.018_{\as} - 0.009_{\as^{2}} - 0.003_{\as^{3}} + \ldots &\simeq  -0.030 \;\;&@\text{NNLL}\nonumber,\\
&\simeq  -0.018_{\as} - 0.009_{\as^{2}} - 0.002_{\as^{3}} + \ldots &\simeq  -0.029 \;\;&@\text{N3LL}\nonumber.\\
\end{align}
The DL and LL results converge fast, while the N3LL resummation and
its N3LO truncation agree at the per-mil level.
From Fig.~\ref{fig:DY-basics} (right), we see that the fiducial cross
section (whether at all orders or N3LO) is essentially insensitive to
transverse momenta below $10\GeV$.

\begin{figure}
  \centering
  \includegraphics[page=2,width=0.49\textwidth]{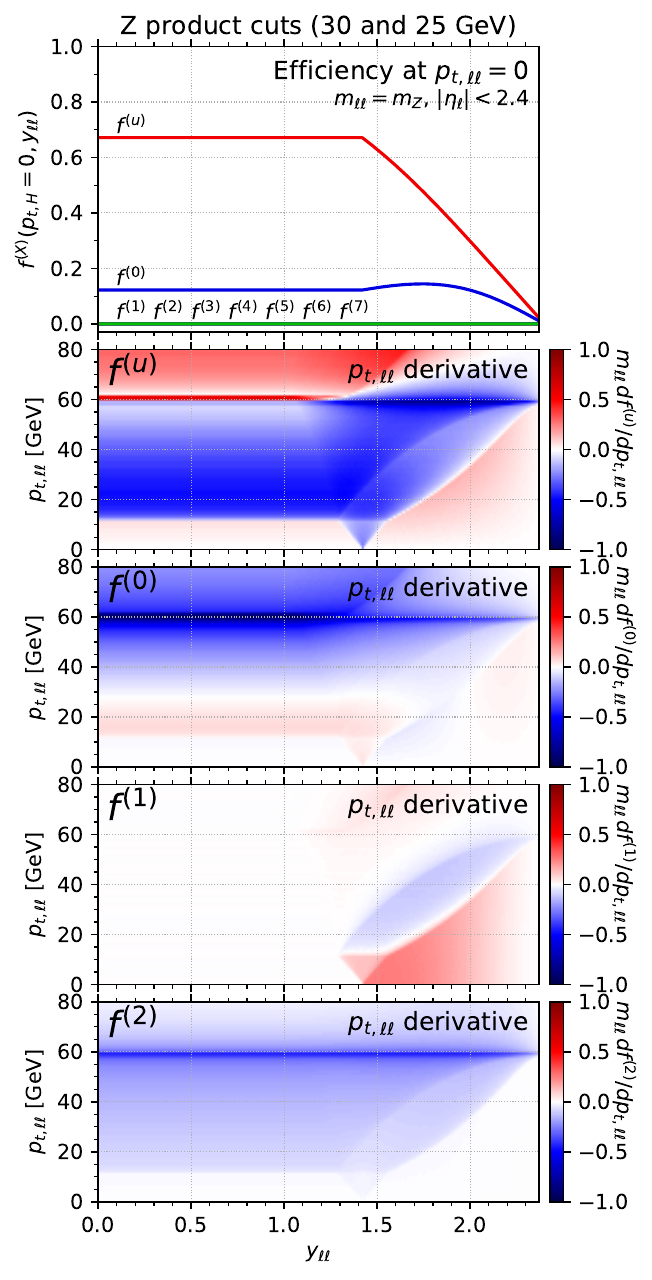}
  \includegraphics[page=3,width=0.49\textwidth]{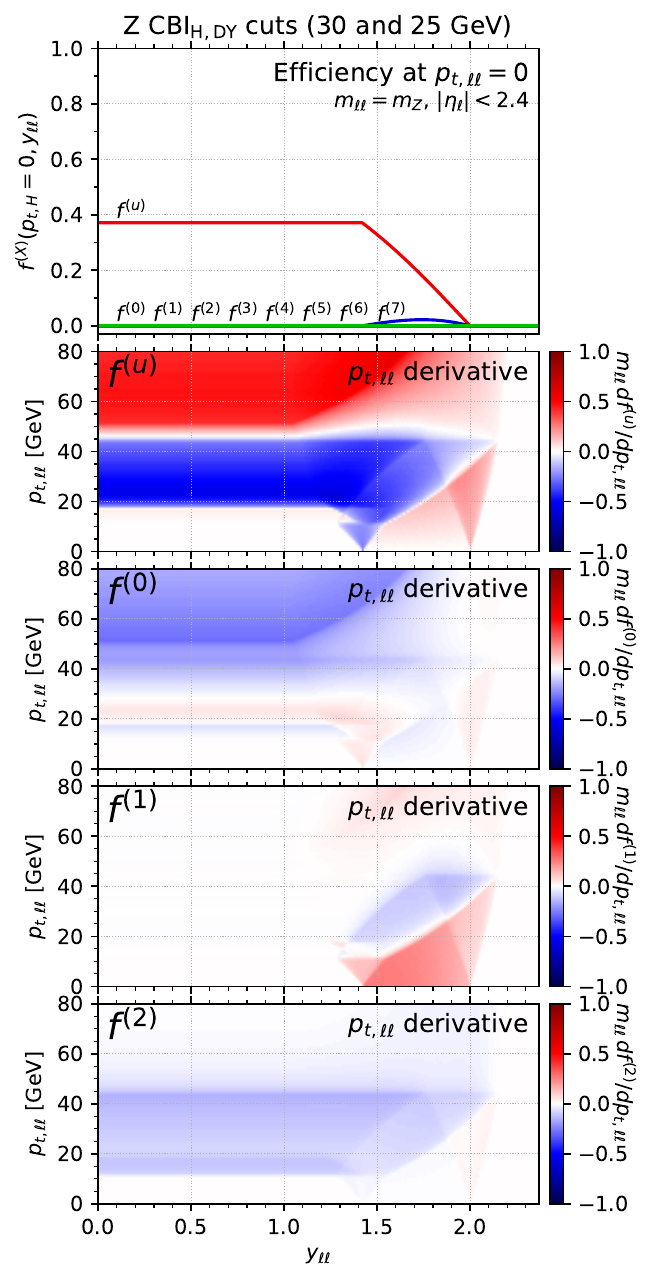}
  \caption{Top panels: $p_{t,\ell\ell}=0$ harmonic acceptances for $Z\to \ell^+\ell^-$,
    as defined in Eq.~(\ref{eq:DY-f-fid-def}), for each of the
    spherical harmonic functions in
    Eq.~(\ref{eq:spherical-harmonics}).
    Lower panels: the $p_{t,\ell\ell}=0$ derivatives of the
    acceptances for the $f^{(\textsc{x})}$ that are non-zero.
    The left-hand column shows results for product cuts, the
    right-hand column for the \CBIH cuts supplemented with the
    condition in Eq.~(\ref{eq:DY-costheta-c1}) (which, together, we
    refer to as \CBIHDY cuts).
    In both figures we use events with a fixed
    $m_{\elll} = m_Z = 91.1876\GeV$.  }
  \label{fig:DY-summary}
\end{figure}

If product cuts are to be a good replacement for standard symmetric
and asymmetric cuts in $Z$-production studies, then one should understand their
interplay with lepton rapidity cuts.
Fig.~\ref{fig:DY-summary} (left) examines the behaviour of the product
cuts when supplemented with a requirement $|y_\ell| < y_{\max} = 2.4$, as used by
CMS~\cite{Sirunyan:2019bzr}.\footnote{Here, we work in the massless
  lepton approximation, where rapidity and pseudorapidity are
  identical. }
The top panel shows the harmonic acceptances for $p_{t,\ell\ell}=0$, as
a function of the lepton-pair rapidity.
The remaining panels show the $p_{t,\ell\ell}$ derivative of the
acceptance for those harmonics that are non-zero.
We comment on two features: firstly, $f^{(u)}$ acquires linear
dependence at
$y_{\ell\ell} = y_t \equiv y_{\max} -
\arccosh(m_{\ell\ell}/2p_{t,\cut}) \simeq 1.42$.\footnote{In
  Fig.~\ref{fig:DY-summary}, the feature at that rapidity is sharp,
  i.e.\ at a single rapidity, because we have fixed $m_{\elll}$ to a
  single value, $m_Z$, rather than integrating it over the $Z$
  resonance shape.
  If we were to consider a wider mass window around the resonance, the
  feature would be smeared out by an amount that can be determined by
  replacing $m_Z$ with $m_Z \pm \Gamma_Z$, leading to a rapidity
  spread of about $\pm 0.036$.
  Alternatively, one could use hardness cuts that are proportional to
  $m_{\elll}$, in analogy with the standard practice for
  $H\to\gamma\gamma$ measurements, e.g.\ using a main hardness cut of
  $0.329m_{\elll}$ and a cut on the softer lepton of $0.274m_{\elll}$.
  In that case, rapidity features would remain sharp, while sharp
  transitions at fixed $p_{t,\elll}$ values would become sharp
  transitions at fixed $p_{t,\elll}/m_{\elll}$ values.
 }
As with the analogous characteristic seen in
sections~\ref{sec:ptrap-cut} and \ref{sec:worked-example}, a suitably
wide rapidity bin around this point will transform this linear
dependence into reasonably tame quadratic dependence.
The second feature concerns $f^{(1)}$, where one sees linear
$p_{t,\elll}$ dependence for $y_{\ell\ell} \ge y_t$.\footnote{This is
  a consequence of the fact that $h_1$ is proportional to $\cos\phi$,
  which breaks the $\phi \to \pi - \phi$ cancellation that normally
  causes the linear term in Eq.~(\ref{eq:onerap-costheta}) to disappear
  under $\phi$ integration.  }
Recalling from our discussion above that $f^{(1)}(p_{t,\elll})$
multiplies $A_1(p_{t,\elll})$ and that the latter goes at most as
$p_{t,\elll}$ for small $p_{t,\elll}$, the net effect on the cross
section will remain quadratic.
Thus we conclude that combining product hardness cuts with standard
lepton rapidity cuts, one obtains quadratic dependence of the full
acceptance on $p_{t,\elll}$, as long as one uses suitably wide
rapidity binning for the lepton pair around $y_t$.
This should ensure that the good perturbative behaviour of product
cuts illustrated in Fig.~\ref{fig:DY-basics} (right) carries over to
the case also with rapidity cuts.

The last question that we touch on concerns the design of cuts for
which the acceptance is independent of $p_{t,\elll}$ at low values of
$p_{t,\elll}$.
While a complete study is beyond the scope of this article, we can
already envisage one complication relative to the Higgs production
case, namely that $f^{(0)}$ is non-zero and multiplies a function
$A_0$ which is zero for $\ptll=0$, but has non-trivial (quadratic)
$p_{t,\elll}$ dependence beyond that point.
One solution to this issue is to design cuts that lead to $f^{(0)} =
0$.
For a $p_{t,\CS}$ cut that corresponds to a constraint
$\cos\theta < c$, a zero value of $f^{(0)}$ can be obtained by placing
an additional requirement
\logbook{}{DY-A0-term.nb, ``main part'' section, aSoln}
\begin{equation}
  \label{eq:DY-costheta-c1}
  \cos\theta > \bar c = \frac{-c_0 + \sqrt{4 - 3 c^2}}{2}\,,
\end{equation}
as can be verified by integrating $h_0$ in the range $\bar c
 < \cos\theta < c$. 
This gives a non-zero range for
$c > \sqrt{1/3}$.
To illustrate what can be achieved, we take the \CBIH procedure,
algorithm~\ref{alg:EBI}, and supplement step~1 with the condition
that the event is also discarded if $\cos \theta_{\CS} < \bar c$,
calling this the \CBIHDY algorithm.
The results are shown in Fig.~\ref{fig:DY-summary} (right).
It is clear that there is a substantial reduction in the acceptance
relative to the product cuts.
However, typically Drell-Yan measurements at low and moderate $\ptll$
are not statistics limited, so one may anticipate that this would not
be an issue.\footnote{A more problematic issue might be that the Born
  acceptance goes to zero at $y_{\elll}\simeq 2$ rather than
  $2.4$.
  However, in the region of $y_{\elll} > y_t \simeq 1.42$, the property of
  independence of the harmonic acceptances on $\ptll$ is anyway lost
  and, as things stand, one might anyway prefer simple product cuts in
  that region. }
The characteristic that we intended to obtain, and that has been
obtained, is that at central $y_{\elll}$ rapidities, $f^{(0,1,2)}$ are
now all identically zero for $\ptll \lesssim 2\Delta$, and $f^{(u)}$
is independent of $\ptll$ in that same range.
This guarantees that the fiducial cross section at those rapidities is
independent of $\ptll$ up to roughly $2\Delta$, within the
approximation that the harmonic decomposition of
Eq.~(\ref{eq:DY-decomposition}) completely describes the cross section.

Clearly there is scope for further investigation of the Drell-Yan
process, both resonant and non-resonant, but the material presented
here provides at least some of the elements that one might wish to
consider and expand on in such a study.

\section{Conclusions}
\label{sec:conclusions}

In this article, we have seen that current widely used cuts for
two-body collider processes can have severe consequences for
perturbation theory, leading to contributions that diverge factorially
as one goes to higher orders.
Unlike the renormalon-induced factorial divergences that are an
expected feature of perturbation theory but set in at very high
orders, the structures that we have observed here set in early, cf.\
Eq.~(\ref{eq:illustrative-asym-expansion}) in a simple approximation
for the all-order structure of the $H\to \gamma\gamma$ fiducial cross
section.
These problems are clearly visible in recent full N3LO fiducial
calculations~\cite{Chen:2021isd,Billis:2021ecs} and they are
associated also with fixed-order calculations' strong sensitivity to
unphysically low momentum scales.
In more general terms, we expect such problems to arise whenever one
integrates a power of some quantity $v$ in a phase-space region where
the perturbative series involves exponentiating double logarithms of
$v$.
It would be valuable to develop a more systematic understanding of how
and where such problems may appear.

In the Higgs and Drell-Yan cases, the poor perturbative behaviour can
be directly traced back to the linear dependence of the
$H\to \gamma\gamma$ acceptance on the Higgs boson transverse momentum
for low $\ptH$ values, which is a feature both of symmetric cuts and
the asymmetric cuts that have come to replace them in many contexts.
One possible solution is to supplement fixed-order calculations with
suitable resummations, as outlined nearly twenty years ago in the
dijet context~\cite{Banfi:2003jj} and advocated recently for the Higgs
case in Ref.~\cite{Billis:2021ecs}.
For legacy fiducial measurements, this is probably the only viable
solution.

\begin{table}
  \centering
  \begin{tabular}{ccccc}
    \toprule
    Cut Type   & cuts on & small-$\ptH$ dependence &  $f_n$ coefficient  & $\ptH$ transition\\
    \midrule 
    \hyperref[sec:symmetric-cuts]{symmetric}   & $p_{t,-}$ & linear    & $+2s_0/(\pi f_0)$   & $\phantom{\sqrt{2}}$none\\[1pt]
    \hyperref[sec:asymmetric-cuts]{asymmetric} & $p_{t,+}$ & linear    & $-2s_0/(\pi f_0)$   & $\phantom{\sqrt{2}}\phantom{2}\Delta$\\[1pt]
    \hyperref[sec:sum-cuts]{sum}               & $\frac12(p_{t,-}+p_{t,+})$& quadratic & $(1+s_0^2)/(4 f_0)$ & $\phantom{\sqrt{2}}2\Delta$\\[1pt]
    \hyperref[sec:product-cuts]{product}       & $\sqrt{p_{t,-}+p_{t,+}}$ & quadratic & $s_0^2/(4 f_0)$     & $\phantom{\sqrt{2}}2\Delta$\\[1pt]
    \hyperref[sec:staggered-cuts]{staggered}   & $p_{t,1} $& quadratic & $s_0^4/(4 f_0^3)$ & $\phantom{2\sqrt{2}}\Delta$\\[1pt]
    \hyperref[sec:cuts-collins-soper]{Collins-Soper} & \hyperref[eq:ptCS]{$p_{t,\CS}$} & none & --- & $\phantom{\sqrt{2}}2\Delta$\\[1pt]
    \hyperref[sec:extended-cs-cuts]{\CBIH} & $p_{t,\CS}$  & none & --- & $2\sqrt{2}\Delta$\\[1pt]
    \midrule
    \hyperref[sec:one-rap-cut]{rapidity}       & $y_\gamma$ & quadratic & $f_0 s_0^2/2$ & \\[1pt]
    \bottomrule
  \end{tabular}
  \caption{Summary of the main hardness cuts, the variable they cut on
    at small $\ptH$, and the small-$\ptH$ dependence of the acceptance.
    For linear cuts $f_n \equiv f_1$ multiplies $\ptH/\mH$, while for
    quadratic cuts $f_n \equiv f_2$ multiplies $(\ptH/\mH)^2$ (in all
    cases there are additional higher order terms that are not shown).
    For a leading threshold of $p_{t,\cut}$, $s_0 = 2p_{t,\cut}/\mH$
    and $f_0 = \sqrt{1-s_0^2}$, while for the rapidity cut $s_0 = 1/\cosh(\yH
    - y_\cut)$.
    For a cut on the softer lepton's transverse momentum of
    $p_{t,-} > p_{t,\cut}-\Delta$, the right-most column indicates the
    $\ptH$ value at which the $p_{t,-}$ cut starts to modify the behaviour of the
    acceptance (additional $\order{\Delta^2/\mH}$ corrections not
    shown).
    For the interplay between hardness and rapidity cuts, see
    sections~\ref{sec:two-rap-cut}, \ref{sec:ptrap-cut} and
    \ref{sec:cbi-hardness-rap}. 
  }
  \label{tab:accceptance-summary}
\end{table}

For future measurements, however, we argue that the choice of cuts
should be revisited, so that one can fully retain the power and
conceptual simplicity of fixed-order calculations.
A summary of the behaviour of different cuts is given in
Table~\ref{tab:accceptance-summary}. 
A straightforward way of eliminating linear $\ptH$ dependence is to
replace a cut on the higher-$p_t$ photon with a cut on the sum or
product of the two photon transverse momenta.
This leaves just a quadratic dependence on $\ptH$, significantly
reducing the problems of convergence and low-$\ptH$ sensitivity in
fixed-order perturbative predictions.
The clearest illustration of the impact is perhaps the comparison of
Fig.~\ref{fig:minpth-impact} for asymmetric cuts with with
Fig.~\ref{fig:minpth-impact-product} for sum and product cuts (all
using an N3LL approximation for the perturbative series).
Ultimately, product cuts seem preferable to sum cuts because their
residual quadratic dependence is smaller.
Combining product (or sum) hardness cuts with rapidity cuts leaves the
conclusions unchanged, so long as any rapidity bins are kept
reasonably wide around certain critical Higgs rapidities, cf.\
Eq.~(\ref{eq:critical-yH}).

It turns out that it is also possible to design cuts whose acceptance
is independent of $\ptH$ at low and moderate values of $\ptH$.
One core element is to replace the usual higher $p_t$ cut with a cut
on the Collins--Soper angle, which is explicitly invariant under
transverse boosts.
Residual $\ptH$ dependence associated with the lower $p_t$ cut and
rapidity cuts can then be addressed with a compensation mechanism
between different regions of decay phase space for each given Higgs
kinematic point, a mechanism that also provides a way to enhance the
high-$\ptH$ acceptance.
The resulting algorithm, dubbed \CBIHR, while not as simple as cutting
on the product of transverse momenta, can be easily encapsulated in
code, which we make available together with this paper.\footnote{It is
  available from \url{https://github.com/gavinsalam/two-body-cuts}.}
Its performance is illustrated as a function of Higgs rapidity and
transverse momentum in Fig.~\ref{fig:CBIHR-performance}, with the
white areas of the lower panel indicating independence on $\ptH$ (this
is to be compared with corresponding plots for asymmetric and product
cuts in Fig.~\ref{fig:all-ATLAS-cuts-eff-derivatives}).
Having an acceptance that is independent of transverse momentum is
potentially valuable not just for the stability of perturbative
predictions, but also for experimental determinations of more
inclusive cross sections, which would then be less reliant on perturbative
or standard-model assumptions about the shape of the $\ptH$
distribution.

We believe that there are several directions where further work could
be of benefit.
One simple question concerns the optimisation of the thresholds used
experimentally.
Here we kept to existing values for the thresholds, which results in
an overall loss of acceptance of the order of a few percent
(backgrounds will also be correspondingly reduced).
If the cuts of this paper are adopted, there may be advantages to adjusting those
thresholds, so as to simultaneously maximise signal significances and
minimise experimental and perturbative systematics (for example,
one might investigate lowering the main cut from $0.35\mH$ or $\mH/3$
to $0.30\mH$, which raises the overall acceptance).
Another avenue for investigation would be to develop a more robust
understanding of the structure of the factorial divergences and of the
power of $(\Lambda/Q)$ that characterises the unavoidable residual
ambiguity in truncated perturbation theory with various cuts.
One might also wish to develop an understanding of how perturbative
and genuine non-perturbative uncertainties scale for matched-resummed
calculations with various cuts.

Finally, we believe it would be worthwhile extending our analyses to a
wider range of processes.
We sketched how this could be done for $Z$ production, where existing
cuts generate small, but relevant, perturbative convergence issues and
where product cuts can bring a significant improvement, cf.\
Fig.~\ref{fig:DY-basics} right.
A clear next step would be to consider continuum $2\to2$ processes,
where the question is not just of cuts, but also of the variables used
to bin the distributions and the interplay with a steeply falling
spectrum.

We look forward to further work on these and other cut-related topics.

\subsection*{Note added}

While this work was being completed it was pointed out to us by
Alexander Huss (private communication), that a cut on the rapidity
difference between the two decay products could also be used as a
hardness cut and that its acceptance is free of both linear and
quadratic $\ptH$ dependence.
At low $\ptH$ we find an acceptance of
$f(\ptH) = f_0 + f_0/(8s_0^2) (\ptH/\mH)^4 + \ord6$.
In combination with a $p_{t,-}$ cut, its acceptance in the
$\ptH = 50{-}100\GeV$ range is intermediate between that of the
product and $p_{t,\CS}$ cuts shown in Fig.~\ref{fig:acceptances-ptCS},
while at yet higher $\ptH$ values, the acceptance approaches saturation
of the values allowed by the $p_{t,-}$ cut.
\logbook{}{see GavinsCode/c++/huss-deltay-checks/deltay-checks.pdf}

\section*{Acknowledgements}

We are grateful to Fabrizio Caola and Pier Monni for numerous
discussions throughout this work.
We also wish to thank to Pier Monni for providing the NNLL and N3LL results
from Refs.~\cite{Bizon:2017rah,Bizon:2018foh},
Alexander Huss for providing results from Ref.~\cite{Chen:2021isd}
and discussions about N3LO results with alternative cuts,
and Tancredi Carli for discussions on the early use of sum cuts,
and all of the above and additionally Silvia Ferrario Ravasio for
helpful comments on the manuscript.
We are also grateful to Marek Sch\"onherr and Frank Siegert for
assistance with the Sherpa program.
Finally, we wish to thank the referee for thoughtful suggestions.

This work was supported
by a Royal Society Research Professorship
(RP$\backslash$R1$\backslash$180112, GPS),
by the European Research Council (ERC) under the European Union’s
Horizon 2020 research and innovation programme (grant agreement No.\
788223, PanScales, GPS and ES), 
by the Science and Technology Facilities Council (STFC) under grant
ST/T000864/1 (GPS),
and, in the context of the KITP programme ``New Physics from Precision
at High Energies'', in part by the National Science Foundation under
Grant No.\ NSF PHY-1748958 (GPS).
%
%

\appendix

\section{Higher-order expansions of small-$\ptH$ acceptances}
\label{sec:higher-order-expansions}

Here, we provide the expansions up to fourth order in $\ptH/\mH$ for
the acceptances for all of the individual cuts discussed in the main
text.
It is convenient to introduce $u = \ptH/\mH$, and to write the results
in terms of the effective Born cut on $\sin\theta$, i.e.\
$\sin \theta > s_0 = 2p_{t,\text{cut}}/\mH$, and the Born acceptance
$f_0 = \sqrt{1-s_0^2}$.
We work with $s_0$ in the range $0<s_0<1$ such that the Born
acceptance is non-zero and not trivially fully inclusive.

For a cut on just the lower-$p_t$ photon, as in a symmetric cut, we have
\logbook{0f0a480}{see rapidity-pt-cuts.nb}
\begin{equation}
  \label{eq:appendix-min}
  f^{(\text{sym})} = 
  f_0
  -\frac{2 s_0}{\pi  f_0} u
  -\frac{s_0^4 }{4 f_0^3}u^2
  -\frac{\left(f_0^6+f_0^4-5 f_0^2+2\right) }{3 \pi  f_0^5 s_0} u^3
  -\frac{3 \left(3 f_0^2+5\right) s_0^6 }{64 f_0^7} u^4
  + \ord5\,.
\end{equation}
For a cut on just the higher-$p_t$ lepton, i.e.\ the relevant part of
the asymmetric cut, we find
\begin{equation}
  \label{eq:appendix-max}
  f^{(\text{asym})} =
  f_0
  +\frac{2 s_0}{\pi  f_0} u
  -\frac{s_0^4 }{4 f_0^3} u^2
  +\frac{\left(f_0^6+f_0^4-5 f_0^2+2\right) }{3 \pi  f_0^5 s_0} u^3
   -\frac{3 \left(3 f_0^2+5\right) s_0^6 }{64 f_0^7} u^4
      + \ord5\,.
\end{equation}
For a cut on the sum of the two lepton transverse momenta we have
\begin{equation}
  \label{eq:appendix-sum}
  f^{(\text{sum})} =
  f_0
  +\frac{1 + s_0^2 }{4 f_0} u^2
    -\frac{\left(9 f_0^6-21 f_0^4+8\right)}{64 f_0^3 s_0^2} u^4
      + \ord6\,,
\end{equation}
while a cut on the product of their transverse momenta leads to
\begin{equation}
  \label{eq:appendix-prod}
  f^{(\text{prod})} =
  f_0
  +\frac{s_0^2}{4 f_0}u^2
  -\frac{\left(9 f_0^6-19 f_0^4-f_0^2+3\right) }{64 f_0^3 s_0^2}u^4
  + \ord6\,.
\end{equation}
The quadratic coefficient in the case of product cuts is always less
than half that in the case of the sum cuts, for any $0<s_0<1$.
The acceptance for a cut on just one of the photons (as with staggered
cuts, chosen in a way
that does not depend on the photon transverse momenta, e.g.\ the one
with larger rapidity), is given by \logbook{0f0a480}{see
  staggered-cuts.nb}
\begin{equation}
  \label{eq:appendix-single}
  f^{(\text{single})} =
  f_0-\frac{s_0^4}{4 f_0^3} u^2 -\frac{3  \left(3
      f_0^2+5\right) s_0^6}{64 f_0^7}u^4 + \ord6\,.
\end{equation}
This corresponds to the low-$\ptH$ behaviour discussed in the case of
staggered cuts, section~\ref{sec:staggered-cuts},
and is equal to the average of Eqs.~(\ref{eq:appendix-min}) and
(\ref{eq:appendix-max}), because half of the time the cut acts on the
harder photon, the other half of the time on the softer one.
Finally a single rapidity cut yields \logbook{0f0a480}{see rapidity-pt-cuts.nb}
\begin{equation}
  \label{eq:appendix-rapidity}
  f^{(<y_\text{cut})} =
  f_0 \left(1+ \frac1{2} s_0^2 u^2+\frac18 \left(3 f_0^4-2 f_0^2-1\right)
  u^4 + \ord6 \right)\,,
\end{equation}
where $s_0 = 1/\cosh(\yH - y_\text{cut})$ and $f_0 = \sqrt{1-s_0^2}$.

\section{Discussion of sensitivity of perturbative series to low
  $\ptH$ values}
\label{sec:interpretations}

In this appendix, we further discuss the basis for our statement that
certain cuts (in particular, those with linear $\ptH$ dependence) result in the
perturbative coefficients having strong sensitivity to unphysical,
non-perturbative transverse momentum scales.
For simplicity, we focus on the case without rapidity cuts.

Let us start with
Eq.~(\ref{eq:sym-sigma-fid-explicit-series-integral}).
In this simple case, there are two ways of viewing the structure of
the result.
One is that the factor $[f^\text{sym}(\ptH) - f^\text{sym}(0)\big]$ in
the integral comes from the expansion of the plus prescription in
Eq.~(\ref{eq:2b}), as would occur in a local subtraction method.
In this interpretation there is a fundamental ambiguity as concerns
what is placed in the integrand.
For example, more generally, one may choose to integrate
\begin{equation}
  \label{eq:appInterp-sub}
  \int_0 d\ptH \left[f(\ptH) \frac{d\sigma^{R}}{d\ptH }- f(0) \frac{d\sigma^{C}}{d\ptH} \right]\,.
\end{equation}
Here, $d\sigma^{R}/d\ptH$ is the ``real'' cross section in some
fixed-order perturbative expansion, by which we mean that it contains
all real and real-virtual contributions for which the Higgs boson has
non-zero $\ptH$.
For a given choice of renormalisation and factorisation scales, it is
uniquely defined at each order of perturbation theory.
Additionally, we have $d\sigma^{C}/d\ptH$, which serves as counterterm
in some local subtraction formalism.
In a Monte Carlo implementation of the subtraction formalism, the
events associated with the $d\sigma^{C}/d\ptH$ counterterm always have
a Higgs boson with zero transverse momentum.
There is considerable freedom in the choice of the distribution of the
counterterm $d\sigma^{C}/d\ptH$, so long as it cancels the
non-integrable divergences of the real cross section for small $\ptH$.
Different choices for the counterterm result in different compensating
additive corrections outside the integral, always proportional to
$f(0)$.
The final result for the fiducial cross section is unique at each
order in the coupling, but the interpretation of what is and is not
part of the integral is not unique.
In this sense, one may worry that any interpretation that the integral
is dominated by low $\ptH$ values would be equally ambiguous.
Note that this would be an interpretation in which we directly
calculate a fiducial cross section.

The second way of viewing the structure of the result is that we have
as a reference point the inclusive (total) cross section,
$\sigma_\text{tot}$, which we assume to have a reasonably well behaved
perturbative series.
The question we then ask is the following: how does the fiducial cross
section differ from the product of total cross section and Born
acceptance?
The answer to this question involves an integral 
\begin{equation}
  \label{eq:appInterp-diff}
  \sigma_\text{fid} - f_0 \sigma_\text{tot} =
  \int_0 d\ptH \left[f(\ptH) \frac{d\sigma^{R}}{d\ptH }- f(0) \frac{d\sigma^{R}}{d\ptH} \right]\,,
\end{equation}
where, for each value of $\ptH$, we ask how the acceptance differs
from the Born acceptance, and weight that difference with the
differential cross section for that $\ptH$.
Consequently we have $\frac{d\sigma^{R}}{d\ptH}$ in both terms in the
integral and there is no ambiguity related to a choice of
counterterm in a subtraction method.
Indeed, at order $\as^n$ (relative to the Born cross section),
Eq.~(\ref{eq:appInterp-diff}) can be evaluated from the N$^{n-1}$LO
cross section for the Higgs plus jet process, without any reference to
the purely virtual ($\ptH=0$) corrections to the inclusive (N$^{n}$LO)
Higgs production cross section.
In a Monte Carlo evaluation of Eq.~(\ref{eq:appInterp-diff}), both
terms are associated with events with the same non-zero $\ptH$.
Accordingly, we can meaningfully bin the contributions to the integral
as a function of $\ptH$ and determine which $\ptH$ ranges contribute
the most; or equivalently, as in Eq.~(\ref{eq:95percent-result}) and
Figs.~\ref{fig:minpth-impact} and \ref{fig:minpth-impact-product}, we
can ask how the integral converges if we place a lower limit on $\ptH$
of
$\epsilon$ and take $\epsilon \to 0$.
All the results that we present here can be interpreted in this second
way, and so we can meaningfully evaluate how much of the contribution
to the difference in Eq.~(\ref{eq:appInterp-diff}) comes from low
$\ptH$ values.

Three further comments are due.
One relates to the observation~\cite{Ebert:2018gsn,Ebert:2020dfc} that
the first corrections to $d\sigma^{R}/d\ptH$, beyond its leading
$1/\ptH$ power structure, are suppressed by two (relative) powers of
$\ptH$, i.e.\ overall they go as $\ptH$.
Let us now suppose that the differential counterterm in
Eq.~(\ref{eq:appInterp-sub}) also satisfies this property, as is the
case by construction in the projection-to-Born method, and as is
effectively the
case also for the integrated counterterm in $\ptH$-based phase-space
slicing methods~\cite{Catani:2007vq}, which normally involve just the
leading-power contribution.
In this situation, at small $\ptH$, the integrand of
Eq.~(\ref{eq:appInterp-sub}) is dominated by the relative order
$\ptH/\mH$ difference of $f(\ptH)$ and $f(\ptH)$, rather than by the
relative order $\ptH^2/\mH^2$ difference of $d\sigma^R/d\ptH$ and
$d\sigma^C/d\ptH$.
In such a situation the small-$\ptH$ dependence of the integrand in
Eq.~(\ref{eq:appInterp-sub}) will be almost identical to that of
Eq.~(\ref{eq:appInterp-diff}), and so conclusions about dependence on
the technical cutoff in the subtraction procedure will coincide with
conclusions about the physical dependence on small scales in
Eq.~(\ref{eq:appInterp-diff}).

Our second comment is that if the integral in
Eq.~(\ref{eq:appInterp-diff}) is sensitive to low scales, then the
proper definition of the perturbative coefficients needs to account
also for the effects of finite quark masses: not just the masses of
the charm and bottom quarks, but also those of the lighter quarks.

Our final comment is that an interpretation as in
Eq.~(\ref{eq:appInterp-diff}) can only be straightforwardly brought
into play if one can define a total cross section, free of fiducial
cuts.
This is not always possible, e.g.\ in $2\to2$ QCD scattering
processes, even though one may expect similar convergence and
low-scale sensitivity issues to be present for such processes.

\section{Remarks on perturbative asymptotics}
\label{sec:remarks-asysmptotics}

When we examined the $\alpha_s$ series expansion for the acceptance at
DL accuracy, e.g.\ Eq.~(\ref{eq:sym-sigma-fid-explicit-series}), it
was straightforward to identify the smallest term and express its
magnitude as a power of $\Lambda / Q$.
Beyond DL accuracy, this becomes more complicated.
Here we give our findings for the impact of two classes of subleading
term that we have identified as potentially being able to modify the
effective power of $\Lambda / Q$: those associated with the running
coupling (referred to as LL in the text) and those that account for
the vector sum of the transverse momenta of multiple emissions.
The analysis in this appendix is in many respects rather elementary,
and intended mainly to highlight where there are indications that a
simple analysis can provide a reliable understanding of the asymptotic
behaviour of the perturbative series and where, instead, further work
is needed.

To help the discussion, it is useful to introduce the shorthand
$\Sigma(L)$ for the fraction of the cross section where the transverse
momentum of a boson of mass $m$ is less than $e^{-L} m$. (In the main
text we used a definition of $e^{-L} m/2$, but the difference only
affects the overall normalisation of the perturbative series for the acceptance, not the
scaling of the smallest term).
For a process with two incoming legs, each carrying a colour factor
$C$, we have
\begin{equation}
  \label{eq:app-DL}
  \Sigma^{\DL}(L) = e^{-R^{\DL}(L)},\qquad R^{\DL} = \frac{2C \as L^2}{\pi}\,.
\end{equation}
The LL result is 
\begin{equation}
  \label{eq:app-LL}
  \Sigma^{\LL}(L) = e^{-R^{\LL}(L)},\qquad R^{\LL} = 2 C L
    r_1(\alpha_s L b_0)\,,
\end{equation}
with $r_1$ given in Eq.~(\ref{eq:dsigmaLL}).
A $p_t$-space resummation of the impact of multiple emission brings a
NLL modification of the result~\cite{Frixione:1998dw}
\begin{equation}
  \label{eq:app-LL-F}
  \Sigma^{\LL_\mathcal{F}}(L) =   \mathcal{F}(R')
  e^{-R^{\LL}(L)}\,,\qquad \mathcal{F}(R') = \frac{\Gamma(1-R'/2)}{\Gamma(1+R'/2)}\,,
\end{equation}
with $R' = \partial_L R^\LL(L)$.
Yet another possibility is to use a Fourier-transform ($b$-space)
approach~\cite{Parisi:1979se}
\begin{equation}
  \label{eq:app-LL-b}
  \partial_L \Sigma^{\LL_b}(L) = e^{-2L} \int_0^\infty bdb J_0(b e^{-L}) \Sigma^\LL(\ln b/\mathcal{B}_0)\,,
\end{equation}
where $\mathcal{B}_0 = 2e^{-\gamma_E}$, with $\gamma_E \simeq 0.577216$ the Euler
number.
For the perturbative expansion of the $b$-space formula we have made
use of the results in Refs.~\cite{Kulesza:2001jc,Bozzi:2005wk}.

\begin{figure}
  \centering
  \includegraphics[scale=0.74,page=1]{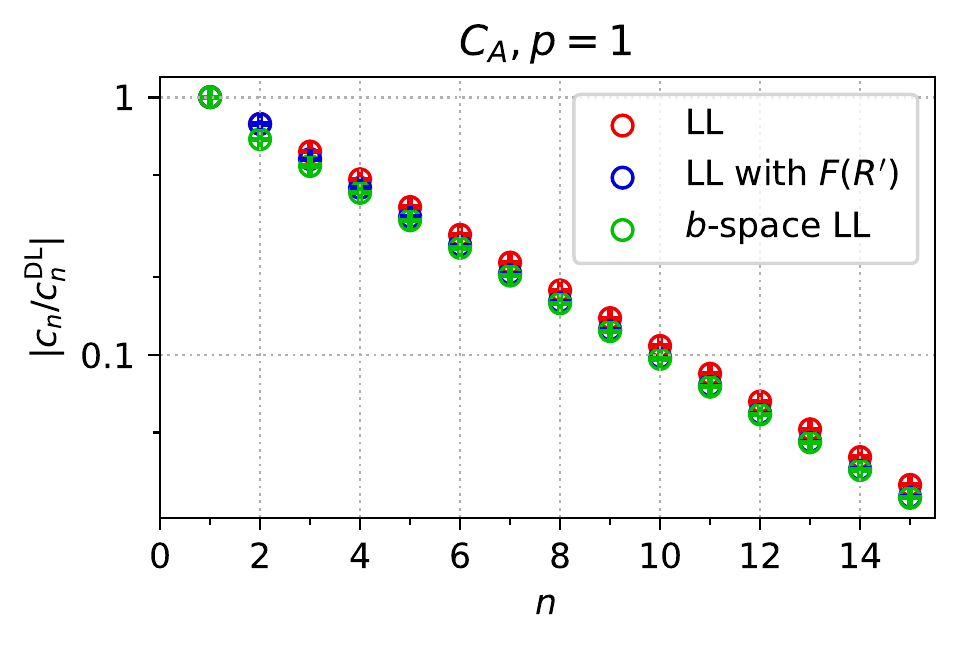}
  \includegraphics[scale=0.74,page=2]{2021-figs/series-behaviours-v2.pdf}
  \includegraphics[scale=0.74,page=3]{2021-figs/series-behaviours-v2.pdf}
  \includegraphics[scale=0.74,page=4]{2021-figs/series-behaviours-v2.pdf}
  \caption{The coefficients of the series expansions for $\delta_p$,
    Eq.~(\ref{eq:app-fiducial-series}), using a variety of
    approximation for $\Sigma$,
    Eqs.~(\ref{eq:app-LL})--(\ref{eq:app-LL-b}), normalised to the
    coefficients from the DL series.
    The four plots show different combinations of colour factor and
    $p$.
    Circles with a central plus indicate a positive ratio, circles
    with a central dot indicate a negative ratio.  }
  \label{fig:various-asymptotics}
\end{figure}

For each of the above approximations for $\Sigma(L)$, we can evaluate
the impact of a $p_t^p$ dependence of the acceptance on the fiducial
cross section, as 
\begin{equation}
  \label{eq:app-fiducial-series}
  \delta_p \equiv \frac{\sigma_{\text{fid}}}{f_0 \sigma_\text{tot}} -1
  = f_p \int_0^\infty dL \, e^{-pL} \partial_L \Sigma(L) = \sum_{n=1}^\infty \as^n c_n\,.
\end{equation}
Fig.~\ref{fig:various-asymptotics} shows the ratio of the $c_n$ as
obtained with each of Eqs.~(\ref{eq:app-LL})--(\ref{eq:app-LL-b}) to
the DL result for $c_n$.
A first observation is that in all cases, asymptotically, the DL
coefficients appear to be effectively rescaled by a factor $1/r^n$
\begin{equation}
  \label{eq:cn-rescaling}
  c_n \sim \frac{c_n^\DL}{r^n}\,,
\end{equation}
where $r$ is some constant.
Below, we will consider how such a rescaling can come about.
For now, however, we examine what we can learn from
Fig.~\ref{fig:various-asymptotics}.

A first point is that with a modification of the series as in
Eq.~(\ref{eq:cn-rescaling}), the expression for the perturbative order
at which one finds the smallest term in the perturbative series becomes
\begin{equation}
  \label{eq:smallest-term}
  n_{\min} \simeq |r| \frac{\pi p^2}{8 \alpha_s C}\,.
\end{equation}
The effective power scaling of the minimal term in the series becomes
\begin{equation}
  \label{eq:app-rescaled-power}
    \left(\frac{\Lambda}{Q}\right)^{|r|\frac{(11 C_A - 2n_f)p^2}{48 C}}\,.
\end{equation}
For $p=1$ (the upper row of Fig.~\ref{fig:various-asymptotics}) the
LL, LL$_\mathcal{F}$ and LL$_{b}$ results all yield compatible values of
$r$.
Specifically for the $p=1$, $C_A$ case we find $r\simeq 1.28$,
corresponding to an effective power of $(\Lambda/Q)^{0.205}$.
\logbook{}{see the usual bessel-games.nb}
For the $p=1$, $C_F$ case we find $r\simeq 2.1$,
corresponding to an effective power of $(\Lambda/Q)^{0.76}$.
The agreement between the different approximations for $\Sigma$
suggests that the resulting rescaling factors and effective powers
have a good chance of being representative of the actual asymptotic
structure of the perturbative series.

In contrast, for $p=2$, we find that the LL, LL$_\mathcal{F}$ and
LL$_{b}$ approximations yield quite different results.
We believe that the underlying reason is that the DL $p=2$ factorial
growth is weaker than for $p=1$, and so there is a greater chance that
artefacts of the various approximations for $\Sigma$ can more easily
come to dominate the expansion.
Specifically, the LL$_\mathcal{F}$ approximation has a well-known
spurious all-order divergence for $R'=2$, and integrating over the
expansion of such a divergence is bound to generate (spurious,
same-sign) factorial growth (careful investigation of
Fig.~\ref{fig:various-asymptotics} shows that $r$ is negative, i.e.\
the observed factorial growth is indeed of the same-sign variety).
The approach of Ref.~\cite{Monni:2016ktx} offers ways of working
around this issue, though the expansion of that approach to high
perturbative orders would require further work.
Meanwhile the LL$_b$ approximation is known to generate spurious
factorial growth in the coefficients of the distribution at finite
$p_t$~\cite{Frixione:1998dw}, a feature that we have explicitly
confirmed by integrating over a region of moderate $L$.%
\logbook{a6cfaa7ea2f3}{math-2021/pt_distribution.out}
As a result, we believe that it is unlikely that the stronger-than-DL
factorial growth seen with the $b$-space approach will be a genuine
feature of the true perturbative series.
Note that the modified logarithms used in Ref.~\cite{Bozzi:2005wk} may
well address this problem, but they bring more complicated expansion
expressions at high orders.
Overall the conclusion is that to obtain a reliable understanding the
asymptotics of the perturbative series for the $p=2$ case, further
more careful work is needed.

As a final point, we examine analytically how a rescaling such as
Eq.~(\ref{eq:cn-rescaling}) can arise.
Consider the coefficient of a term $\as^n L^{2n-\ell}$.
For $\ell=0$, the coefficient scales as $1/n!$, which is promoted to
$(2n)!/n!$ after the integration in
Eq.~(\ref{eq:app-fiducial-series}).
Naively one might expect terms with $\ell >0$ to be suppressed
because the integration over $L$ will only give an enhancement
$(2n-\ell)!$.
However this assumes that the coefficient of $\as^n L^{2n-\ell}$ has
the same factorial scaling as that of $\as^n L^{2n}$.
To see why this is not the case, we supplement the DL approximation
with just the term in the series expansion of $r_1(\lambda)$ in
Eq.~(\ref{eq:app-LL}) that is proportional to $b_0$ (and its
exponentiation), i.e.\ we consider the DL formula with the
replacement
\begin{equation}
  \label{eq:app-replacement}
  \frac{2C\as L^2}{\pi} \to
    \frac{2C\as L^2}{\pi} \left(1 + \frac{4}{3}\as b_0 L\right)\,.
\end{equation}
Exponentiating this, after some straightforward manipulation,%
\logbook{}{cuts-2021-parts-removed.tex}
one
obtains
\begin{equation}
  \label{eq:app-series-with-one-b0}
  \Sigma =
  \sum_{n=0}^\infty
           \frac{1}{n!}\left(-\frac{2C\as }{\pi}\right)^{n}
           \sum_{\ell=0}^{n/2}
           \frac{n!}{\ell! (n-2\ell)!}
           \left(-\frac{2\pi}{3} \frac{b_0}{C }\right)^\ell L^{2n-\ell}\,.
\end{equation}
The result for $\delta_p$ is then
\begin{equation}
  \label{eq:app-one-b0-deltap}
  \delta_p = 
    \frac{1}{2^p} \sum_{n=0}^\infty
    \frac{(2n)!}{n!}\left(-\frac{2C\as }{\pi p^2}\right)^{n}
    \sum_{\ell=0}^{n/2}
    \frac{1}{\ell!}
    \frac{n!}{ (n-2\ell)!}
    \frac{(2n-\ell)!}{(2n)!}
    \left(-\frac{2 \pi}{3} \frac{p b_0}{C }\right)^\ell \,.
\end{equation}
Next we observe that when $n \gg \ell$,
\begin{equation}
  \label{eq:4}
    \frac{n!}{ (n-2\ell)!}
    \frac{(2n-\ell)!}{(2n)!}
    \simeq
    \frac{n^{2\ell}}{(2n)^\ell}
    = \left(\frac{n}{2}\right)^\ell\,,
\end{equation}
from which we obtain
\begin{subequations}
  \begin{align}
    \delta_p
    &\simeq
      \frac{1}{2^p} \sum_{n=0}^\infty
      \frac{(2n)!}{n!}\left(-\frac{2C\as }{\pi p^2}\right)^{n}
      \sum_{\ell=0}^{n/2}
      \frac{1}{\ell!}
      \left(-\frac{n \pi}{3} \frac{p b_0}{C }\right)^\ell  \,,
    \\
    &\simeq
      \frac{1}{2^p} \sum_{k=0}^\infty
      \frac{(2k)!}{k!}\left(-\frac{2C\as }{\pi p^2}\right)^{k}
      \exp\left(-\frac{k \pi}{3} \frac{p b_0}{C }\right)\,,
  \end{align}
\end{subequations}
which is precisely of the form of Eq.~(\ref{eq:cn-rescaling}) with
\begin{equation}
  \label{eq:7}
  r = \exp\left(\frac{\pi}{3} \frac{p b_0}{C }\right)\,.
\end{equation}
For $p=1$ and $C = C_A$ this gives $r \simeq 1.24$, which is in the
same ballpark as the complete numerical result given above of
$r\simeq 1.28$.
For $p=1$ and $C = C_F$, one finds $r\simeq 1.6$, which is not so close
to the observed numerical result of $r=2.1$.
In both cases, however, we have made quite a number of approximations,
such as taking just the first non-trivial $b_0$ term as our starting
point, Eq.~(\ref{eq:app-replacement}), so it is perhaps unsurprising
that the analysis does not yield the full structure of the series.
Still, we believe that the analysis \emph{is} sufficient to motivate the
observed structural form of the scaling as written in
Eq.~(\ref{eq:cn-rescaling}) and observed in
Fig.~\ref{fig:various-asymptotics}.
%
%
%

\section{Remarks on defiducialisation}
\label{sec:defid-remark}

An alternative approach to the elimination of artefacts from cuts is
to adopt a defiducialisation procedure~\cite{Glazov:2020gza}.
In the case of Higgs production and decay, this is particularly simple
and one may write a defiducialised cross section as
\begin{subequations}
    \label{eq:defid}
  \begin{align}
    \label{eq:defid-a}
    \sigma_\text{defid}
    &=
      \int_{-\yH^\text{max}}^{+\yH^\text{max}}
      d\yH  \int_0^{\ptH^\text{max}} d\ptH
      \frac{d\sigma^{\text{fid}}}{d\yH d\ptH} \frac{1}{f(\yH,\ptH)}\,,
    \\
    \label{eq:defid-b}
    &\equiv 
      \int_{-\yH^\text{max}}^{+\yH^\text{max}}
      d\yH  \int_0^{\ptH^\text{max}} d\ptH
      \frac{d\sigma}{d\yH d\ptH}
      \,,
  \end{align}
\end{subequations}
where $d\sigma^{\text{fid}}/{d\yH d\ptH}$ is the differential 
cross section with some specific set of fiducial cuts, and
$f(\yH,\ptH$) is the acceptance with those cuts.
The meaning of Eq.~(\ref{eq:defid-a}) is that each event that passes
the cuts is binned with a weight $1/f(\yH,\ptH)$. 
Such an approach effectively yields a bin of a simplified template
cross section~\cite{Berger:2019wnu}, as made evident from
Eq.~(\ref{eq:defid-b}).
Care is needed with the choice of $\yH^\text{max}$ and
$\ptH^\text{max}$ so as to avoid regions where the acceptance is zero
or close to zero.
This implies that that $\yH^\text{max}$ should be kept well away from
the upper limit on the photon acceptance (for example, for a maximum
photon pseudorapidity of $2.37$, one might choose
$\yH^\text{max}=2$).
With cuts that remove slices of (pseudo)rapidity for the photons, a
$\ptH^\text{max}$ restriction may be necessary to avoid the
high-$\ptH$ region where one or other of the photons from the Higgs
decay is highly likely to be collimated into the slice.
Alternatively, one may remove the slice regions from the rapidity
integral.

To evaluate Eq.~(\ref{eq:defid-a}) in practice, $f(\yH, \ptH)$ can be
pre-tabulated.
If this is considered too cumbersome, one could instead evaluate
\begin{equation}
  \label{eq:defid-phi-sum}
  \sigma_\text{defid}
  =
  \int_{-\yH^\text{max}}^{+\yH^\text{max}}
  d\yH  \int_0^{\ptH^\text{max}} d\ptH
  \int d\phi_\CS
  \frac{d\sigma^{\text{fid}}}{d\yH d\ptH d\phi_\CS}
  \left[ \frac14 \sum_{i=1}^4 f(\yH,\ptH,\phi_i) \right]^{-1}\,.
\end{equation}
This equation is to be interpreted as follows: for each event, one
determines $\yH$, $\ptH$ and the Collins-Soper azimuth $\phi_\CS$.
One then evaluates the quantity in square brackets across the four
$\phi_i$ values as given in Eq.~(\ref{eq:phi-mirror-points}) and uses
this to determine the weight when binning the event.
The logic of this approach is that the quantity $f(\yH,\ptH,\phi_i)$
can be evaluated exactly, and relatively efficiently, with the help of
the code supplied with this article.
The sum over four $\phi_i$ values serves to avoid large fluctuations
in event weights (for example due to the region of low
$\phi_{\CS}$ values in the right-hand, $\ptH=100\GeV$, panel of
Fig.~\ref{fig:EBI-phasespace}, where the non-negotiable $p_{t,-}$ cut
causes the acceptance to be very close to zero).

A further variant is to defiducialise \emph{just} the $\ptH$
dependence.
In our view, the simplest such approach is the
following,\footnote{Inspired by a suggestion from the referee.}
\begin{subequations}
  \label{eq:defid-ptH}
  \begin{align}
    \label{eq:defid-ptH-a}
    \sigma_{\text{defid},\ptH}
    &=
      \int_{-\yH^\text{max}}^{+\yH^\text{max}}
      d\yH  \int_0^{\ptH^\text{max}} d\ptH
      \frac{d\sigma^{\text{fid}}}{d\yH d\ptH}  \frac{f(\yH,0)}{f(\yH,\ptH)}\,,
    \\
    \label{eq:defid-ptH-b}
    &\equiv 
      \int_{-\yH^\text{max}}^{+\yH^\text{max}}
      d\yH  \int_0^{\ptH^\text{max}} d\ptH
      \frac{d\sigma}{d\yH d\ptH} f(\yH,0)
      \,,
  \end{align}
\end{subequations}
where the weight assigned to each observed event is now
$\frac{f(\yH,0)}{f(\yH,\ptH)}$.
As with full defiducialisation, this may also be adapted to use a
weight that is $\phi$-dependent.

Note that defiducialisation is a rigorous and straightforward
procedure only for scalar decays.
Applications to vector-boson decays (as in the original proposal)
encounter the complication the angular distribution of the decay
products depends on the kinematics and production mechanism of the
vector boson in a non-trivial manner.
One might consider exploring an approximate defiducialisation, e.g.\
based just on the unpolarised acceptance, however it is not clear to
what extent this would be superior to the simple use of product cuts.

\bibliographystyle{JHEP} \bibliography{cuts}

\end{document}